\documentclass[prd,amsmath,amssymb,aps,twocolumn]{revtex4-1}

\usepackage{subcaption}
\usepackage[dvipsnames]{xcolor}
\usepackage{graphicx}
\usepackage{xfrac}
\usepackage{todonotes}
\usepackage{comment}
\usepackage[colorlinks=true,citecolor=blue]{hyperref}
\usepackage{ulem}
\usepackage{soul}
\setstcolor{Green}
\usepackage{cancel}
\usepackage{mathrsfs}
\DeclareMathAlphabet{\mathcalligra}{T1}{calligra}{m}{n}
\usepackage[utf8]{inputenc}

\newcommand{\be}{\begin{equation} }
	\newcommand{\ee}{\end{equation}}
\newcommand{\bes}{\begin{equation*} }
	\newcommand{\ees}{\end{equation*}}
\newcommand{\bea}{\begin{eqnarray} }
	\newcommand{\eea}{\end{eqnarray}}
\newcommand{\beas}{\begin{eqnarray*} }
	\newcommand{\eeas}{\end{eqnarray*}}
\newcommand{\ba}{\begin{align} }
	\newcommand{\ea}{\end{align} }
\newcommand{\bas}{\begin{align*} }
	\newcommand{\eas}{\end{align*} }

\usepackage{upgreek}
\usepackage{stmaryrd}
\usepackage{caption}

\usepackage{mathtools}

\begin{document}
	\title{Vacuum polarization and renormalized stress-energy tensor of spherical thin shells}
	\author{Julio Arrechea}
	\email{julio.arrechea@sissa.it}
    \affiliation{SISSA, International School for Advanced Studies,
     via Bonomea 265, 34136 Trieste, Italy}
    \affiliation{INFN Sezione di Trieste,
     via Valerio 2, 34127 Trieste, Italy} 
    \affiliation{IFPU, Institute for Fundamental Physics of the Universe, via Beirut 2, 34014 Trieste, Italy}
	
	\author{Cormac Breen}
	\email{cormac.breen@tudublin.ie}
	\affiliation{School of Mathematics and Statistics, Technological University Dublin, Grangegorman, Dublin 7, Ireland}

\author{Adrian Ottewill}
	\email{adrian.ottewill@ucd.ie}
	\affiliation{School of Mathematics and Statistics, University College Dublin, Belfield, Dublin 4, Ireland}

 \author{Lorenzo Pisani}
	\email{lorenzo.pisani2@mail.dcu.ie}
	\affiliation{
		Center for Astrophysics and Relativity, School of Mathematical Sciences, Dublin City University, Glasnevin, Dublin 9, Ireland}
	
\author{Peter Taylor}
	\email{peter.taylor@dcu.ie}
	\affiliation{
		Center for Astrophysics and Relativity, School of Mathematical Sciences, Dublin City University, Glasnevin, Dublin 9, Ireland}
		
\begin{abstract}
    We provide a thorough study of the properties of the Boulware vacuum in the spacetime of a spherical, static thin shell with a Minkowski interior. To this end, we calculate the renormalized vacuum polarization and stress-energy tensor of massless scalar fields via the extended-coordinate prescription, paying particular attention to their scaling as the shell approaches the black hole limit. Near the surface of the thin shell, we obtain the expected leading-order singular behavior of both quantities via two independent methods: a high-frequency approximation for the modes, and a weak-field approximation. At the center of the shell we find non-local, Casimir-like contributions that remain finite in the black hole limit, and whose backreaction effects we compute via the semiclassical Einstein equations. Away from these regions amenable to analytic treatment, we obtain numerical results for a wide range of shell compactnesses and field couplings. In the black hole limit, we show that the vacuum polarization and renormalized stress-energy tensor outside the shell quickly approach the ones generated by a Schwarzschild black hole, suggesting a possible universality in the vacuum outside highly compact horizonless objects. This work addresses the conceptual and technical aspects necessary for computing renormalized expectation values in matter configurations, laying the foundations for future explorations on the subject. 
\end{abstract}
\maketitle

\section{Introduction}

Quantum fields have profound implications in our understanding of expanding universes~\cite{Parker1968} and black holes~\cite{Hawking:1975}. Even in their vacuum states, quantum fields propagating on curved spacetimes exhibit zero-point energies that cannot be fully renormalized away. These zero-point energies acquire the form of vacuum polarization clouds and energy/particle fluxes~\cite{ChristensenFulling1977,Fulling:1977jm,Candelas1980}, phenomena that have been studied in detail in the last decades~\cite{CandelasHoward:1984,AHS1995, LeviOri:2016b}. In eternal spherical black holes, for example, three vacuum states---Hartle-Hawking~\cite{HartleHawking:1976}, Unruh~\cite{Unruh:1976}, and Boulware~\cite{Boulware:1975}---are usually considered. Each of them describes a different physical situation, leading to distinct arrangements of vacuum polarization and energy fluxes throughout the spacetime, across the event horizon, and at infinity.  

The Hartle-Hawking state is stationary and regular across the event horizon, but it fills the spacetime with a thermal bath that extends to infinity, requiring the black hole to be enclosed within an isolating surface or box~\cite{York1984}. The Unruh state reproduces the late-time limit of a black hole formed via gravitational collapse. It is thus regular across the future event horizon and describes a thermal flux of particles at future null infinity. The Boulware vacuum, being the state consistent with asymptotic flatness (it reduces to the Minkowski vacuum at infinity) and staticity, becomes singular at the event horizon. For this reason, it is identified as the natural vacuum state in horizonless equilibrium configurations like stars~\cite{Hiscock1988, Numajiri:2024qgh} (see the recent work~\cite{Saffin:2026tvg}) and thin shells. Due to the greater complexities introduced by classical matter, little attention has been devoted to analyzing the properties of the Boulware vacuum in its natural, self-consistent setting. In addition, the Boulware vacuum violates all pointwise energy conditions on the Schwarzschild exterior~\cite{Visser1996b}, which has led to the study of horizonless alternatives to black holes sustained by the repulsion induced by their own vacuum polarization effects~\cite{Barcelo:2009tpa}. Recently, these were also seen to
contribute to the relaxation of general-relativistic compactness bounds~\cite{Arrechea:2023upd,Arrecheaetal2023,Reyes2023}, allowing matter to withstand collapse in situations that are classically forbidden. The present work aims to set in motion investigations in this direction by examining the simplest scenario in which matter can be placed arbitrarily close to the black hole limit: a spherical thin shell in equilibrium with an empty (Minkowski) interior. To this end, we will apply the recently developed extended-coordinate prescription~\cite{taylorbreen:2017,arrecheaetal2025} to calculate the expectation value in the Boulware vacuum of the renormalized vacuum polarization and stress-energy tensor of scalar fields.

Doing quantum field theory in spacetimes with thin-layer boundaries immediately brings up problems. As was shown in~\cite{Deutsch:1978sc,Kennedy:1979ar}, the presence of sharp boundaries induces divergences in expectation values that depend (to leading order and in flat spacetime) on the intrinsic properties of the boundary: its extrinsic curvature (tensor and scalar), induced metric, normal vectors, and the boundary condition satisfied by the field. Similarly, a distributional thin shell separating two spacetime regions introduces an arbitrarily small scale that coexists with the continuous medium that is the quantum field. The sensitivity of the field to arbitrarily small scales makes it so that all modes feel the presence of the boundary. While sharp boundaries are perfectly well-defined in the classical theory, they generate contributions to the quantum stress-energy that diverge as the boundary is approached and that cannot be renormalized away, as they are not the standard ultraviolet divergences to be subtracted by covariant renormalization prescriptions~\cite{Christensen:1976}. Sharp boundaries are ubiquitous in general relativistic matter models, like constant-density stars~\cite{Schwarzschild:1916ae}, the Oppenheimer-Snyder dust collapse model~\cite{Oppenheimer:1939ue}, and some black hole mimicker proposals like gravastars~\cite{Mazur:2001fv}.
Thus, a complete understanding of the pathologies they induce in quantum theory is necessary to postulate more realistic extensions of these models, as attempted for example in~\cite{Calmet:2025sck}. 

The renormalized stress-energy tensor (RSET) is the central element of the semiclassical theory of gravity. This theory considers the backreaction effects of the RSET  on the classical metric $g_{\alpha\beta}$ via the semiclassical Einstein equations
\begin{align}\label{eq:semieqs}
	G_{\alpha\beta}+\Lambda\,g_{\alpha\beta}+\alpha_{1} H^{(1)}_{\alpha\beta}+\alpha_{2}\,H^{(2)}_{\alpha\beta}=8\pi\,\left(T^{\textrm{(cl)}}_{\alpha\beta}+\langle\hat{T}_{\alpha\beta}\rangle_{\textrm{ren}}\right),
	\end{align}
where $\Lambda$ is the cosmological constant, $G_{\alpha\beta}$ is the Einstein tensor and $\langle\hat{T}_{\alpha\beta}\rangle_{\textrm{ren}}$ is the RSET. The tensors $H^{(1)}_{\alpha\beta}$, $H^{(2)}_{\alpha\beta}$ are geometrical and include terms that are quadratic in the curvature; their inclusion arises through the point-splitting regularization process that yields the RSET. This regularization process allows for the calculation of the finite physical RSET alongside an infinite renormalization of the constants $\Lambda$, $\alpha_{1}$ and $\alpha_{2}$. A divergence in the RSET indicates the breakdown of semiclassical gravity as a valid approximation. In the case of thin shells, we expect this divergence to be halted by endowing the shell with a finite thickness or by letting it fluctuate quantum-mechanically~\cite{Ford:1998he}. 

The RSET combines both local and non-local contributions, although this split is not easy to identify due to the complex form that $\langle\hat{T}_{\alpha\beta}\rangle_{\textrm{ren}}$ takes in most situations. 
Roughly speaking, local contributions include, but are not limited to, terms that depend on an arbitrary lengthscale~\cite{BrownOttewill:1986}, being intrinsically ambiguous. Non-local contributions encode the state dependence, can become relevant in regions where local curvature invariants are small, and their form is independent of any renormalization ambiguities. The most robust predictions stemming from semiclassical gravity will be obtained in situations where said ambiguous contributions are either absent or negligible. Fortunately, this is the case for the thin shell spacetimes considered in this paper, since renormalization ambiguities vanish in the Schwarzschild and Minkowski spacetimes for massless fields. Furthermore, inside the shell, the entirety of local RSET contributions vanish and only purely non-local vacuum polarization effects remain. Near the center of the shell, these effects are analogous to the Casimir stress-energies produced inside spheres~\cite{Bordag:2001qi,Fulling:2003zx, Milton:2011kz}, with the differences that i) our thin shell is transparent and ii) it has a maximum compactness limit given by the Schwarzschild radius. We will show how, as the shell approaches its gravitational radius, all RSET components remain bounded at the center of the sphere. Fortunately, this prevents observers from being decimated by the backreaction of vacuum polarization effects caused by arbitrarily distant matter. It also suggests that the divergences characteristic of the Boulware state originate at the exterior of the shell, rather than propagating towards the interior region, although a more thorough analysis of the Boulware vacuum in the black hole limit would require going beyond the thin shell models considered in this work. 

This article is intended to be as self-contained as possible, serving as a guideline for future extensions to more realistic models. For readers interested in the technical framework required to calculate the RSET in thin shell spacetimes, we point them to Sections~\ref{sec:extcoord}---\ref{sec:num_app}. Readers more interested in results and physical lessons can focus on Sections~\ref{sec:analytic}---\ref{sec:conclusions} instead.
In particular, Section~\ref{sec:extcoord} presents the extended coordinate prescription adapted to Boulware states. Section~\ref{sec:thinshells} applies the thin shell formalism to the case of a spherical shell matching a Minkowski interior and a Schwarzschild exterior. Section~\ref{sec:num_app} details how we calculate the Euclidean modes on this thin shell spacetime and tests the convergence of the extended-coordinate prescription. Section~\ref{sec:analytic} deals with analytically approximating the near-surface divergences in the RSET and renormalized vacuum polarization (RVP) through two independent methods: a Wentzel–Kramers–Brillouin (WKB) expansion of the Euclidean modes, and a weak-field approximation. Section~\ref{sec:results} contains numerical results for the RVP and RSET. We validate numerically the analytic results from the previous section, and analyze these quantities both inside and outside the shell, and in the black hole limit. We conclude with some discussion and outlook in Section~\ref{sec:conclusions}.

\section{The extended coordinate prescription}
\label{sec:extcoord}

In this section, we summarize the key elements of the extended coordinate method for computing the renormalized vacuum polarization and the renormalized stress-energy tensor for a quantum scalar field in the Boulware vacuum state propagating on an arbitrary static, spherically symmetric and asymptotically flat background. This extended coordinate method is a mode-sum renormalization prescription, first developed for renormalizing quantum expectation values of field observables in the Hartle-Hawking state \cite{taylorbreen:2016, taylorbreen:2017, taylorbreenottewill:2021} and then later adapted to the case where the field is in the Boulware state \cite{arrecheaetal2025}. The computations are most conveniently performed using the Wick-rotated line-element obtained by complexifying the static time coordinate $t\to -i\,\tau$, yielding a line-element with Riemannian signature usually referred to as the Euclideanized metric. Moreover, the static nature of the metric guarantees that the Euclidean two-point function for the quantum scalar field satisfying appropriate boundary conditions is equivalent to the Feynman propagator computed on the Lorentzian line-element \cite{CandelasRaine1977}. 

In $(\tau, r,\theta,\phi)$ coordinates, we take the Euclidean line-element to be 
\begin{equation}\label{eq:sphsymmetric_Euclidean}
    ds^{2}=f(r)d\tau^2+h(r)dr^2+r^2d\Omega^2,
\end{equation}
where $f(r)$ and $h(r)$ are arbitrary functions of $r$ and $d\Omega^{2}=d\theta^{2}+\sin^{2}\theta\,d\phi^{2}$ is the line-element for the two-sphere $\mathbb{S}^{2}$. If the line-element (\ref{eq:sphsymmetric_Euclidean}) describes the geometry of a static black hole with event horizon located at $r=r_{+}$, then the Euclidean geometry would possess a conical singularity at $r=r_{+}$ unless the Euclidean time $\tau$ was wrapped up and identified modulo the inverse Hawking temperature $\beta=1/T_{\textrm{H}}$, that is, we would have to impose the further condition $\tau=\tau+\beta$ in order to have a regular geometry at the event horizon. This periodicity in imaginary time is inherited in the two-point function for the quantum scalar field on the Euclidean geometry resulting in a discrete frequency spectrum, which is a significant computational advantage. The Green's function which satisfies regularity boundary conditions at the event horizon then corresponds to the Hartle-Hawking state \cite{KayWald1991, HartleHawking:1976}, which corresponds to the quantum field in a thermal state at the Hawking temperature of the black hole, in other words, the black hole and the quantum field are in thermal equilibrium. This is simply the KMS condition \cite{Kubo1957, MartinSchwinger1959} in a black hole context.

In this paper, we are instead interested in quantizing a scalar field in the Boulware state. This is the state which corresponds to a vacuum as seen by a static observer in the asymptotically flat region of spacetime. Another viewpoint for the Boulware state is that it is a zero temperature state. This viewpoint is particularly useful since it again permits the use of Euclidean techniques. The difference between Boulware states and Hartle-Hawking states when defined on the Euclidean section is that the Euclidean time $\tau$ is not wrapped up and identified for the Boulware state. Moreover, this will mean that the frequency spectrum is continuous rather than discrete. 

For clarity of exposition, we focus here on the computation of the renormalized expectation of the vacuum polarization for a quantized scalar field $\hat{\phi}$ in the Boulware state, which we write as $\langle \textrm{B}| \hat{\phi}^{2}|\textrm{B}\rangle_{\textrm{ren}}$. For further details, see Ref.~\cite{arrecheaetal2025}. On the Lorentzian line-element, we can write the vacuum polarization for the field $\hat{\phi}(x)$ in the state $|\textrm{B}\rangle$ as
\begin{align}
\langle \textrm{B}| \hat{\phi}^{2}|\textrm{B}\rangle_{\textrm{ren}}=-i \lim_{x'\to x} \left\{\mathcal{G}_{\textrm{B}}(x,x')-\mathcal{K}(x,x')\right\}
\end{align}
where $\mathcal{G}_{\textrm{B}}(x,x')$ is the Feynman Green function for the scalar wave operator satisfying the Boulware boundary conditions and $\mathcal{K}(x,x')$ is a Hadamard parametrix for the scalar wave operator which is required to renormalize the Feynman Green function in the coincidence limit $x'\to x$. Expressing these in terms of the corresponding Euclidean quantities gives
\begin{align}
\label{eq:VPdef}
\langle \textrm{B}| \hat{\phi}^{2}|\textrm{B}\rangle_{\textrm{ren}}=\lim_{x'\to x} \left\{G_{\textrm{B}}(x,x')-K(x,x')\right\}
\end{align}
where $G_{\textrm{B}}(x,x')$ is the Euclidean two-point function satisfying appropriate boundary conditions for the Boulware state and $K(x,x')$ is a Hadamard parametrix for the Euclidean wave-operator. The Euclidean two-point function satisfies the Euclidean wave equation
\begin{align}
   ( \Box-\mu^{2}-\xi\,R)G_{\textrm{B}}(x,x')=-\delta(x,x'),
\end{align}
where $\Box$ is the d'Alembertian operator on the Euclidean line-element (\ref{eq:sphsymmetric_Euclidean}), $\mu$ is the field mass, $\xi$ is the coupling strength of the field to the background curvature and $R$ is the Ricci scalar. The inhomogeneous term is the covariant Dirac delta distribution. The solution admits a mode-sum representation
\begin{align}
	\label{eq:Gmodesum}
	G_{\textrm{B}}(x,x')=\frac{1}{8 \pi^2}\sum_{l=0}^{\infty}(2l+1)P_l(\cos\gamma)\nonumber\\
    \times\int_{-\infty}^{\infty}d\omega\,e^{-i\omega (\tau'-\tau)}g_{\omega l}(r,r'),
\end{align}
where $\cos\gamma=\cos\theta\cos\theta'+\sin\theta\,\sin\theta'\,\cos(\phi'-\phi)$ and $P_l(z)$ is the Legendre polynomial of the first kind. The bi-distribution  $g_{\omega l}(r,r')$ is the one-dimensional radial Green's function satisfying
\begin{align}
    \left\{\frac{d}{dr}\left(\frac{r^{2}\sqrt{f}}{\sqrt{h}}\frac{d}{dr}\right)-\sqrt{h\,f}\left[\frac{\omega^{2}r^{2}}{f}+V_{l}(r)\right]\right\}g_{\omega l}(r,r')\nonumber\\
    =-\delta(r-r'),
\end{align}
with
\begin{align}
     V_{l}(r)=l(l+1)+r^{2}(\mu^{2}+\xi\,R).
\end{align}
The solution is constructed by an ordered normalized product of homogeneous solutions
\begin{align}
    g_{\omega l}(r,r')=\frac{p_{\omega l}(r_{<})q_{\omega l}(r_{>})}{N_{\omega l}},
\end{align}
where $p_{\omega l}(r)$ is a homogeneous solution satisfying a boundary condition on the left boundary of the radial domain and $q_{\omega l}(r)$ is a homogeneous solution satisfying a boundary condition on the right boundary of the radial domain (usually infinity). We have adopted the ordering notation $r_{<}=\min\{r,r'\}$, $r_{>}=\max\{r,r'\}$. The normalization constant $N_{\omega l}$ is obtained by the Wronskian condition
\begin{align}
\label{eq:normalization}
    N_{\omega l}&=-\frac{r^{2}\sqrt{f}}{\sqrt{h}}\mathcal{W}\{p_{\omega l},q_{\omega l}\}\nonumber\\
    &=-\frac{r^{2}\sqrt{f}}{\sqrt{h}}\left(\frac{d\,q_{\omega l}}{dr}p_{\omega l}-\frac{d\,p_{\omega l}}{dr}q_{\omega l}\right).
\end{align}
We discuss the details of the computation of the radial modes $p_{\omega l}(r)$ and $q_{\omega l}(r)$ and of their normalization constant $N_{\omega l}$ in the context of thin shells in Subsec. \ref{subsec:eucl_modes}. 

We will denote the coordinate separation of the points $x'$ and $x$ by $\Delta x=x'-x$. Now it is clear that in the coincidence limit $\Delta x\to0$, or equivalently $x'\to x$, the two-point function $G_{\textrm{B}}(x,x')$ diverges; it involves, by definition, the square of an operator valued distribution $\hat{\phi}(x)$ evaluated at the same spacetime point. This is mathematically ill-defined and this is precisely why it is necessary to renormalize the two-point function by subtracting the Hadamard parametrix $K(x,x')$. In terms of the mode-sum (\ref{eq:Gmodesum}), this divergence does not manifest at the level of the individual modes but rather in the fact that the sum over $l$ and integral in $\omega$ do not converge whenever $x'=x$.

On the other hand, the divergence in the Hadamard parametrix $K(x,x')$ is explicitly geometrical. For a general Riemannian manifold in four dimensions, the Hadamard parametrix for the wave equation will have the following universal form \cite{WaldBook:1994}
\begin{align}
\label{eq:HadamardForm}
    K(x,x')=\frac{1}{8\pi^{2}}\!\!\left(\frac{\Delta^{1/2}(x,x')}{\sigma(x,x')}+V(x,x')\log\Big(\frac{2\sigma(x,x')}{\ell^{2}}\Big)\!\!\right)\!,
\end{align}
where $\sigma(x,x')$ is Synge's world function, while $\Delta^{1/2}(x,x')$ and $V(x,x')$ are symmetric, regular biscalars. These quantities only depend on the local geometry and the field parameters, and do not carry any global information, such as the details of the quantum state. The parameter $\ell$ is an arbitrary lengthscale required to make the argument of the logarithm dimensionless and is associated with part of the renormalization ambiguity. The terms involving $\Delta^{1/2}/\sigma$ and $V\log\sigma$ on the right hand side of Eq.~\eqref{eq:HadamardForm} are usually referred to as, respectively, the direct and tail parts of the Hadamard parametrix. 

The biscalar $\sigma(x,x')$ satisfies the differential equation
\begin{align}
\label{eq:sigmaeqn}   2\sigma=\sigma_{;\alpha}\sigma^{;\alpha},
\end{align}
where throughout we use the notation $\sigma_{;\alpha}\equiv \nabla_\alpha \sigma$. For $x'\ne x$, the world function $\sigma$ is  positive definite on a Riemannian manifold and zero only at coincidence.
The biscalar $\Delta^{1/2}(x,x')$ is the Van Vleck–Morette determinant, defined by the transport equation
\begin{align}
    \label{eq:Ueqn}    2\sigma^{;\alpha}\nabla_\alpha\Delta^{1/2}=(4-\Box\sigma)\Delta^{1/2},
\end{align}
with boundary condition
\begin{align}
    \lim_{x'\to x}\Delta(x,x')=1.
\end{align}
Finally, the biscalar $V(x,x')$ satisfies the homogeneous wave equation 
\begin{align}
\label{eq:Veqn}
    (\Box-\mu^2-\xi^2 R)V(x,x')=0.
\end{align}
Introducing the Hadamard ansatz allows us to construct a series solution of \eqref{eq:Veqn}
\begin{align}	V(x,x')=\sum_{p=0}^{\infty}V_{p}(x,x')\,\sigma^{p},
\end{align}
and substituting into \eqref{eq:Veqn}, we obtain the recurrence relations for $V_p$
\begin{subequations}
  \label{eq:Vkeqn}  
\begin{multline}
	(p+1)(2p+4)V_{p+1}+2(p+1)\sigma^{;\alpha}\nabla_{\alpha}V_{p+1}\\
	-2(p+1)V_{p+1}\Delta^{-\frac{1}{2}}\sigma^{;\alpha}\nabla_{\alpha}\Delta^{\frac{1}{2}}\\
	+(\Box-\mu^{2}-\xi\,R)V_{p}=0,
\end{multline}
together with the boundary condition
\begin{multline}
	2 V_{0}+2\sigma^{;\alpha}\nabla_{\alpha}V_{0}-2 V_{0}\Delta^{-\frac{1}{2}}\sigma^{;\alpha}\nabla_{\alpha}\Delta^{\frac{1}{2}}\\
	+(\Box-\mu^{2}-\xi\,R)U=0.
\end{multline}
\end{subequations}

The computational challenge of applying the Hadamard renormalization in practice, that is to say taking the limit in (\ref{eq:VPdef}), is that the divergence encoded in the Hadamard parametrix is purely geometrical: $\sigma(x,x')\to 0$ as $x'\to x$, whereas the divergence in the two-point function $G_{\textrm{B}}(x,x')$ manifests in the non-convergence of the modes. The extended coordinate method addresses the problem of subtracting the parametrix from the two-point function in a way that this coincidence limit can be meaningfully computed. Moreover, the resultant expression is rapidly convergent making it a very efficient approach.

The principle of the method is to write the Hadamard parametrix as a mode-sum in the same form as that of the two-point function and then perform the subtraction mode by mode \cite{taylorbreen:2016,taylorbreen:2017,taylorbreenottewill:2021,arrecheaetal2025}. This renormalized mode-sum will converge in the coincidence limit. The details have been presented and explained elsewhere so here we give only a brief overview. The method necessitates the introduction of a carefully chosen set of ``extended coordinates'' in which to expand the parametrix, which in this case are simply $\{s^{2}, \Delta \tau^{2},\Delta r\}$,  where
\begin{align} 
s^{2}=
 f(r)\,\Delta\tau^{2}+2 r^{2}(1-\cos\gamma).
\end{align}
Note that the dependence on the angles is contained in $s^{2}$ through $\gamma$. The choice of coordinates is adapted to the symmetry of the background metric. Then if we expand the Hadamard parametrix in these extended coordinates and take the partial coincidence limit $r'=r$ (or $
\Delta r=0$) for simplicity, we get an expansion of the form:
\begin{align}
\label{eq:hadamard_mode_sum_1}
    K(x,x')&=\frac{1}{8\pi^{2}}\Bigg(\sum_{a=0}^{m}\sum_{b=0}^{a}\mathcal{D}_{ab}^{(\textrm{r})}(r)\frac{\Delta\tau^{2a+2b}}{s^{2b+2}}\nonumber\\
&+\sum_{a=1}^{m}\sum_{b=1}^{a}\mathcal{D}_{ab}^{(\textrm{p})}(r)\Delta\tau^{2a-2b}s^{2b-2}\nonumber\\
&+\sum_{a=1}^{m-1}\sum_{b=0}^{a-1}\mathcal{T}^{(\textrm{r})}_{ab}(r)\frac{\Delta\tau^{2a+2b+2}}{s^{2b+2}} \nonumber \\
& +\sum_{a=0}^{m-1}\sum_{b=0}^{a}\mathcal{T}_{ab}^{(\textrm{l})}(r)s^{2a-2b}\Delta\tau^{2b}\log\left(\frac{s^{2}}{\ell^{2}}\right)\nonumber\\
&+\sum_{a=1}^{m-1}\sum_{b=0}^{a}\mathcal{T}_{ab}^{(\textrm{p})}(r)s^{2a-2b}\Delta\tau^{2b}\Bigg)+\mathcal{O}(\epsilon^{2m}\log\epsilon).
\end{align}
The integer $m$ here fixes the order of the expansion. Higher order expansions are necessary to produce rapidly convergent mode-sum representations of the renormalized expectation values. The minimum order of this expansion to guarantee a regular vacuum polarization is $m=1$. The expansion of the parametrix has been divided into five separate parts: two terms which are rational in the extended coordinates, two terms that are polynomial in the extended coordinates and terms involving $\log(s^{2}).$ The first few coefficients appearing here can be found in the appendix. What we are trying to achieve is a mode-sum representation of the terms appearing in this expansion. Focusing on the rational terms, we adopt an \textit{ansatz} of the form
\begin{align}
\label{eq:directmodesum}
    \frac{\Delta\tau^{2a+2b}}{s^{2+2b}}=\sum_{l=0}^{\infty}(2l+1)P_{l}(\cos\gamma)\int_{-\infty}^{\infty}d\omega \,e^{i\omega\Delta\tau}\Psi_{\omega l}(a,b|r),
\end{align}
for some yet to be determined functions $\Psi_{\omega l}(a,b|r)$ which we call regularization parameters. To determine the regularization parameters we invert this relationship using the completeness relations for the Legendre polynomials and the Fourier modes. It turns out that these can be obtained in closed form in terms of Special Functions. Again the details of the derivation are given in Refs.~\cite{taylorbreen:2016,taylorbreen:2017,taylorbreenottewill:2021,arrecheaetal2025}. Similarly, we can take as an \textit{ansatz} for the logarithmic terms
\begin{align}
\label{eq:logterms}
\Delta\tau^{2b}s^{2a-2b}&\log(s^{2})=\nonumber\\
&\sum_{l=0}^{\infty}(2l+1)P_{l}(\cos\gamma)\int_{-\infty}^{\infty}d\omega\,e^{i\omega\Delta\tau}\chi_{\omega l}(a,b|r),
\end{align}
where $\chi_{\omega l}(a,b|r)$ are the regularization parameters for the logarithmic terms. As before, we can invert this relationship using completeness relations and derive closed-form representations for $\chi_{\omega l}(a,b|r)$. Explicit expressions for the regularization parameters are deferred to Appendix \ref{app:reg_par}. The terms that are polynomial in the extended coordinates are kept in closed form and do not require a mode-sum representation; they cannot affect the high $l$ or high $\omega$ behavior of the renormalized mode-sum.

Putting the details together, we obtain the following mode-sum representation of the Hadamard parametrix (\ref{eq:hadamard_mode_sum_1}) expanded to order $m=1$
    \begin{align}
\label{eq:hadamard_mode_sum_2}
    K(x,x')&=\frac{1}{8\pi^{2}}\int_{-\infty}^{\infty}d\omega\,e^{i\omega\Delta\tau}\sum_{l=0}^{\infty}(2l+1)P_{l}(\cos\gamma)k_{\omega l}^{(1)}(r)\nonumber\\
    &+\frac{1}{8\pi^{2}}\Bigg\{\mathcal{D}_{11}^{(\textrm{p})}(r)+\frac{\mathcal{D}_{10}^{(\textrm{r})}}{f(r)}+\frac{\mathcal{D}_{11}^{(\textrm{r})}(r)}{f^{2}(r)}\nonumber\\
    &+\left(\log\left(\frac{f(r)}{\ell^{2}\lambda^{2}}\right)-2\gamma_{\textrm{E}}\right)\mathcal{T}_{00}^{(\textrm{l})}(r)\Bigg\}
    +\mathcal{O}(\epsilon^{2}\log\epsilon),
\end{align}
where $k_{\omega l}^{(1)}(r)$ captures the large $l$, large $\omega$ behavior of the modes $g_{\omega l}(r):=g_{\omega l}(r,r)$; the explicit form for $k_{\omega l}^{(1)}(r)$ is found in Appendix \ref{app:reg_par}. The parameter $\lambda$ appearing in the expression above is an arbitrary inverse lengthscale associated with an infra-red cut-off in the frequency integrals. The renormalized expectation values themselves are independent of the particular choice \cite{arrecheaetal2025}. For massive fields, it is natural to take $\lambda=\mu$, the mass of the field.

The renormalized vacuum polarization (RVP) is then computed by subtracting mode-by-mode the Hadamard parametrix from the Green function and then taking the coincidence limit yielding
\begin{align}
    \langle \textrm{B}|&\hat{\phi}^{2}|\textrm{B}\rangle_{\textrm{ren}}\coloneqq\lim_{x'\to x}\left(G_{\textrm{B}}(x,x')-K(x,x')\right)\nonumber\\
    =&\frac{1}{8\pi^{2}}\int_{-\infty}^{\infty}\sum_{l=0}^{\infty}(2l+1)\left(g_{\omega l}(r)-k_{\omega l}^{(1)}(r)\right)d\omega\nonumber\\
     &-\frac{1}{8\pi^{2}}\Bigg[\mathcal{D}_{11}^{(\textrm{p})}(r)+\frac{\mathcal{D}_{10}^{(\textrm{r})}(r)}{f}+\frac{\mathcal{D}_{11}^{(\textrm{r})}(r)}{f^2}\nonumber\\
     &+\left(\log\left(\frac{f(r)}{\ell^{2}\lambda^{2}}\right)-2\gamma_{\textrm{E}}\right)\mathcal{T}_{00}^{(\textrm{l})}(r)\Bigg].
\end{align}
The modes $k_{\omega l}^{(1)}(r)$ captures the leading and subleading large $l$ and large $\omega$ behavior of the modes $g_{\omega l}(r):=g_{\omega l}(r,r)$ so that the mode-sum above converges, albeit slowly. We can accelerate the convergence by taking higher-order expansions in the parametrix. This has the overall effect of introducing some additional analytical terms. The result for general expansion order is
\begin{align}
\label{eq:vacpolren}
    \langle \textrm{B}|\hat{\phi}^{2}|\textrm{B}\rangle_{\textrm{ren}}&=\frac{1}{8\pi^{2}}\int_{-\infty}^{\infty}\sum_{l=0}^{\infty}(2l+1)\left(g_{\omega l}(r)-k_{\omega l}^{(m)}(r)\right)d\omega\nonumber\\
    &-\frac{\sqrt{\pi}}{8\pi^{2}}\sum_{a=1}^{m-1}\sum_{b=0}^{a}\mathcal{T}_{ab}^{(\textrm{l})}(r)\,\frac{(a-1)!2^{2a}f^{a-b}}{\Gamma(-a+\tfrac{1}{2})\lambda^{2a}}\nonumber\\
     &-\frac{1}{8\pi^{2}}\Bigg[\mathcal{D}_{11}^{(\textrm{p})}(r)+\frac{\mathcal{D}_{10}^{(\textrm{r})}(r)}{f}+\frac{\mathcal{D}_{11}^{(\textrm{r})}(r)}{f^2}\nonumber\\
     &+\left(\log\left(\frac{f(r)}{\ell^{2}\lambda^{2}}\right)-2\gamma_{\textrm{E}}\right)\mathcal{T}_{00}^{(\textrm{l})}(r)\Bigg],
\end{align}
where $k^{(m)}_{\omega l}(r)$ now captures the large $l$, large $\omega$ behavior to any desired order, so that any order of convergence in the mode-sums could be in principle achieved. In practice, computing the regularization parameters for large $m$ can be computationally expensive and so there is a goldilocks zone which balances the rapidity of convergence and the computational cost of the regularization parameters. The explicit expression for $k^{(m)}_{\omega l}(r)$ is given in Appendix \ref{app:reg_par}.

This procedure is now straightforwardly extended to the computation of the renormalized stress-energy tensor (RSET) $\langle \textrm{B}|\hat{T}^{\alpha}{}_{\beta}|\textrm{B}\rangle_{\textrm{ren}}$. If we start by defining
\begin{align}
    W(x,x'):=G_{\textrm{B}}(x,x')-K(x,x'),
\end{align}
then the coincidence limit of $W(x,x')$ is just the vacuum polarization which we will now label as
\begin{align}
    \mathsf{w}(r)=\lim_{x'\to x}W(x,x')=\langle \textrm{B}|\hat{\phi}^{2}|\textrm{B}\rangle_{\textrm{ren}}.
\end{align}
Let us further denote the coincidence limit of two covariant derivatives of $W(x,x')$ at the point $x$ as
\begin{align}
       \mathsf{w}_{\alpha\beta}(x)\equiv\lim_{x' \to x}\left[W(x,x')_{;\alpha\beta}\right].
\end{align}
Then the components of the RSET can be written as \cite{taylorbreenottewill:2021}
\begin{align}
	\label{eq:RSETDef}
	\langle\textrm{B}| \hat{T}^{\alpha}{}_{\beta}|\textrm{B}\rangle_{\textrm{ren}} & = - \mathsf{w}^{\alpha}{}_{\beta}   -   (\xi-\tfrac{1}{2})\mathsf{w}^{;\alpha}{}_{;\beta}+ (\xi-\tfrac{1}{4}) \square \mathsf{w} \,\delta^{\alpha}{}_{\beta}  \nonumber\\
	&\qquad +\xi R^{\alpha}{}_{\beta} \mathsf{w} -  \frac{1}{8\pi^2} v_1 \delta^{\alpha}{}_{\beta},
\end{align}
where
\begin{align}
	v_1 &=\tfrac{1}{720}R_{\alpha\beta\gamma\delta}R^{\alpha\beta\gamma\delta}- \tfrac{1}{720}R_{\alpha\beta}R^{\alpha\beta}- \tfrac{1}{24}(\xi-\tfrac{1}{5})\square R\nonumber\\
	&\quad  + \tfrac{1}{8}(\xi-\tfrac{1}{6})^2R^2 + \tfrac{1}{4}\mu^2(\xi-\tfrac{1}{6}) R +\tfrac{1}{8}\mu^4,
\end{align}
is a local geometrical term necessary to include in the expression (\ref{eq:RSETDef}) in order to ensure conservation of the RSET~\cite{WaldBook:1994}.

It is clear from this expression that in order to compute the components of the RSET, all that remains is to be able to compute the components $\mathsf{w}_{\alpha}{}_{\beta}$. The time and angular components $\mathsf{w}^{\tau}{}_{\tau}$ and $\mathsf{w}^{\theta}{}_{\theta}=\mathsf{w}^{\phi}{}_{\phi}$ are most easily expressed in terms of derivatives of $\mathsf{w}(r)$ and mixed derivatives of $W(x,x')$ using Synge's rule~\cite{taylorbreenottewill:2021}
\begin{align}
 \mathsf{w}_{\alpha\beta}(x)=  \tfrac{1}{2} \mathsf{w}_{;\alpha\beta}(x) - \lim_{x \to x'}	\left[W(x,x')_{;\alpha'\beta}\right],
\end{align}  
where the mixed time and mixed angular derivatives are obtained by appropriately differentiating the mode sum expression (\ref{eq:hadamard_mode_sum_2}) to obtain two additional mode sums: 
\begin{align}\label{eq:gttren}
	[g^{\tau \tau'}&W_{,\tau\tau'}]=\nonumber\\&\frac{1}{8\pi^{2} f}\Bigg\{\int_{-\infty}^{\infty}\sum_{l=0}^{\infty}(2l+1)\omega^2\left(g_{\omega l}-k_{\omega l}^{(m)}\right)d\omega\nonumber\\
&-\sqrt{\pi}\sum_{\substack{a=0\\a\ne 1}}^{m-1}\sum_{b=0}^{a}\mathcal{T}_{ab}^{(\textrm{l})}(r)\,\frac{a!2^{2a}f^{a-b}}{\Gamma(\tfrac{1}{2}-a)(a-1)\lambda^{2a-2}}\Bigg\}\nonumber\\&
	+\frac{1}{4\pi^{2}}\Bigg[\mathcal{T}_{10}^{(\textrm{p})}+\mathcal{D}_{22}^{(\textrm{p})}+\frac{1}{f}(\mathcal{T}_{11}^{(\textrm{p})}+\mathcal{D}_{21}^{(\textrm{p})}) \nonumber\\&+\frac{\mathcal{D}_{20}^{(\textrm{r})}}{f^2}+\frac{\mathcal{D}_{21}^{(\textrm{r})}}{f^3}+\frac{\mathcal{D}_{22}^{(\textrm{r})}}{f^4}+\frac{\mathcal{T}_{10}^{(\textrm{r})}}{f^2} \nonumber\\&+(\mathcal{T}_{10}^{(\textrm{l})}+\frac{\mathcal{T}_{11}^{(\textrm{l})}}{f})(\log( f /(\ell^{2}\lambda^{2}))-2\gamma_{\textrm{E}}+3)\Bigg],
\end{align}
\begin{align}\label{eq:gphiphiren}
	[g^{\phi \phi'}&W_{,\phi\phi'}]=\nonumber\\&\frac{1}{16\pi^{2}r^{2}}\Bigg\{\int_{-\infty}^{\infty}\sum_{l=0}^{\infty}(2l+1)l(l+1)\left(g_{\omega l}-k_{\omega l}^{(m)}\right)d\omega\nonumber\\&+\sqrt{\pi}\sum_{a=2}^{m-1}\sum_{b=0}^{a}\mathcal{T}_{ab}^{(\textrm{l})}\,\frac{r^2(a-b)(a-2)!2^{2a}f^{a-b-1}}{\Gamma(\tfrac{3}{2}-a)\lambda^{2a-2}}\Bigg\}\nonumber\\&
    +\frac{1}{4\pi^{2}}\Bigg[\mathcal{D}_{22}^{(\textrm{p})}+\mathcal{T}_{10}^{(\textrm{p})}-\frac{\mathcal{D}_{20}^{(\textrm{r})}}{f^2}-2\frac{\mathcal{D}_{21}^{(\textrm{r})}}{f^3}-3\frac{\mathcal{D}_{22}^{(\textrm{r})}}{f^4}\nonumber\\&-\frac{\mathcal{T}_{10}^{(\textrm{r})}}{f^2} +\frac{\mathcal{T}_{11}^{(\textrm{l})}}{f} +\mathcal{T}_{10}^{(\textrm{l})}(\log( f /(\ell^{2}\lambda^{2}))-2\gamma_{\textrm{E}}+1)\Bigg].
\end{align}

Finally, we need to be able to compute $\mathsf{w}^{r}{}_{r}$. By using the wave equation satisfied by $W(x,x')$ we can write $\mathsf{w}^{r}{}_{r}$ as~\cite{BrownOttewill:1986}
\begin{align}
	\mathsf{w}^{r}{}_{r}  = -\mathsf{w}^{\tau}{}_{\tau}-\mathsf{w}^{\theta}{}_{\theta}-\mathsf{w}^{\phi}{}_{\phi} +  \xi R \mathsf{w}  +\mu^2 \mathsf{w}  -\frac{3}{4\pi^2}v_1.
\end{align}
Since we have mode-sum representations of all the terms on the right-hand side, except for the $v_{1}$ term which is a trivial geometrical expression, we have all the ingredients required to efficiently and accurately compute the RSET components.

\section{Thin shells in equilibrium}
\label{sec:thinshells}
\subsection{General formalism}
The spacetime we will consider is that of a spherical and static thin shell that connects a Minkowski interior with a Schwarzschild exterior, though much of the essential details can also be applied to non-vacuum interior spacetimes such as stars. The Minkowski interior is of particular interest as any curvature-dependent terms in the RVP and RSET will vanish, isolating the purely non-local nature of the contribution coming from the boundary conditions defining the quantum state and contributions coming from the matching across the shell. 

To make this article as self-contained as possible, we briefly review here the thin shell formalism presented in~\cite{Israel:1966rt, Poisson:2009pwt}. We will specialize for the case under consideration at the end of this Section. We start by considering two spherically symmetric spacetimes 
\begin{align}
    ds^{2}_{-}=&
    -f_{-}(r)dt_{-}^{2}+h_{-}(r)dr^{2}+r^2d\Omega^{2},\quad r < r_{0}
    \nonumber\\
    ds^{2}_{+}=&
    -f_{+}(r)dt_{+}^{2}+h_{+}(r)dr^{2}+r^2d\Omega^{2}, \quad r > r_{0}
\end{align}
matched at the spherical timelike boundary $r=r_{0}$. In order for this layer to be a solution to the Einstein equations, it must satisfy Israel-Darmois junction conditions~\cite{Darmois1927,Israel:1966rt}. We introduce the geometrical quantities that allow us to characterize the junction conditions. We define coordinates intrinsic to our hypersurface, $\zeta^{A}=\{\uptau,\theta,\phi\}$ where $\uptau$ here is a proper time coordinate (not to be confused with Euclidean time $\tau$). The basis vectors tangent to the hypersurface of the boundary are thus
\begin{equation}
e^{\alpha(\pm)}_{A}=\frac{\partial x^{\alpha}_{\pm}}{\partial \zeta^{A}}.
\end{equation}
Explicitly, we have
\begin{align}
    e^{t(\pm)}_{A}=
    &
    \left(\frac{\partial t_{\pm}}{\partial \uptau},0,0\right)=\left(\frac{1}{\sqrt{f_{\pm}(r)}},0,0\right)\nonumber\\
    e^{r(\pm)}_{A}=
    &
    \left(0,0,0\right),\quad e^{\theta(\pm)}_{A}=
    \left(0,1,0\right),\quad
    e^{\phi(\pm)}_{A}=
    \left(0,0,1\right).
\end{align}
The unit spacelike vectors normal to this surface are 
\begin{equation}
n_{\alpha}^{(\pm)}=\left(0,\sqrt{h_{\pm}(r)},0,0\right).
\end{equation}

Now the first Israel-Darmois junction condition demands continuity of the induced metric \mbox{$h_{AB}^{(\pm)}=g_{\alpha\beta}e^{\left(\pm\right)\alpha}_{A}e^{\left(\pm\right)\beta}_{B}$} across the boundary, i.e.,
\begin{align}
   \llbracket h_{AB}\rrbracket =h^{(+)}_{AB}-h^{(-)}_{AB}=0
\end{align}
were here and throughout we have adopted the notation
\begin{align}
    \llbracket \mathcal{O}\rrbracket\equiv \mathcal{O}^{(+)}-\mathcal{O}^{(-)}.
\end{align}
This condition translates into the following relation between the interior and exterior time coordinates
\begin{equation}
\label{eq:timecoord}
    \left(\frac{dt_{-}}{d\uptau}\right)^{2}=\frac{f_{+}(r_{0})}{f_{-}(r_{0})}\left(\frac{dt_{+}}{d\uptau}\right)^{2},
\end{equation}
making $g_{tt}$ continuous at $r=r_{0}$, but not differentiable. The line-element on the hypersurface is then
\begin{align}
ds^{2}=h_{AB}d\zeta^{A}d\zeta^{B}=-d\uptau^{2}+r_{0}^{2}d\Omega^{2}.
\end{align}

Since the first junction condition is a constraint on the induced metric, the function $h(r)$ can still exhibit jumps. Constraints on the jump in $h(r)$ is specified by the second junction condition. First we define the extrinsic curvature of the hypersurface as
\begin{equation}
K^{(\pm)}_{AB}=n_{\alpha;\beta}e^{(\pm)\alpha}_{A}e^{(\pm)\beta}_{B},\quad K^{(\pm)}=K^{(\pm)}_{AB}h^{AB}.
\end{equation}
Explicitly, we have
\begin{align}
    K_{\uptau\uptau}^{(\pm)}=
    &
    -\frac{1}{2}\frac{f_{\pm}'(r_{0})}{f_{\pm}(r_{0})\sqrt{h_{\pm}(r_{0})}},\nonumber\\
    K_{\theta\theta}^{(\pm)}=
    &
    \frac{r_{0}}{\sqrt{h_{\pm}(r_{0})}},\nonumber\\
    K_{\phi\phi}^{(\pm)}=
    &
    \frac{r_{0} \sin^{2}{\theta}}{\sqrt{h_{\pm}(r_{0})}},\nonumber\\
    K^{(\pm)}=
    &
    \frac{1}{2}\frac{f_{\pm}'(r_{0})}{f_{\pm}(r_{0})\sqrt{h_{\pm}(r_{0})}}+\frac{2}{r_{0}}\frac{1}{\sqrt{h_{\pm}(r_{0})}}.
\end{align}
The discontinuity of the extrinsic curvature across the boundary is 
\begin{align}
\label{eq:Kab}
     \llbracket K^{\uptau}_{\ \uptau}\rrbracket =
     &
     \frac{1}{2} \frac{f'_{+}(r_{0})}{f_{+}(r_{0}) \sqrt{h_{+}(r_{0})}} -  \frac{1}{2} \frac{f'_{-}(r_{0})}{f_{-}(r_{0}) \sqrt{h_{-}(r_{0})}}, \\ 
    \llbracket K^{\theta}_{\ \theta}\rrbracket =
    &
    \llbracket K^{\phi}_{\ \phi}\rrbracket=
    \frac{1}{r_{0} \sqrt{h_{+}(r_{0})}} - \frac{1}{r_{0} \sqrt{h_{-}(r_{0})}} ,  \\
     \llbracket K \rrbracket =
     &
     \frac{1}{2} \left( \frac{f'_{+}(r_{0})}{f_{+}(r_{0}) \sqrt{h_{+}(r_{0})}} - \frac{f'_{-}(r_{0})}{f_{-}(r_{0}) \sqrt{h_{-}(r_{0})}} \right) \nonumber\\
    &
    + \frac{2}{r_{0}} \left[ \frac{1}{ \sqrt{h_{+} (r_{0})}} - \frac{1}{\sqrt{ h_{-} (r_{0})}} \right].
\end{align}
The second Israel-Darmois junction condition identifies the behavior in the extrinsic curvature across the boundary with the stress-energy  associated to the boundary itself. In particular, discontinuity of the extrinsic curvature across the boundary indicates a thin shell with intrinsic stress-energy given by the Lanczos equation~\cite{Sen1924,Lanczos1924} 
\begin{align}
\label{eq:SET_shell}
    8  \pi S_{AB} = - \left( \llbracket K_{AB} \rrbracket -  \llbracket K\rrbracket h_{AB}\right).
\end{align}
To express the Ricci scalar in the four-dimensional spacetime in terms of the distributional stress-energy tensor on the hypersurface, we work in coordinates where the continuity of the metric is explicit across the shell $r=r_{0}$, i.e., as well as the new time coordinate $\uptau$ introduced earlier, we would also introduce a proper radial coordinate $z$ by $dz/dr=\sqrt{h_{\pm}(r)}$ for $r\gtrless r_{0}$ whence the metric is
\begin{align}
    ds^{2}=-d\uptau^{2}+dz^{2}+r^{2}(z)d\Omega^{2}
\end{align}
where $r$ here is given implicitly as a function of $z$. Then, it is straightforwardly shown that
\begin{align}
\label{eq:RicciShell}
    R=-8\pi S\,\delta(z-z_{0})=-8\pi S\,\frac{\delta(r-r_{0})}{\sqrt{h_{\pm}(r_{0})}}=\bar R\,\delta(r-r_0),
\end{align}
where $S=h^{AB}S_{AB}$ and $z_{0}=z(r_{0})$, and we have introduced the relation
\begin{align}
    \bar R=-\frac{8\pi S}{\sqrt{h_\pm(r_0)}}.
\end{align}

On the other hand, if $f(r)$ is $C^{1}$ and $h(r)$ is $C^{0}$ at the boundary, as occurs, for example, at the surfaces of many fluid stars~\cite{Schwarzschild:1916ae, Urbano:2018nrs}, the jump in extrinsic curvature vanishes and hence there is no stress-energy associated with the boundary itself. Nevertheless, such boundaries are sharp enough as to cause a jump in Ricci scalar across the surface, which leaves an imprint in the propagation of field modes. 

It is interesting to express the stress-energy tensor $S_{AB}$ as that of a perfect fluid,
\begin{align}
    S_{AB} = \rho\, u_A u_B + P \left( h_{AB} + u_A u_B \right),
\end{align}
where $\rho$ and $P$ represent the surface energy density and the surface tangential pressure. Specifically, we have $S^{\uptau}{}_{\uptau} = - \rho$ and $S^{\theta}{}_{\theta} = S^{\phi}{}_{\phi}=P$ where
\begin{align}
     \rho & = \frac{1}{4 \pi r_{0}} \left( \frac{1}{\sqrt{h_{-}(r_{0})}} - \frac{1}{\sqrt{h_{+}(r_{0})}} \right), \\
     P & = \frac{1}{8 \pi r_{0}} \left[ \frac{1}{ \sqrt{h_{+} (r_{0})}} - \frac{1}{\sqrt{ h_{-} (r_{0})}} \right] \nonumber\\
     &+ \frac{1}{16 \pi} \left( \frac{f'_{+}(r_{0})}{f_{+}(r_{0}) \sqrt{h_{+}(r_{0})}} - \frac{f'_{-}(r_{0})}{f_{-}(r_{0}) \sqrt{h_{-}(r_{0})}} \right).
    \label{Eq:DensitiesApp}
\end{align}

The complete stress-energy tensor of the surface layer is then~\cite{Israel:1966rt, Poisson:2009pwt}
\begin{equation}
    \label{eq:distr_set_shell}  T^{\alpha\beta}=S^{AB}e^{\alpha}_{~A}e^{\beta}_{~B}\,\frac{\delta(r-r_{0})}{\sqrt{h_{\pm}(r_{0})}}=S^{\alpha \beta}\,\frac{\delta(r-r_{0})}{\sqrt{h_{\pm}(r_{0})}}.
\end{equation}

\subsection{Matching Minkowski to Schwarzschild spacetimes }
    Now we specialize to the case where the interior spacetime is Minkowski with line element
    \begin{align}
        ds_{-}^{2}=-dt_{-}^{2}+dr^{2}+r^{2}d\Omega^{2}
    \end{align}
    and the exterior spacetime is Schwarzschild with line element
    \begin{align}
        ds_{+}^{2}=-\left(1-\frac{2M}{r}\right)dt_{+}^{2}+\left(1-\frac{2M}{r}\right)^{-1}dr^{2}+r^{2}d\Omega^{2}.
    \end{align}
    
    Now the continuity condition (\ref{eq:timecoord}) implies \mbox{$dt_{-}^{2}=(1-2M/r_{0})dt_{+}^{2}$} so we express both metrics in terms of the Schwarzschild time and drop the subscript $+$ for typographical convenience,
\begin{align}\label{eq:shellmetric}
    ds^{2}_{-}=
    &
    -\left(1-\frac{2M}{r_{0}}\right)dt^{2}+dr^{2}+r^{2}d\Omega^{2},\\
    ds^{2}_{+}=
    &
    -\left(1-\frac{2M}{r}\right)dt^{2}+\left(1-\frac{2M}{r}\right)^{-1}dr^{2}+r^{2}d\Omega^{2}. \nonumber
\end{align}

The Ricci scalar of the shell is given by Eq.~(\ref{eq:RicciShell}), with
\begin{equation}\label{eq:TraceS}
    S=S^{AB}h_{AB}=\frac{1}{4\pi r_{0}}\left[\frac{2-3M/r_{0}}{\sqrt{1-\frac{2M}{r_{0}}}}-2\right]. 
\end{equation}
In terms of the surface energy density and tangential pressure, we have
\begin{align}
    \rho=
    &
    \frac{1}{4\pi r_{0}}\left(1-\sqrt{1-\frac{2M}{r_{0}}}\right)\nonumber\\
    P=
    &
    -\frac{1}{8\pi r_{0}}\left(1-\sqrt{1-\frac{2M}{r_{0}}}\right)^{-1}-\frac{1}{8\pi r_{0}^2}\frac{M}{\sqrt{1-\frac{2M}{r_{0}}}}.
\end{align}
It is worth noting that the trace $S$ has the following asymptotic limits
\begin{align}
    S\sim
    &
    \frac{1}{8\pi\sqrt{2Mr_{0}\left(1-\frac{2M}{r_{0}}\right)}} &r_{0}\to 2M\nonumber\\
    S\sim
    &
    -\frac{M}{4\pi r_{0}^2} &r_{0}\to\infty,
\end{align}
i.e., it is positive (and divergent) in the black hole limit and negative in the large-shell limit. Finally, we note that the trace $S$ can also be written as 
\begin{equation}
    S=\frac{\left(1-3\sqrt{1-\frac{2M}{r_{0}}}\right)\rho}{2\sqrt{1-\frac{2M}{r_{0}}}}, 
\end{equation}
showing that it vanishes exactly at the Buchdahl compactness limit $2M/r_{0}=8/9$~\cite{PhysRev.116.1027}, the maximum compactness for constant-density stars in general relativity. 

\section{Numerical Approach}
\label{sec:num_app}
\subsection{Euclidean modes}
\label{subsec:eucl_modes}
One of the advantages of the extended coordinate prescription described in Sec.~\ref{sec:extcoord} is that it makes use of Euclidean techniques which simplify the calculation of the field's radial modes. In particular, the radial modes on the Lorentzian spacetime are generally oscillatory solutions (except for possible bound state modes) whereas on the Euclidean geometry, the radial modes are generally exponentially growing or decaying functions of $r$. Computing the Euclidean modes is computationally easier.

The coincidence limit of the one-dimensional radial Green function appearing in the calculation of the vacuum polarization and the RSET is
\begin{align}
    g_{\omega l}(r)=\frac{p_{\omega l}(r)\,q_{\omega l}(r)}{N_{\omega l}}
\end{align}
where $p_{\omega l}(r)$ and $q_{\omega l}(r)$ are linearly independent solutions to the homogeneous second-order linear differential equation
\begin{align}
\label{eq:radialhomogeneous}
    \frac{d}{dr}\left(\frac{r^{2}\sqrt{f}}{\sqrt{h}}\frac{d\,X_{\omega l}}{dr}\right)-\sqrt{h\,f}\Bigg(\frac{\omega^{2}r^{2}}{f}+l\left(l+1\right)\nonumber\\+r^{2}(\mu^{2}+\xi R)\Bigg)X_{\omega l}
=0
\end{align}
and the normalization constant $N_{\omega l}$ is given by Eq.~(\ref{eq:normalization}).

The Boulware quantum state can be defined as a zero temperature Euclidean state and can be specified by imposing appropriate boundary conditions on these modes. The modes $p_{\omega l}(r)$ are obtained by imposing boundary conditions on the left boundary of the region of interest and integrating outward in $r$ while $q_{\omega l}(r)$ are obtained by imposing boundary conditions on the right boundary of the region of interest and integrating inwards in $r$. 

In our thin shell model, spacetime is separated into two different patches, hence the solutions to~\eqref{eq:radialhomogeneous} will have to be suitably connected through the shell. We denote by $\{p_{\omega l}^{\rm M},q_{\omega l}^{\rm M}\}$ a basis of Minkowski modes, i.e., linearly independent solutions of (\ref{eq:radialhomogeneous}) with \mbox{$h(r)=1$} and \mbox{$f(r)=(1-2M/r_{0})$} a constant. Similarly, we denote by  $\{p_{\omega l}^{\rm S},q_{\omega l}^{\rm S}\}$ a basis of Schwarzschild modes, i.e., linearly independent solutions of  with \mbox{$f(r)=[h(r)]^{-1}=(1-2M/r)$}. Once these bases of modes are obtained independently in each region, the Euclidean modes associated to the thin shell spacetime correspond to
\begin{align}\label{eq:pModeShell}
    p_{\omega l}^{\text{shell}}=
    &
    \begin{cases}
        p_{\omega l}^{\rm M}, ~\qquad\qquad\qquad r < r_0\\
        \alpha_{\omega l} p_{\omega l}^{\rm S}+\beta_{\omega l}q_{\omega l}^{\rm S}, \quad r > r_0
    \end{cases}
\end{align}
and
\begin{align}\label{eq:qModeShell}
    q_{\omega l}^{\text{shell}}=
    &
    \begin{cases}
        \gamma_{\omega l} p_{\omega l}^{\rm M}+\delta _{\omega l}q_{\omega l}^{\rm M}, \quad r < r_0\\
        q_{\omega l}^{\rm S}, ~\qquad\qquad\qquad r > r_0
    \end{cases}
\end{align}
and equivalently for their radial derivatives. The coefficients $\left\{\alpha_{\omega l},\beta_{\omega l},\gamma_{\omega l},\delta_{\omega l}\right\}$ are found numerically by matching the Schwarzschild and Minkowski modes bases at $r=r_0$. 
Finally, the Wronskian constant can be related to the respective one in these mode bases via
\begin{align}
\label{eq:wronsk_shell}
    N_{\omega l}^{\text{shell}}=
    &
    \begin{cases}
        \delta_{\omega l}N_{\omega l}^{\rm M}, \quad r < r_0\\
        \alpha_{\omega l} N_{\omega l}^{\rm S}, \quad r > r_0.
    \end{cases}
\end{align}

The next two subsections contain details about how these two mode bases are calculated including the boundary conditions and the matching conditions. As we will see, the matching results in discontinuous radial derivatives for the modes at $r=r_0$.

\subsection{Schwarzschild modes}
Outside the shell, the spacetime is Schwarzschild and the homogeneous radial equation is
\begin{equation}\label{eq:ModeSchw}
   \left\{ \frac{d}{dr}\left(r^{2}-2 M r\right)\frac{d}{dr}-l\left(l+1\right)-\mu^{2}r^{2}-\frac{\omega^{2}r^{2}}{1-\frac{2M}{r}}\right\}X_{\omega l}=0.
\end{equation}
This equation has a regular singular point at $r=2 M$ and an irregular singular point at $r=\infty$. We take $p_{\omega l}^{\textrm{S}}(r)$ to be the modes regular at $r=2M$. This solution can be expressed in terms of the regular confluent Heun function as
\begin{align}
\label{eq:pSchwHeun}
	p_{\omega l}^{\rm S}(r)&=\frac{e^{-2M\omega+\Omega\, r}}{\left(2M\right)^{2M\omega}}\left(\frac{r}{2M}-1\right)^{2M\omega}\mathsf{H}\left(a_1,a_2,a_3,a_4,a_5;z\right),
\end{align}
where we have labelled
\begin{align}
\label{eq:Heunparameters}
 a_1= &
 l(l+1)-2M\left\{\omega+\Omega-2M\left[\mu^{2}+2\omega\left(\omega-\Omega\right)\right]\right\},\nonumber\\
 a_2 = 
 &
 4M\left[M\left(\omega-\Omega\right)^{2}-\Omega\right],\nonumber\\
 a_3 =
 &
 1+4M\omega,\nonumber\\
 a_4 =
 &
 1,\nonumber\\
 a_5=
 &
 -4M\Omega,\nonumber\\
 \Omega = &\sqrt{\mu^{2}+\omega^{2}},\nonumber\\
 z = 
 &
 1-\frac{r}{2M}.
 \end{align}
In our notation, the confluent Heun function $\mathsf{H}(a_1,a_2,a_3,a_4,a_5;z)$ is the function which satisfies 
\begin{align}
    z(z-1) \mathsf{H}''(z)+ (a_3 (z-1)+a_4 z+z(z-1)a_5) \mathsf{H}'(z)\nonumber\\
    +(a_2 z-a_1)\mathsf{H}(z)=0
\end{align}
and is normalized to unity at $z=0$. The advantage of expressing our solution in terms of confluent Heun functions is that an efficient routine, $\mathsf{HeunC}$, for computing this function is built into the software package Mathematica.

We take as our $q_{\omega l}^{\textrm{S}}(r)$ modes the solutions to (\ref{eq:ModeSchw}) that are regular at $r=\infty$. While these solutions can still be expressed in terms of confluent Heun functions, they correspond to Heun functions of the logarithmic kind which are not yet implemented in Mathematica. We instead proceed by obtaining the $q_{\omega l}^{\textrm{S}}(r)$ mode through direct integration of Eq.~\eqref{eq:ModeSchw}. To obtain these modes, most of the difficulty lies in reliably generating the boundary conditions at $r=\infty$. Since this is an  irregular singular point, the solution for large $r$ is an asymptotic series solution and therefore non-converging. So care is needed in truncating this series to obtain a sufficiently accurate starting value for the numerical integration. We obtain this asymptotic series solution by first factoring out the leading-order exponential behavior from the $q_{\omega l}^{S}(r)$ modes. We write
\begin{equation}\label{eq:qModeS}
    q_{\omega l}^{\rm S}(r)=e^{-\Omega r}\tilde{q}_{\omega l}(r),
\end{equation}
 where $\tilde{q}_{\omega l}$ satisfies
\begin{align} 
    &
    r\left(r-2M\right)^{2}\frac{d^{2}\tilde{q}_{\omega l}}{dr^{2}}\nonumber\\
    &
    +2\left(r-2M\right)\left[2\Omega M r-M-\left(1-\Omega r\right)r\right]\frac{d\tilde{q}_{\omega l}}{dr}\nonumber\\
    &
    +\left[2\mu^{2}Mr^{2}-l\left(l+1\right)\left(r-2M\right)\right.\nonumber\\
    &
    \left.-2\Omega\left(2M^2-3Mr+r^2\right)-4\Omega^2 M \left(r-M\right)r\right]\tilde{q}_{\omega l}=0.
\end{align}
Assuming the series expansion
\begin{equation}\label{eq:SeriesY}
    \tilde{q}_{\omega l} \sim r^{-\upsilon}\sum_{j=0}^{\infty}\frac{b_{j}}{r^{j}},
\end{equation}
we obtain 
\begin{align}
    \upsilon =
    &
    1+\frac{\mu^{2} M}{\Omega}+\frac{2 M \omega^2}{\Omega},
\end{align}
and the following three-term recurrence relation for the $b_{j}$ coefficients
\begin{align}
    &b_{j}=   
    \left\{4M^2\left(\upsilon+j-3\right)^2 b_{j-3}\right.\nonumber\\
    & \left.+2M\left[l\left(l+1\right)-\left(2\upsilon+2j-5\right)\left(\upsilon+j-2-2\Omega M\right)\right]b_{j-2}\right.\nonumber\\
    &
    +\left.
    \left[\left(\upsilon-l+j-2\right)\left(\upsilon+l+j-1\right)
    \right.\right.\nonumber\\
    &
    \left.\left.+4\Omega^2 M^2-2\Omega M \left(4\upsilon+4j-7\right)\right]b_{j-1}\vphantom{\left(\upsilon+j-3\right)^2}\right\}\nonumber\\
    &
    \times  
    \left\{2\left[2\Omega^2 M-\mu^2 M- \Omega\upsilon-2\Omega\left(j-1\right)\right]\right\}^{-1},
\end{align}
assuming the initial conditions $b_{-2}=b_{-1}=0$ and \mbox{$b_{0}=1$.}

We found it necessary to adapt the radial position of the outer boundary of our integration domain for each mode according to the specific values of $l$ and $\omega$ to ensure that the wave equation is satisfied to our desired tolerance. Specifically, the radius at which the outer boundary conditions are imposed must be increased with both $\omega$ and $l$.
Once this outer boundary is selected, we integrate Eq.~\eqref{eq:SeriesY} towards the horizon and obtain the $q_{\omega l}^{\rm S}$ mode from~\eqref{eq:qModeS}. 

With this setup, we proceed with our mode calculation ensuring that the Wronskian condition~\eqref{eq:normalization} is satisfied everywhere up to 30 decimal places. Maintaining this degree of precision makes it very inefficient to obtain modes in the frequency range $\omega\gtrsim 20/M$. In most cases, the contribution of such high-$\omega$ modes to the RSET is negligible, with the exception of the near-shell region. Near the surface of the shell, we can employ the WKB approximation to capture the high frequency asymptotics, as we will show later.

\subsection{Minkowski modes}
The mode equation in the shell interior becomes
\begin{equation}
   \left\{ \frac{d}{dr}\left(r^{2}\frac{d}{dr}\right)-\left[l\left(l+1\right)+\tilde{\Omega}^{2}r^{2}\right]\right\}X_{\omega l}=0,
\end{equation}
where $\tilde{\Omega}^{2}=\mu^{2}+\omega^{2}/\left(1-2M/r_{0}\right)$. Note that the frequency appearing here is different to that appearing in the usual treatment of a quantum field in Minkowski spacetime since our time coordinate on the Minkowski interior is scaled so that the metric is continuous across the shell. Apart from this trivial rescaling of the time coordinate (and hence the frequency), the linearly independent solutions are the usual modified spherical Bessel functions. We take $p^{\textrm{M}}_{\omega l}(r)$ to be the solution regular at the origin $r=0$ and $q^{\textrm{M}}_{\omega l}(r)$ to be the solution regular at infinity. Hence we have
\begin{align}
\label{eq:ModesMink}
    p_{\omega l}^{\rm M}(r)=
    &
    \sqrt{2}\left(\frac{\tilde{\Omega}}{2}\right)^{-l}\Gamma\left(l+\frac{3}{2}\right)\frac{I_{l+\frac{1}{2}}\left(\tilde{\Omega} r\right)}{\sqrt{\tilde{\Omega} r}},\nonumber\\
    q_{\omega l}^{\rm M}(r)=
    &
    \frac{1}{\sqrt{2\left(1-\frac{2M}{r_{0}}\right)}}\left(\frac{\tilde{\Omega}}{2}\right)^{l}\sqrt{\frac{\tilde{\Omega}}{r}}\frac{K_{l+\frac{1}{2}}\left(\tilde{\Omega} r\right)}{\Gamma\left(l+\frac{3}{2}\right)},
\end{align}
where we have normalized the solutions so that $N_{\omega l}^{\textrm{M}}=1$.

\subsection{Matching conditions at the surface}
Let $X^{\textrm{shell}}_{\omega l}(r)$ represent either $p_{\omega l}^{\textrm{shell}}(r)$ or $q_{\omega l}^{\textrm{shell}}(r)$. Now we impose that our modes are continuous across the boundary which means
\begin{align}\label{eq:condmode}
    \llbracket X_{\omega l}^{\textrm{shell}}\rrbracket=0.
\end{align}
In terms of the Minkowski and Schwarzschild modes, this implies
\begin{align}
\label{eq:modematch}
    p^{\textrm{M}}(r_{0})&=\alpha\,p^{\textrm{S}}(r_{0})+\beta\,q^{\textrm{S}}(r_{0}),\nonumber\\
    q^{\textrm{S}}(r_{0})&=\gamma\,p^{\textrm{M}}(r_{0})+\delta\,q^{\textrm{M}}(r_{0}),
\end{align}
where we are suppressing the dependence of the mode numbers $\omega$ ad $l$ for convenience.

Notice that condition~\eqref{eq:condmode} implies the shell is transparent for the scalar field, i.e., that we do not force the field itself, its radial derivative, or a combination both, to vanish at the surface. Conducting shells can be modelled by imposing Dirichlet, Neumann, or Robin boundary conditions to the field, which affect the near-shell divergences of the RVP and RSET~\cite{Deutsch:1978sc}. We leave aside such situations here, focusing on the purely gravitational effect of the shell on the modes.

Now the derivatives of the modes across the shell are more complicated because of the Ricci scalar term appearing in Eq.~(\ref{eq:radialhomogeneous}). Integrating this equation over the interval $(r_{0}-\epsilon,r_{0}+\epsilon)$ and then taking the limit $\epsilon\to 0$ gives
\begin{align}
    \left[\!\!\left[ \frac{1}{\sqrt{h(r)}}\frac{dX_{\omega l}^{\textrm{shell}}(r)}{dr}\right]\!\!\right]=-8\pi\,\xi\,S\,X_{\omega l}^{\textrm{shell}}(r_{0}).
\end{align}
So we see that the derivatives of the modes are not continuous across the shell. In fact, even for the case of minimal coupling $\xi=0$, it is $X_{\omega l}'(r)/\sqrt{h(r)}$ that is continuous and not the derivative of the modes themselves. In terms of the Schwarzschild and Minkowski modes, this condition implies
\begin{align}
\label{eq:derivmatch}
    \sqrt{1-\frac{2M}{r_{0}}}\left(\alpha\,p^{\textrm{S}}{}'(r_{0})+\beta\,q^{\textrm{S}}{}'(r_{0})\right)-p^{\textrm{M}}{}'(r_{0})\nonumber\\
    =-8\pi\,\xi\,S\,p^{\textrm{M}}(r_{0}),\nonumber\\
    \sqrt{1-\frac{2M}{r_{0}}} q^{\textrm{S}}{}'(r_{0})-\left(\gamma\,p^{\textrm{M}}{}'(r_{0})+\delta\,q^{\textrm{M}}{}'(r_{0})\right)\nonumber\\
    =-8\pi\,\xi\,S\,q^{\textrm{S}}(r_{0}),
\end{align}
where as before we are suppressing the dependence on the mode numbers.

Now solving Eq.~(\ref{eq:modematch}) and (\ref{eq:derivmatch}) gives
\begin{align}
    \alpha_{\omega l}&=\frac{1}{N_{\omega l}^{\textrm{S}}}\Bigg[r^{2}\sqrt{f}\left(q^{\textrm{S}}_{\omega l}\frac{d p^{\textrm{M}}_{\omega l}}{dr}-8\pi\xi S p^{\textrm{M}}_{\omega l}q^{\textrm{S}}_{\omega l}\right)\nonumber\\
  &\qquad\qquad\qquad\qquad\qquad  -r^{2}f\,p^{\textrm{M}}_{\omega l}\frac{d q^{\textrm{S}}_{\omega l}}{dr}\Bigg]_{r=r_{0}}\nonumber\\
     \beta_{\omega l}&=-\frac{1}{N_{\omega l}^{\textrm{S}}}\Bigg[r^{2}\sqrt{f}\left(p^{\textrm{S}}_{\omega l}\frac{d p^{\textrm{M}}_{\omega l}}{dr}-8\pi\xi S p^{\textrm{M}}_{\omega l}p^{\textrm{S}}_{\omega l}\right)\nonumber\\
   &\qquad\qquad\qquad\qquad\qquad -r^{2}f\,p^{\textrm{M}}_{\omega l}\frac{d p^{\textrm{S}}_{\omega l}}{dr}\Bigg]_{r=r_{0}}\nonumber\\
    \gamma_{\omega l}&=\frac{1}{N_{\omega l}^{\textrm{M}}}\Bigg[r^{2}\sqrt{f}q^{\textrm{M}}_{\omega l}\frac{d q^{\textrm{S}}_{\omega l}}{dr}+8\pi\xi S r^{2}q^{\textrm{M}}_{\omega l}q^{\textrm{S}}_{\omega l}\nonumber\\
    &\qquad\qquad\qquad\qquad\qquad-r^{2}\,q^{\textrm{S}}_{\omega l}\frac{d q^{\textrm{M}}_{\omega l}}{dr}\Bigg]_{r=r_{0}}\nonumber\\
     \delta_{\omega l}&=-\frac{1}{N_{\omega l}^{\textrm{M}}}\Bigg[r^{2}\sqrt{f}p^{\textrm{M}}_{\omega l}\frac{d q^{\textrm{S}}_{\omega l}}{dr}+8\pi\xi S r^{2}p^{\textrm{M}}_{\omega l}q^{\textrm{S}}_{\omega l}\nonumber\\
   &\qquad\qquad\qquad\qquad\qquad -r^{2}\,q^{\textrm{S}}_{\omega l}\frac{d p^{\textrm{M}}_{\omega l}}{dr}\Bigg]_{r=r_{0}}.
\end{align}

The above coefficients were obtained numerically from the Schwarzschild and Minkowski modes for a large range of $\omega$ and $l$ values. For fixed $l$ and large $\omega$, each of these coefficients is an exponentially growing function of $\omega$. This of course presents a numerical issue especially when computing ratios of these coefficients for large $\omega$.
It is possible to approximate these coefficients for large $\omega$ using the WKB approximation for the Schwarzschild modes. This way, we obtain analytical large-frequency approximations for these matching coefficients.
We will make use of the WKB approximation in the next Section to estimate the rate of divergence of the RVP and the RSET near the surface of the shell.

\subsection{Tests of convergences}

We now have all the ingredients required to calculate the renormalized vacuum polarization and the renormalized stress-energy tensor of thin shell spacetimes. As we will see in this Subsection, only approximate numerical results can be attained as $r\to r_0$ due to a loss of convergence in the integrals and mode sums appearing in the extended-coordinate prescription. In this section, we numerically analyze the rate of convergence of the renormalized mode-sums inside and outside the thin shell and in the limit in which the surface of the shell is approached.

The renormalized integrand $g_{\omega l }-k_{\omega l}^{(m)}(r)$ can be expressed in terms of the pure Minkowski and Schwarzschild modes using expressions~\eqref{eq:qModeShell} and~\eqref{eq:pModeShell}, and the Wronskian relation~\eqref{eq:wronsk_shell}.
We find
\begin{align}
\label{eq:gshell}
    g_{\omega l}^{\text{shell}}=\frac{p_{\omega l}^{\text{shell}}q_{\omega l}^{\text{shell}}}{N_{\omega l}^{\text{shell}}}=
    &
    \begin{cases}
        \frac{p_{\omega l}^{\rm M}q_{\omega l}^{\rm M}}{N_{\omega l}^{\rm M}}+\frac{\gamma_{\omega l}}{\delta_{\omega l}}\frac{\left(p_{\omega l}^{\rm M}\right)^{2}}{N_{\omega l }^{\rm M}}, \quad r < r_0\\ \\
        \frac{p_{\omega l}^{\rm S}q_{\omega l}^{\rm S}}{N_{\omega l}^{\rm S}}+\frac{\beta_{\omega l}}{\alpha_{\omega l}}\frac{\left(q_{\omega l}^{\rm S}\right)^{2}}{N_{\omega l }^{\rm S}}, \quad r > r_0.
    \end{cases}
\end{align}
The first terms on the right-hand side of these expressions are the Green's functions in Minkowski and Schwarzschild spacetime $g_{\omega l}^{\rm M}$ and $g_{\omega l}^{\rm S}$, respectively, whose large-$\omega$ and $l$ behaviour is exactly cancelled by the regularization parameters $k_{\omega l}^{(m)}$. The second terms, however, are a consequence of the discontinuity in the derivative of the modes induced by the shell. {These boundary terms are exponentially convergent in both $\omega$ and $l$ everywhere for $r\neq r_0$. However, the rate of convergence is not uniform in $r$, and indeed becomes very slow as $r\to r_0$, so that an unreasonable amount of modes could be required to compute the renormalized quantities. Exactly at $r=r_0$, these terms do not converge.} In the following, we show numerical evidence for this statement.

\begin{figure}
\includegraphics[width=\linewidth]{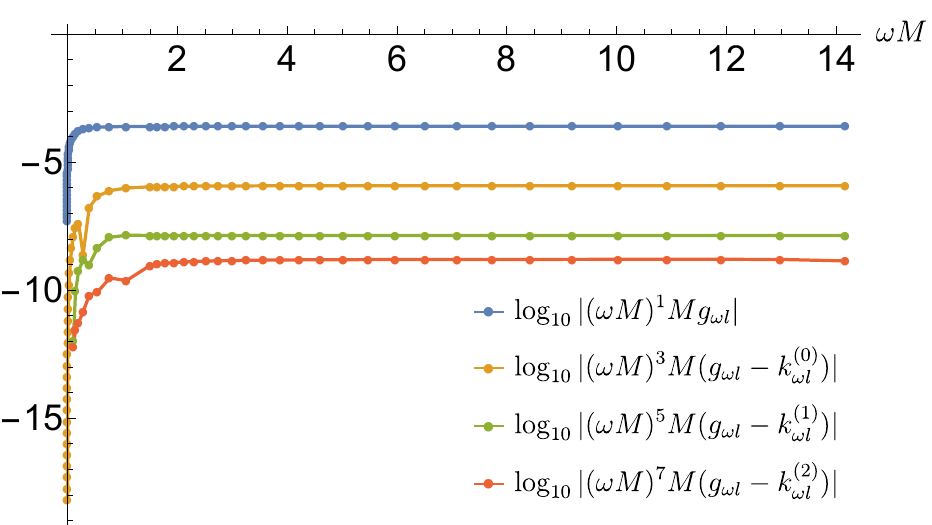}
\caption{Analysis of the convergence of $g_{\omega l}(r)-k_{\omega l}^{(m)}(r)$ for a massless scalar field on the exterior of the thin shell for $r\gg r_0$ at different orders of expansion $m$ for fixed $l$ and large $\omega$. Results obtained with the set of parameters: $r_0=2.001 M$, $r=5M$ and $l=1$.}
\label{fig:large_om_ext}
\end{figure}

\begin{figure}
\includegraphics[width=\linewidth]{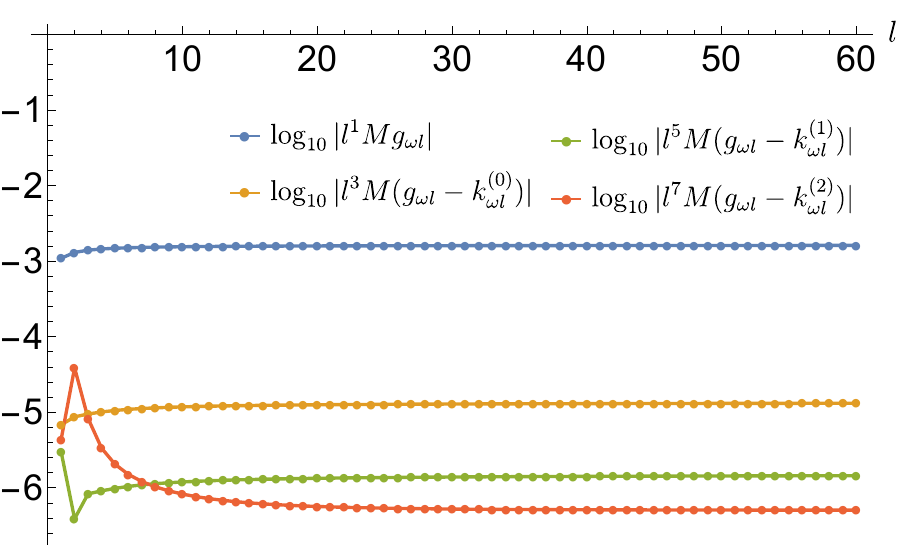}
\caption{Analysis of the convergence of $g_{\omega l}(r)-k_{\omega l}^{(m)}(r)$ for a massless scalar field on the exterior of the thin shell for $r\gg r_0$ at different orders of expansion $m$ for fixed $\omega$ and large $l$. Results obtained with the set of parameters: $r_0=2.001 M$, $r=5M$ and $\omega M=0.01$.}
\label{fig:large_L_ext}
\end{figure}

In the extended coordinate prescription, the rate of convergence of the mode-sums is directly related to the order $m$ of the expansion of the Hadamard parametrix. In general, the convergence of the mode-sums can be improved by subtracting higher-order terms of the Hadamard parametrix. The behavior of the renormalized integrand $g_{\omega l}(r)-k_{\omega l}^{(m)}(r)$ at large $\omega$ and large $l$ depends on the order of the expansion $m$ as follows:
\begin{align}
\label{eq:large_om_L_behavior}
    g_{\omega l}(r)-k_{\omega l}^{(m)}(r)&=\begin{cases}\sim \omega^{-2m-3},&\textrm{large $\omega$, fixed $l$},\\
\sim l^{-2m-3}, & \textrm{large $l$, fixed $\omega$}. \end{cases}
\end{align} 
Let us consider, at first, a radial point on the exterior region and far enough from the shell. Fig.~\ref{fig:large_om_ext} reports the large-frequency behavior of $g_{\omega l}(r)-k_{\omega l}^{(m)}(r)$ for a fixed $l$-mode, while Fig.~\ref{fig:large_L_ext} presents its large-$l$ behavior at fixed $\omega$. The logarithmic scale in the Figures is introduced for visualization purposes. Since the quantities $\log_{10}|(\omega M)^{2m+3}M(g_{\omega l}-k_{\omega l}^{(m)})|$ in Fig.~\ref{fig:large_om_ext} asymptote to horizontal lines in the large frequency regime, it follows that $M(g_{\omega l}-k_{\omega l}^{(m)})\sim(\omega M)^{-2m-3}$ at large frequency, where the $M$ factors are introduced to render all the quantities dimensionless. To simplify the notation, we omit them in the remainder of the discussion. Similarly, by inspecting the curves in Fig.~\ref{fig:large_L_ext}, one can verify that the behavior at large $l$ for fixed $\omega$ is $M(g_{\omega l}-k_{\omega l}^{(m)})\sim l^{-2m-3}$. These convergence rates are in agreement with those expected from Eq.~\eqref{eq:large_om_L_behavior}.

\begin{figure}
\includegraphics[width=\linewidth]{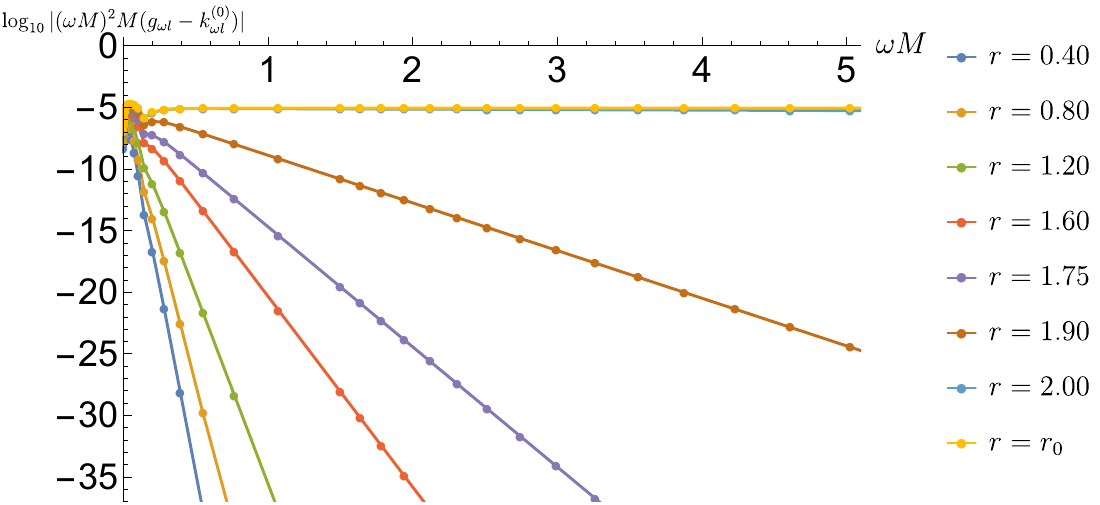}
\caption{Analysis of the convergence of $g_{\omega l}(r)-k_{\omega l}^{(0)}(r)$ for a massless scalar field on the Minkowski interior of the thin shell at different radial points for fixed $l$ and large $\omega$. Results obtained with the set of parameters: $r_0=2.001M$ and $l=1$. The figure shows how the order of convergence decreases as $r\to r_0$. On the surface, the function $g_{\omega l}(r_0)-k_{\omega l}^{(0)}(r_0)$ scales as $\omega^{-2}$ at large $\omega$}
\label{fig:large_om_diff_radii}
\end{figure}

\begin{figure}
\includegraphics[width=\linewidth]{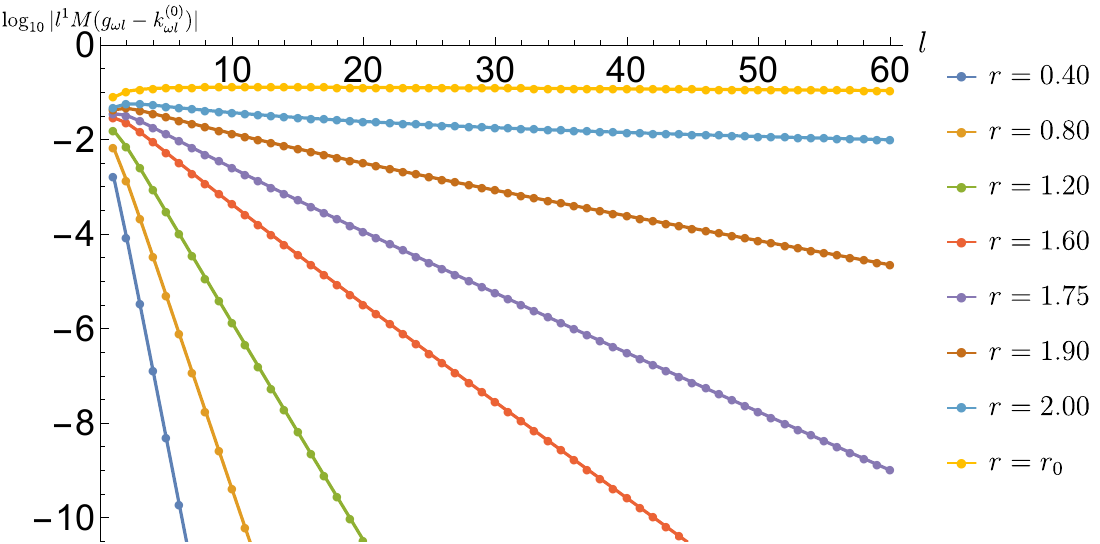}
\caption{Analysis of the convergence of $g_{\omega l}(r)-k_{\omega l}^{(0)}(r)$ for a massless scalar field on the Minkowski interior of the thin shell at different radial points for fixed $\omega$ and large $l$. Results obtained with the set of parameters: $r_0=2.001M$ and $\omega M=0.01$. The figure shows how the order of convergence decreases as $r\to r_0$. On the surface, the function $g_{\omega l}(r_0)-k_{\omega l}^{(0)}(r_0)$ scales as $l^{-1}$ at large $l$.}
\label{fig:large_L_diff_radii}
\end{figure}

The interior of the shell (in case of a massless field) represents an exception to the convergence pattern in~\eqref{eq:large_om_L_behavior}. In particular, the higher-order terms in the expansion of the Hadamard parametrix, which involve derivatives of the radial functions $f(r)$ and $h(r)$, vanish in Minkowski spacetime for a massless field. Thus, in this particular case, the Hadamard parametrix is fully encoded in the expansion up to order $m=0$. Therefore, the Minkowski interior with a massless field provides a useful scenario for studying the convergence as the surface is approached, allowing us to isolate any loss of convergence due to the boundary.
To this end, Figures \ref{fig:large_om_diff_radii} and \ref{fig:large_L_diff_radii} show that the behavior of the renormalized integrand in the interior region for a massless scalar field at large frequency and large $l$ strongly depends on the radial distance from the shell. As $r\to r_0$, the decay rate in $\omega$ and $l$ of the renormalized integrand decreases rapidly as the shell surface is approached. In Figures~\ref{fig:large_om_diff_radii} and \ref{fig:large_L_diff_radii}, lines with a negative slope indicate a convergence rate that is faster than $\omega^{-2}$ and $l^{-1}$, respectively. Clearly, as $r\to r_0$, the convergence rate decreases. On the surface, $g_{\omega l}(r_0)-k_{\omega l}^{(0)}(r_0)$ scales as $\omega^{-2}$ for fixed $l$ at large $\omega$ (as shown in Fig.~\ref{fig:large_om_diff_radii}), but as $l^{-1}$ for fixed $\omega$ at large $l$ (see Fig.~\ref{fig:large_L_diff_radii}). As a result, the $\omega$-integral converges, but the $l$-sum diverges on the shell surface, thus making the mode-sum divergent there. Furthermore, radial points near the surface require truncating the $\omega$-integral and the $l$-sum at increasingly large $\omega$ and $l$ values to obtain accurate results, making the numerical method challenging to apply. 

In Section \ref{sec:analytic}, to study more effectively the region near the shell surface, we investigate the properties of the renormalized vacuum polarization and RSET using analytic methods.

\section{Analytical Approximations}
\label{sec:analytic}
\subsection{Vacuum polarization and renormalized stress-energy tensor near the surface}\label{Subsec:WKB}
In this section we will employ the WKB approximation in order to calculate the leading order divergence in the RSET components as the surface is approached.

From inspection of $ g_{\omega l}^{\text{shell}}$ in Eq.~\eqref{eq:gshell} we see that the presence of the shell is encoded in the terms:
\begin{align}
    \frac{\gamma_{\omega l}}{\delta_{\omega l}}\frac{(p^{M}_{\omega l} (r))^2}{N^{M}_{\omega l}},~~~~~~\frac{\beta_{\omega l}}{\alpha_{\omega l}}\frac{(q^{S}_{\omega l} (r))^2}{N^{S}_{\omega l}}.
\end{align}    
Therefore, these terms must be the source of the near shell divergence that is evident in our numerical calculations. In order to calculate these divergences we take the following WKB approximation to the mode functions in the interior and exterior: 
\begin{align}
    p^{M}_{\omega l} (r) \approx \frac{1}{\sqrt{2 r^2 W_M(r)} } \exp\left[\int \frac{W_M}{\sqrt{f(r_0)}}dr\right]\\
    q^{M}_{\omega l} (r) \approx \frac{1}{\sqrt{2 r^2 W_M(r)} } \exp\left[\int -\frac{W_M}{\sqrt{f(r_0)}}dr\right]\\
    q^{S}_{\omega l} (r) \approx \frac{1}{\sqrt{2 r^2 W_{S}(r)} } \exp\left[-\int \frac{W_{S}}{f(r)}dr\right]\\
     p^{S}_{\omega l} (r) \approx \frac{1}{\sqrt{2 r^2 W_{S}(r)} } \exp\left[\int \frac{W_{S}}{f(r)}dr\right]
\end{align}
where we have taken:
\[W_M^2(r)=\lambda^2 f(r_0)/r^2 + \omega^2, ~~~W_S^2(r)=\lambda^2 f(r)/r^2 + \omega^2,\]
with $\lambda=l+1/2$.
\begin{widetext}
We will outline the calculation in the interior and simply present the exterior results, as the method is very similar for both. 
Inserting the WKB approximation and introducing a new variable $x=\lambda^2 f(r_0)/r_0^2 + \omega^2$, we obtain:
\begin{align}
    \frac{\gamma_{\omega l}}{\delta_{\omega l}}\frac{(p^{M}_{\omega l} (r))^2}{N^{M}_{\omega l}}\approx\frac{r_0 \left(x+\omega
   ^2\right)-r_0 \sqrt{f(r_0)} \left(16 \pi  \xi  r_0 S x+x+\omega ^2\right)-M \left(x+3 \omega ^2\right)}{r_0 \sqrt{f(r_0)} \left(16 \pi  \xi  r_0 S x+x+\omega ^2\right)+M \left(x+3 \omega ^2\right)-r_0
   \left(x+\omega ^2\right) -4 r_0^2 x^{3/2}}\nonumber\\
  \times \frac{1}{2 r^2 W_{M}(r)} \exp\left[\frac{2}{\sqrt{f(r_0)}}\left(\int W_{M}(r)dr -\int W_{M}(r)dr\bigg{|}_{r=r_0}\right)\right].
\end{align}
The final term in the denominator will dominate for both large $\omega$ and $\lambda$ (large $x$), so we ignore the other terms in the denominator. We then expand the $r$ dependent terms  in the near surface limit to obtain:
\begin{align}
\label{eq:approx}
    \frac{\gamma_{\omega l}}{\delta_{\omega l}}\frac{(p^{M}_{\omega l} (r))^2}{N^{M}_{\omega l}}\approx\frac{r_0 \sqrt{f(r_0)} \left(16 \pi  \xi  r_0 S x+x+\omega ^2\right)+M \left(x+3 \omega ^2\right)-r_0 \left(x+\omega
   ^2\right)}{ 4 r_0^2 x^{3/2}}\nonumber\\
  \times\left(\frac{1}{2 r_0^2 \sqrt{x}} + \frac{\varepsilon  \left(x+\omega ^2\right)}{2 r_0^3 x^{3/2}}\right)\exp\left[-\frac{2 \sqrt{x}}{\sqrt{f(r_0)}}\varepsilon\right]\left(1+\sqrt{\frac{x}{r_0^2 f(r)}}\left(\frac{\omega^2}{x}-1\right) \varepsilon^2\right),
\end{align}
where $\varepsilon$ is the proper radial distance from the shell ($r_0-r$ in interior).
The leading order terms in $\varepsilon$ are sufficient to calculate the leading order divergence in the RVP, however we require the higher order terms to correctly capture the divergence in the RSET.
In order to calculate these divergences we must evaluate the following double mode sums:
\end{widetext}
\begin{align}
\label{eq:modesums}
 \int_{0}^\infty \sum_{l=0}^{\infty} (2l+1)  \frac{\gamma_{\omega l}}{\delta_{\omega l}}\frac{(p^{M}_{\omega l} (r))^2}{N^{M}_{\omega l}} d \omega,\nonumber\\
\int_{0}^\infty \sum_{l=0}^{\infty} (2l+1)  \omega^2\frac{\gamma_{\omega l}}{\delta_{\omega l}}\frac{(p^{M}_{\omega l} (r))^2}{N^{M}_{\omega l}} d \omega,\nonumber\\
\int_{0}^\infty \sum_{l=0}^{\infty} (2l+1)\frac{l(l+1)}{2}  \frac{\gamma_{\omega l}}{\delta_{\omega l}}\frac{(p^{M}_{\omega l} (r))^2}{N^{M}_{\omega l}} d \omega,
\end{align}
which is achieved by employing our approximate expression in Eq. (\ref{eq:approx}). We will focus on the first mode sum; the procedure for the others is very similar. To evaluate the sum over $l$, we use the Watson Sommerfield formula to convert the sum into two integrals, with the leading order behavior coming from:
\begin{align}
\int_{0}^\infty \sum_{l=0}^{\infty} (2l+1)  \frac{\gamma_{\omega l}}{\delta_{\omega l}}\frac{(p^{M}_{\omega l} (r))^2}{N^{M}_{\omega l}} d \omega\nonumber\\
\approx\int_{0}^\infty \int_{0}^\infty  2 \lambda \frac{\gamma_{\omega \lambda}}{\delta_{\omega \lambda}}\frac{(p^{M}_{\omega \lambda} (r))^2}{N^{M}_{\omega \lambda}}d \lambda d \omega. 
\end{align}
Using the approximate expression in Eq. (\ref{eq:approx}), this integral can be computed in closed form in terms of the exponential integral function. Hence, we can write down the leading order divergence for the RVP in the interior:
\begin{align}
\label{eq:vpolshell}
 \langle\hat\phi^2\rangle^- =  \frac{\left(\xi -\frac{1}{6}\right) S}{2\pi \varepsilon} +O(1).
\end{align}
 Similarly, we are able to calculate the leading order behavior to the other modes sums in Eq.(\ref{eq:modesums}), yielding the following expressions for  the leading order divergence for the RSET components in the interior:
\begin{align}
\label{eq:int}
     &   \langle\hat{T}^{t}_{~t}\rangle^-  =\frac{3  \llbracket K^\tau_\tau\rrbracket -4\pi S}{1440\pi^2\varepsilon^3}+\frac{\left(\xi -\frac{1}{6}\right)^2 S}{\pi  \varepsilon ^3}+\mathcal{O}(\varepsilon^{-2})\nonumber\\
      &      \langle\hat{T}^{\theta}_{~\theta}\rangle^- =\frac{3  \llbracket K^\theta_\theta\rrbracket -4\pi S}{1440\pi^2\varepsilon^3}+\frac{\left(\xi -\frac{1}{6}\right)^2 S}{\pi  \varepsilon ^3}+\mathcal{O}(\varepsilon^{-2}) \nonumber\\
        &\langle\hat{T}^{r}_{~r}\rangle^-= \frac{3  \llbracket K^\theta_\theta\rrbracket -4\pi S}{1440\pi^2 r_0 \varepsilon^2}+\frac{\left(\xi -\frac{1}{6}\right)^2 S}{ r_0\pi  \varepsilon ^2}+O(\varepsilon^{-1}),\nonumber\\
\end{align}
where $\llbracket K^\tau_\tau\rrbracket $ and $\llbracket K^\theta_\theta\rrbracket $ are defined in Eq. (\ref{eq:Kab}). We note that $\langle\hat{T}^{\mu}_{~~\nu}\rangle^{-}$ is a conserved quantity, relative to the interior metric, to leading order in $\varepsilon$.

The analysis for the exterior proceeds in a similar manner and yields the following expressions:
\begin{align}
\label{eq:ext}
 &\langle\hat\phi^2\rangle^{+} =\frac{\left(\xi -\frac{1}{6}\right) S}{2\pi \varepsilon} +O(1),\nonumber\\
      &   \langle\hat{T}^{t}_{~t}\rangle^+ =\frac{3  \llbracket K^\tau_\tau\rrbracket -4\pi S}{1440\pi^2\varepsilon^3}+\frac{\left(\xi -\frac{1}{6}\right)^2 S}{\pi  \varepsilon ^3}+\mathcal{O}(\varepsilon^{-2})\nonumber\\
      &      \langle\hat{T}^{\theta}_{~\theta}\rangle^+=\frac{3  \llbracket K^\theta_\theta\rrbracket -4\pi S}{1440\pi^2\varepsilon^3}+\frac{\left(\xi -\frac{1}{6}\right)^2 S}{\pi  \varepsilon ^3}+\mathcal{O}(\varepsilon^{-2}) \nonumber\\
        &\langle\hat{T}^{r}_{~r}\rangle^+= \frac{
        (3 r_0 f'(r_0)-4)(360\left(\xi -\frac{1}{6}\right)^2 -1)S}{1440 \pi r_0 \sqrt{f(r_0)} \varepsilon^2 }\nonumber\\
        &-\frac{(3r_0f'(r_0) \llbracket K^\tau_\tau\rrbracket  +f(r_0)\llbracket K^\theta_\theta\rrbracket  )} {1920\pi^2 r_0 \sqrt{f(r_0)}\varepsilon^2}+O(\varepsilon^{-1})\nonumber\\
\end{align}
where $\varepsilon$ is now the proper spatial distance outside the surface defined as 
\begin{align}
\varepsilon=\int_{r_0}^rf(r')^{-1/2} = f(r_0)^{-1/2}(r-r_0) +O\bigl((r-r_0)^2\bigr).
\end{align}
Once more $\langle\hat{T}^{\mu}_{~~\nu}\rangle^{+}$ is a conserved quantity to leading order in $\varepsilon$,  relative to the exterior Schwarzchild  metric.

Finally, comparing the expressions in (\ref{eq:ext}) with Eq. (\ref{eq:int}) and Eq. (\ref{eq:vpolshell}) we observe the following relationships between the leading order interior and exterior near shell divergences
\begin{align}
&\langle\hat\phi^2\rangle^{+}_{1}=\langle\hat\phi^2\rangle^{-}_{1} ,~~~\langle\hat{T}^{t}_{~t}\rangle^{+}_{3}=\langle\hat{T}^{t}_{~t}\rangle^{-}_{3},~~~\langle\hat{T}^{\theta}_{~\theta}\rangle^{+}_{3}=\langle\hat{T}^{\theta}_{~\theta}\rangle^{-}_{3} 
\end{align}
where, for example, $\langle\hat\phi^2\rangle^{\pm}_{1}$ denotes the coefficient of $\varepsilon^{-1}$ in $\langle\hat\phi^2\rangle^{\pm}$. We also note that the RSET has the correct trace in both the interior and exterior, however this is guaranteed from the construction of the RSET in Eq. (\ref{eq:RSETDef}).

When expressed in terms of the extrinsic curvature of the thin shell, as in Eqs.~\eqref{eq:int} and \eqref{eq:ext}, the divergence structure of the RSET is found to be in agreement with the divergence obtained for a spacetime with a boundary, as studied by Deutsch and Candelas in Ref.~\cite{PhysRevD.20.3063}, where they derived the leading-order divergences of the components of the RSET and their coefficients as the boundary is approached using a Green's function approach and conservation arguments. In principle, a stronger divergence proportional to $\varepsilon^{-4}$ could be present, as is found in~\cite{PhysRevD.20.3063}. Such a term appears if the shell enforces Dirichlet, Neumann, or Robin boundry conditions to the field. In our case, the shell is transparent, hence the singularities appearing in the RSET are weaker and due to gravitational effects only. 

\subsection{Vacuum polarization and renormalized stress-energy tensor at the center}
Here we present a calculation of the vacuum polarization and the RSET components at $r=0$. The starting point for this calculation is to point-split the Euclidean Green function in the radial direction only and place the inner-most point at $r=0$. The reason for this choice is apparent from consideration of the Frobenius series for the $p^{\text{shell}}_{\omega l}$ mode about the regular singular point $r=0$:
\begin{equation}
\label{eq:frobienus}
p^{\text{shell}}_{\omega l}= a_0 r^{l} + O(r^{l+1}),
\end{equation}
where $a_0$ is an arbitrary constant which we choose to set to unity.
Hence, placing the innermost radial point (the argument of $p^{\text{shell}}_{\omega l}$) at $r=0$, we observe that the only contribution to the Green function comes from the $l=0$ mode. Explicitly:
\begin{align}
\label{eq:Gh}
 & \{W(x,x')\} = \frac{1}{8\pi^2} \int_{-\infty}^{\infty}\bar{q}^{\text{ shell}}_{\omega 0}(r) d \omega-\frac{1}{4 \pi^2 r^2},
\end{align}
where the notation $\{\}$ denotes that the partial coincidence limit $\tau\to \tau',\theta\to\theta',\phi\to\phi'$ has been taken and we have defined:
\[\bar{q}^{\text{ shell}}_{\omega l}=\frac{q^{\text{ shell}}_{\omega l}}{N^{\text{shell}}_{\omega l}}.\]
Similarly, by taking the required derivatives of the point separated Green function, letting $\tau\to \tau',\theta\to\theta',\phi\to\phi'$ and finally placing the innermost radial point at $r=0$, we obtain:
\begin{align}
       &   \{\nabla_{\theta}\nabla^{\theta}W(x,x')\} = \frac{1}{8\pi^2 r} \int_{-\infty}^{\infty}\frac{d \bar{q}^{\text{ shell}}_{\omega 0}(r)}{dr} d \omega +\frac{1}{2 \pi^2 r^4}\nonumber\\
            &   \{\nabla_{\tau}\nabla^{\tau'}W(x,x')\} = -\{\nabla_{\tau}\nabla^{\tau}W(x,x')\} \nonumber\\
            &=\frac{1}{8\pi^2} \int_{-\infty}^{\infty}\tilde{\omega} ^2 \bar{q}^{\text{ shell}}_{\omega 0}(r) d \omega -\frac{1}{2 \pi^2 r^4}\nonumber\\
\end{align}
where we have used the connection coefficients $\Gamma^{r}_{\tau\tau}=0$ and $\Gamma^{r}_{\theta \theta}=r$.
For the remaining derivatives taken at both points, we require the $l=1$ modes. To see this, we note that:
  \begin{align}
       &   \{\nabla_{\theta}\nabla^{\theta'}W(x,x')\} = \frac{1}{8 \pi^2} \frac{1}{r r'}\int_{-\infty}^{\infty} \sum_{l=0}^{\infty}(2l+1) \frac{l(l+1)}{2} \nonumber\\
      & \times (r')^l \bar{q}^{\text{ shell}}_{\omega l}(r) d\omega +\mathcal{O}\left((r')^{l+1}\right)-\frac{1}{2 \pi^2 (r-r')^4},
\end{align}
where we have used Eq (\ref{eq:frobienus}) and the bi-vector of parallel transport
$$g^{\theta'}_{~\theta}=\frac{1}{r r'}.$$ 
In limit $r' \to 0$, we see that only the $l=1$ contributes to the sum over $l$, giving:
\begin{align}
\label{eq:Gththp}
       &   \{\nabla_{\theta}\nabla^{\theta'}W(x,x')\} = \frac{3}{8 \pi^2 r}\int_{-\infty}^{\infty} \bar{q}^{\text{ shell}}_{\omega 1}(r) d\omega-\frac{1}{2 \pi^2 r^4}.
\end{align}
Similarly, we obtain:
\begin{align}
\label{eq:Grrp}
       &   \{\nabla_{r}\nabla^{r'}W^{}(x,x')\} = \frac{3}{8 \pi^2}\int_{-\infty}^{\infty} \frac{d \bar{q}^{\text{~shell}}_{\omega 1}(r_0)}{dr} d\omega+\frac{3}{2 \pi^2 r^4}.
\end{align}
To perform these integrals, we require expressions for the $\bar{q}^{\text{ shell}}_{\omega 0}$ and $\bar{q}^{\text{ shell}}_{\omega 1}$ modes valid in the vicinity of $r=0$. To obtain these, we exploit the fact that, for $r\leq r_0$ $p^{\text{shell}}_{\omega l}(r)=p^{M}_{\omega l}(r)$, given by Eq.~(\ref{eq:ModesMink}).
We may then integrate the Wronksian condition to obtain an expression for $\bar{q}^{\text{ shell}}_{\omega l}$ valid for $r\leq r_0$:
\begin{align}
  \bar{q}^{\text{ shell}}_{\omega l}(r)= p^{\text{shell}}_{\omega l}(r)\bigg(\int_{r}^{r_0}\frac{1}{r'^2 \sqrt{f(r_0)} (p^{\text{M}}_{\omega l}(r'))^2} dr'\nonumber\\ +  \int_{r_0}^{\infty}\frac{1}{r'^2 f(r') (p^{\text{shell}}_{\omega l}(r'))^2} dr'\bigg)
\end{align}
We are able to perform the first integral exactly and noting the the second integral is a constant in $r$, we obtain the result:
\begin{align}
  \bar{q}^{\text{ shell}}_{\omega l}(r)= q^{-}_{\omega l}(r)+ \alpha_{\omega l} p_{\omega l}(r),
\end{align}
with
\begin{align}
\label{eq:q-}
   q^{-}_{\omega l}(r)= q^{\text{M}}_{\omega l}(r) -  \frac{I_{l+\frac{1}{2}}(  \tilde{\omega} r)}{I_{l+\frac{1}{2}}(\tilde{\omega} r_0) }q^{\text{M}}_{\omega l}(r_0).
\end{align}
Since $q^{-}_{\omega l}(r_0)=0$ by definition, we must have that:
\begin{align}
\label{eq:q}
  \bar{q}^{\text{ shell}}_{\omega l}(r)= q^{-}_{\omega l}(r)+ \frac{\bar{q}^{\textrm{ S}}_{\omega l}(r_0)}{p^{\textrm{M}}_{\omega l}(r_0)} p^{\text{shell}}_{\omega l}(r),
\end{align}
where we have made use of $p^{\text{shell}}_{\omega l}(r_0)=p^{M}_{\omega l}(r_0)$, $q^{\text{shell}}_{\omega l}(r_0)=q^{\textrm{S}}_{\omega l}(r_0)$.

For $l=0$, we are able to perform the required integrals over $\omega$ of $q^{-}_{\omega 0}$. For example, inserting Eq.(\ref{eq:q}) into Eq.(\ref{eq:Gh}) yields:
\begin{align}
&\{W(x,x')\}= \frac{\cot \left(\frac{\pi r}{2 r_0}\right)}{8 \pi  r r_0}-\frac{1}{4 \pi^2 r^2}+\frac{1}{4\pi^2} \int_{0}^{\infty} \frac{\bar{q}^{\textrm{ S}}_{\omega0}(r_0)}{p^{\textrm{M}}_{\omega 0}(r_0)} d\omega.
\end{align}
Expanding about $r=0$, we see that the divergent terms cancel yielding the following result for the RVP at $r=0$:
\begin{align}
\label{eq:phisq}
\langle\hat{\phi}^2\rangle_{ren}=&-\frac{1}{48 r_0^2}+\frac{1}{4\pi^2} \int_{0}^{\infty} \frac{\bar{q}^{\textrm{ S}}_{\omega0}(r_0)}{p^{\textrm{M}}_{\omega 0}(r_0)} d\omega.
\end{align}
As $\bar{q}^{\text{ shell}}_{\omega l}(r_0)$ must be computed numerically, the second integral in the above expression can only be calculated numerically.  Similarly, we obtain the results at $r=0$:
\begin{align}
       & \mathsf{w}_{~\theta}^{\theta}= -\frac{\pi^2}{1440 r_0^4}+\frac{1}{12\pi^2} \int_{0}^{\infty}\tilde{\omega}^2 \frac{\bar{q}^{\textrm{ S}}_{\omega0}(r_0)}{p^{\textrm{M}}_{\omega 0}(r_0)} d\omega\nonumber\\
            & \mathsf{w}_{~\tau}^{\tau'}= -\mathsf{w}_{~\tau}^{\tau}=3\mathsf{w}_{~\theta}^{\theta}\nonumber\\
\end{align}
where the square brackets indicate that the full coincidence limit has been taken and we have made use of:
$$\frac{d p^{\textrm{M}}_{\omega 0}(r_0)}{dr}\bigg|_{r=0}=\frac{1}{3f(r_0)},$$ 
which follows from the $r=0$ expansion of $p^{\textrm{M}}_{\omega 0}$.
For the integrals involving the $l=1$ modes, we are unable to carry out the required integrals over omega of $q^{-}_{\omega 1}$. However, we note from Eq. (\ref{eq:q-}) that the $r \to 0$ divergence in $q^{-}_{\omega 1}$ comes entirely from $q^{\text{M}}_{\omega 1}$. Hence we can extract the divergent parts of the required integrals through the results:
\begin{align}
  &  \frac{3}{8 \pi^2 r}\int_{-\infty}^{\infty} q^{\text{M}}_{\omega 1}(r) d\omega=\frac{1}{2\pi^2 r^4},\\
  &  \frac{3}{8 \pi^2 r}\int_{-\infty}^{\infty}\frac{d q^{\text{M}}_{\omega 1}(r)}{dr} d\omega=-\frac{3}{2\pi^2 r^4}.
\end{align}
Inserting these results into Eqs. (\ref{eq:Gththp}) and (\ref{eq:Grrp}) and taking the $r=0$ limit, yields:
\begin{align}
       \mathsf{w}_{~\theta}^{\theta'} &=  \mathsf{w}_{~r}^{r'}
    =\frac{\mathscr{I}}{r_0^4}+\frac{3}{4 \pi^2 }\int_{0}^{\infty} \frac{\bar{q}^{\textrm{ S}}_{\omega 1}(r_0)}{p^{\textrm{M}}_{\omega 1}(r_0)} d\omega,
\end{align}
where $\mathscr{I}$ is given by:
\begin{align}
  \mathscr{I}=  -\frac{1}{12 \pi^2 }\int_{0}^{\infty} \frac{e^{-x} x^3 (x+1)}{x \cosh (x)-\sinh (x)} dx\approx -0.03955 .
\end{align}
The final required derivative of the Green function is obtained via the wave equation satisfied by $W(x,x') $ for a massless field:
\begin{align}
\mathsf{w}_{~r}^{r}&=-\mathsf{w}_{~\tau}^{\tau}-2 \mathsf{w}_{~\theta}^{\theta}=\mathsf{w}_{~\theta}^{\theta}.
\end{align}
Hence, we see that for massless fields, the RSET at $r=0$ depends only on $[ \nabla_{\theta}\nabla^{\theta}G_{\textrm{B}}^{}(x,x')]$ and $[ \nabla_{\theta}\nabla^{\theta'}G_{\textrm{B}}^{}(x,x')]$. Drawing together our results above, we arrive at the following expression for the RSET components at $r=0$ for a massless field:
\begin{align}
\label{eq:RSET}
&\langle \hat{T}^{r}_{~r}\rangle=\langle \hat{T}^{\theta}_{~\theta}\rangle= \frac{\pi^2}{2880 r_0^4} (3-8\xi) +\frac{\mathscr{I}}{2r_0^4}(8 \xi-1)\nonumber\\
&+\frac{1}{4\pi^2}\int_{0}^{\infty}\bigg[\frac{(8\xi-3)\tilde{\omega}^2}{6}\frac{\bar{q}^{\textrm{ S}}_{\omega0}(r_0)}{p^{\textrm{M}}_{\omega 0}(r_0)} +\frac{3(8\xi-1) }{2}\frac{\bar{q}^{\textrm{ S}}_{\omega 1}(r_0)}{p^{\textrm{M}}_{\omega 1}(r_0)}\bigg]d\omega,\nonumber\\
&\langle \hat{T}^{t}_{~t}\rangle= -\frac{\pi^2}{960 r_0^4} (1-8\xi) +\frac{3\mathscr{I}}{2r_0^4}(4 \xi-1)\nonumber\\
&+\frac{1}{4\pi^2}\int_{0}^{\infty}\bigg[\frac{(1+4\xi)\tilde{\omega}^2}{2}\frac{\bar{q}^{\textrm{ S}}_{\omega0}(r_0)}{p^{\textrm{M}}_{\omega 0}(r_0)} +\frac{9(4\xi-1)}{2}\frac{\bar{q}^{\textrm{ S}}_{\omega 1}(r_0)}{p^{\textrm{M}}_{\omega 1}(r_0)}\bigg]d\omega.
\end{align}
The above expressions and the equivalent expression for $\langle \hat{\phi}^2\rangle_{\textrm{ren}}$ in Eq~(\ref{eq:phisq}) are in excellent agreement with the numerical $r>0$ results.

In the black hole limit $r_0 \to 2M$, we observe that the numerical integrals present in the $r=0$ expressions for the vacuum polarisation and the RSET components tend to $0$ for all parameter values considered. To examine this further, we first note that the integrand at $\omega=0$ diverges logarithmically as $r_0 \to 2M$. Therefore, in order to meaningfully consider the $r_0 \to 2M$ limit, we consider instead:
\begin{align}
    \int_{0}^{\infty} \left( \frac{\bar{q}^{\textrm{ S}}_{\omega l }(r_0)}{p^{\textrm{M}}_{\omega l}(r_0)} - \frac{\bar{Q}_{l}\left(\frac{r_0}{M}-1\right)}{p^{\textrm{M}}_{\omega l}(r_0)} \right) d \omega + \int_{0}^{\infty} \frac{\bar{Q}_{l}\left(\frac{r_0}{M}-1\right)}{ p^{\textrm{M}}_{\omega l}(r_0)} d \omega.
\end{align}
Here we have made use of the fact that for $\omega=0$, the radial equation in Schwarzchild reduces to the Legendre equation, here $\bar{Q}_l(x)$ denotes the Legendre function of the second kind that has been appropriately normalised to ensure agreement with $\bar{q}^{\textrm{ S}}_{0l}(r_0)$ .  $p^{\textrm{M}}_{\omega l}(r_0)$ is a strictly positive and increasing function of $\omega$, hence the first integral is uniformly convergent in $r_0$ (at least in the neighbourhood of $2M$) and we may take the  $r_0 \to 2M$ limit inside the integral. For $l = 0$ and $l = 1$ we are able to carry out the
second integral before taking the $r_0 \to 2M$ limit. In both
cases this second integral is $\mathcal{O}\left(\sqrt{r_0-2M}\log(r_0-2M)\right)$ and hence vanishes in the black hole limit.

Returning to the first integral, for fixed $l,\omega$ and as $r_0 \to 2M$:
\begin{align}
    &p^{\textrm{M}}_{\omega l}(r_0)\sim\sqrt{r_0-2M}~e^{\tfrac{(2 M)^{3/2} \omega }{\sqrt{r_0-2M}}}\nonumber\\
    &\bar{q}^{\textrm{ S}}_{\omega l}(r_0)\sim (r_0-2M)^{-2 M \omega}.\nonumber
\end{align}
Hence, for fixed $l,\omega$, the integrand tends to $0$ as $r_0 \to 2M$. For large $\omega$ the integrand decays exponentially in $\omega$ for $r_0>2M$.  Therefore we arrive at the, somewhat surprising, result that the vacuum polarisation and the RSET at $r=0$ all remain finite in the black hole limit and are given by the expressions:
\begin{align}
 & \langle\hat\phi^2\rangle_{\textrm{ren}}=-\frac{1}{192 M^2}  \nonumber\\
&  \langle \hat T^{r}_{~r}\rangle=\langle \hat T^{\theta}_{~\theta}\rangle= \frac{\pi^2}{46080 M^4} (3-8\xi) +\frac{\mathscr{I}}{32 M^4}(8 \xi-1)\nonumber\\
&\langle \hat T^{t}_{~t}\rangle= -\frac{\pi^2}{15360 M^4} (1-8\xi) +\frac{3\mathscr{I}}{32 M^4}(4 \xi-1).
\end{align}
We will exploit these results in Subsec.~\ref{subsec:backreaction} to calculate backreaction effects near $r=0$ in the black hole limit.

\subsection{Weak field approximation}
In this section, under the assumption of small shell compactness ($M/r_0\ll 1$), we employ a weak field approximation to analyze the behavior of the renormalized vacuum polarization and RSET in the near-shell limit. Indeed, there exist closed-form analytical expressions for renormalized expectation values on the weak-field approximated spacetime~\cite{Dalvit1994,Satz2005vacuum,boasso2025vacuum}. So, in what follows, we start simply deriving the weak-field approximation before considering a quantum field on the spacetime.

In the weak field approximation, one assumes that the metric $g_{\alpha\beta}$, which in this context satisfies the Einstein equations sourced by the stress-energy tensor of a static, spherical thin shell
\begin{align}
    R_{\alpha\beta}-\frac12 g_{\alpha\beta}R=8\pi \,T_{\alpha\beta}^{\textrm{(shell)}},
\end{align}
can be expressed as a linear perturbation around Minkowski spacetime (see, for instance, \cite{Satz2005vacuum}) $g_{\alpha\beta}\sim\eta_{\alpha\beta}+\epsilon \,g_{\alpha\beta}^{(1)}$, where $\eta_{\alpha\beta}$ is the Minkowski metric and $\epsilon$ is simply a bookkeeping parameter. This approximation is valid under the assumption that $|g^{(1)}_{\alpha\beta}|\ll 1$, which holds for small values of shell compactness: $M/r_0\ll 1$. We employ similar expansions for
\begin{align}
    &R_{\alpha\beta}\sim R^{(0)}_{\alpha\beta}+\epsilon R^{(1)}_{\alpha\beta},\nonumber\\
    &R\sim R^{(0)}+\epsilon R^{(1)},\nonumber\\
    &T_{\alpha\beta}\sim T^{(0)}_{\alpha\beta}+\epsilon \,T^{(1)}_{\alpha\beta},
\end{align}
where $R_{\alpha\beta}^{(0)}=0$, $R^{(0)}=0$ and $T^{(0)}_{\alpha\beta}=0$. We employ an expansion in $\epsilon$ to formally derive the zeroth and linear order Einstein equations. In addition, whenever explicit expressions for the quantities at first order in the $\epsilon$ expansion are required, namely $g^{(1)}_{\alpha\beta}$, $T^{(1)}_{\alpha\beta}$, $R^{(1)}_{\alpha\beta}$ and $R^{(1)}$, a further expansion in the compactness $M/r_0$ is performed for consistency. Hence, the first-order terms in the $\epsilon$ expansion are linear in $M/r_0$ \cite{boasso2025vacuum}. For instance, $T^{(1)}_{\alpha\beta}$ is obtained as a linear order expansion in $M/r_0$ of the distributional stress-energy tensor of the shell in \eqref{eq:distr_set_shell}
\begin{align}
\label{eq:set_shell_linear}
    T^{\tau~(1)}_{~\tau}=-\frac{M}{4 \pi r_0^2}\delta(r-r_0),\nonumber\\
    T^{r~(1)}_{~r}=T^{\theta~(1)}_{~\theta}=T^{\phi~(1)}_{~\phi}=0.
\end{align}
At linear order in $\epsilon$ we obtain for the Einstein equations
\begin{align}
\label{eq:EEs_linear}
   G_{\alpha\beta}^{(1)}= R^{(1)}_{\alpha\beta}-\frac12\eta_{\alpha\beta}R^{(1)}=8\pi \,T^{(1)}_{\alpha\beta}
\end{align}
where the Einstein tensor at linear order is
\begin{align}
\label{eq:einst_tens_linear}
    G_{\alpha\beta}^{(1)}=\frac12\Big[-\bar\Box\, g_{\alpha\beta}^{(1)}+2\bar\nabla_\gamma\bar\nabla_{(\alpha} \,g_{\beta)}^{\gamma~(1)}-\bar\nabla_\alpha\bar\nabla_\beta g^{(1)}\nonumber\\
    -\eta_{\alpha\beta}(\bar\nabla_\gamma \bar\nabla_\delta g^{\gamma\delta~(1)}-\bar\Box g^{(1)})\Big],
\end{align}
and $g^{(1)}=\eta^{\alpha\beta}g_{\alpha\beta}^{(1)}$ is the trace of the linear order metric. Within this section, we adopt the notation $\bar\Box$ and $\bar\nabla$ to highlight that the covariant derivatives are with respect to the flat background spacetime. To compute the perturbation to the metric at linear order $g_{\alpha\beta}^{(1)}$, we assume
\begin{align}
\label{eq:metr_linear_ansatz}
    g_{\alpha\beta}^{(1)}=\textrm{diag}(\mathcal{A}(r),\mathcal{B}(r),0,0),
\end{align}
which is justified by staticity and spherical symmetry. Introducing this ansatz and the stress-energy tensor of the shell at linear order (see Eq.~\eqref{eq:set_shell_linear}) into Eqs.~\eqref{eq:EEs_linear} and \eqref{eq:einst_tens_linear} yields for the $\tau\tau$ and $rr$ components of the Einstein equations
\begin{align}
    &\frac{\partial_r(r\mathcal{B}(r))}{r^2}=\frac{2M}{r_0^2}\delta(r-r_0),\\
    &-\frac{\mathcal{B}(r)}{r^2}+\frac{\mathcal{A}'(r)}{r}=0.
\end{align}
Integrating the first equation gives 
\begin{align}
\label{eq:grr_linear}
    g^{r~(1)}_r&=\mathcal{B}(r)=\frac{2M}{r}\Theta(r-r_0)+\frac{C_1}{r}\nonumber\\
    &=\frac{2M}{r}\Theta(r-r_0),
\end{align}
where we have set the integration constant $C_1=0$, so that the interior spacetime remains Minkowski and the exterior metric retrieves the expected Schwarzschild metric at linear order. Similarly, integrating the second equation, we obtain
\begin{align}
\label{eq:gtautau_linear}
    g^{\tau~(1)}_\tau=&\mathcal{A}(r)=\left(-\frac{2M}{r}+\frac{2M}{r_0}\right)\Theta(r-r_0)+C_2\nonumber\\
    =&\left(-\frac{2M}{r}+\frac{2M}{r_0}\right)\Theta(r-r_0)-\frac{2M}{r_0}.
\end{align}
We have set the integration constant $C_2=-2M/r_0$ to reflect asymptotic flatness as $r\to\infty$. Note also that the $\tau\tau$ component of the metric receives a shift of $-2M/r_0$ (as expected from \eqref{eq:shellmetric}) in the interior, ensuring that $g^{\tau~(1)}_\tau$ is continuous across the shell. Effectively, the perturbation on the metric generates a spacetime that is Schwarzschild (at linear order in $M/r$) for $r>r_0$ and that remains Minkowksi in the interior $r<r_0$. Note that integrating the linearized Einstein equations yields a metric that indeed satisfies the assumption $g_{\alpha\beta}^{(1)}\ll 1$ (if $M/r_0\ll 1$), thereby making the perturbation theory consistent despite the presence of delta distributions in the shell's stress-energy tensor. 

Introducing the metric components  \eqref{eq:grr_linear} and \eqref{eq:gtautau_linear} into the equation for the Ricci tensor at linear order
\begin{align}
\label{eq:ricci_tens_linear}
    R_{\alpha\beta}^{(1)}=\frac12\Big[-\bar\Box\, g_{\alpha\beta}^{(1)}+2\bar\nabla_\gamma\bar\nabla_{(\alpha} \,g_{\beta)}^{\gamma~(1)}-\bar\nabla_\alpha\bar\nabla_\beta g^{(1)}\Big],
\end{align}
we find
\begin{align}
\label{eq:ricci_tens_distr_linear}
    &R^{\tau~(1)}_{~\tau}=-\frac{M}{r_0^2}\delta(r-r_0)=\bar R^{\tau~(1)}_{~\tau}\delta(r-r_0),\nonumber\\
    &R^{r~(1)}_{~r}=\frac{M}{r_0^2}\delta(r-r_0)=\bar R^{r~(1)}_{~r}\delta(r-r_0),\nonumber\\
    &R^{\theta~(1)}_{~\theta}=\frac{M}{r_0^2}\delta(r-r_0)=\bar R^{\theta~(1)}_{~\theta}\delta(r-r_0),
\end{align}
and
\begin{align}
\label{eq:ricci_scal_distr_linear}
    R^{(1)}=R^{\alpha~(1)}_{~\alpha}&=\frac{2M}{r_0^2}\delta(r-r_0)=\bar R^{(1)}\,\delta(r-r_0)\nonumber\\
    &=-8\pi  S^{(1)}\,\delta(r-r_0),
\end{align}
where we have defined for later convenience
\begin{align}
\label{eq:linear_bar_R}
    \bar R^{r~(1)}_{~r}&=
    \bar R^{\theta~(1)}_{~\theta}=-\bar R^{\tau~(1)}_{~\tau}=\frac{M}{r_0^2},\nonumber\\
    \bar R^{(1)}&=-8\pi S^{(1)}=\frac{2M}{r_0^2}.
\end{align}
The linearized expressions for the Ricci scalar and tensor satisfy the contracted Bianchi identities with respect to a flat background.

The RVP and the RSET of a massless scalar field in the Boulware state can be obtained from the one-loop effective action computed using covariant perturbation theory \cite{barvinsky1987beyond,barvinsky1990covariant,avramidi1991covariant}. In the weak field regime and at linear order in $R^{(1)}$, these quantities can be expressed as \cite{Dalvit1994,satzetal2005,boasso2025vacuum}
\begin{align}
\label{eq:phi2_weak}
    \langle\hat\phi^2\rangle_\textrm{ren}=\frac{1}{16 \pi^2}\left(\xi-\frac16\right) \log\left(\frac{-\bar\Box}{\mu^2}\right)R^{(1)}
\end{align}
and
\begin{align}
\label{eq:Tmunu_weak}
	\langle \hat T^\alpha_{~\beta}\rangle_\textrm{ren}= & -\frac{\left(\xi-\frac{1}{6}\right)^{2}}{16\pi^{2}}(\bar\nabla^{\alpha}\bar\nabla_{\beta}-\delta^{\alpha}_{~\beta}\bar\Box) \log\left(\frac{-\bar\Box}{\mu^2}\right)R^{(1)}(x) \nonumber\\
	&- \frac{1}{5760\pi^{2}}\Bigg[ (2\bar\nabla^{\alpha}\bar\nabla_{\beta}+\delta^{\alpha}_{~\beta}\bar\Box) \log\left(\frac{-\bar\Box}{\mu^2}\right)R^{(1)}(x) \nonumber\\
    &\qquad\qquad\quad - 6\bar\Box\log\left(\frac{-\bar\Box}{\mu^2}\right)R^{\alpha~(1)}_{~\beta}(x) \Bigg],
\end{align}
where the nonlocal operator $\log(-\bar\Box)$ acting on a time-independent scalar function can be expressed as \cite{boasso2025vacuum}
\begin{equation}
	\log\left(\frac{-\bar\Box}{\mu^2}\right)\varphi(\vec{x}) = -\frac{1}{2\pi}\int\mathrm{d}^{3}\vec{x}'\, \frac{\varphi(\vec{x}')}{|\vec{x}-\vec{x}'|^{3}}\, ,
	\label{eq:log_op}
\end{equation}
with $\varphi$ denoting either $R^{(1)}$ or $R^{\alpha~(1)}_{~\beta}$ and assumed to vanish at infinity. In this Section, we consider massless scalar fields, so $\mu$ here is simply an arbitrary parameter. Note that the local terms dependent on $\log(\mu^2)$ have been dropped since $R^{(1)}(x)=0$ and $R^{\alpha~(1)}_\beta(x)=0$ at spacetime points away from the shell.

What we need for deriving the RVP and the RSET is to compute the terms of the form $\log\left(\frac{-\bar\Box}{\mu^2}\right)R^{(1)}$ and $\log\left(\frac{-\bar\Box}{\mu^2}\right)R^{\alpha~(1)}_{~\beta}(x)$. We leave the details of this computation to Appendix~\ref{app:weak_field}, and just present the results here.

Introducing the expression \eqref{eq:logR_results} into \eqref{eq:phi2_weak} yields the expression for the renormalized vacuum polarization
\begin{align} \label{eq:VP_weakfield}      \langle\hat\phi^2\rangle_\textrm{ren}=\begin{cases}-\frac{1}{8\pi^2}\left(\xi-\frac16\right)\frac{\bar R^{(1)} \,r_0}{(r_0-r)(r+r_0)}, \qquad \;r<r_0,\\
-\frac{1}{8\pi^2}\left(\xi-\frac16\right)\frac{\bar R^{(1)}\,r_0^2}{r(r-r_0)(r+r_0)}, \qquad r>r_0. \end{cases}
\end{align}
Thus, taking the near-surface limit gives on the interior
\begin{align}
\label{eq:phi2_weak_int}
    \langle\hat\phi^2\rangle^-_\textrm{ren}=-\frac{\left(\xi-\frac16\right)\bar R^{(1)}}{16\pi^2\varepsilon}+\mathcal{O}(1),
\end{align}
and on the exterior
\begin{align}
\label{eq:phi2_weak_ext}
    \langle\hat\phi^2\rangle^+_\textrm{ren}=-\frac{\left(\xi-\frac16\right)\bar R^{(1)}}{16\pi^2\varepsilon}+\mathcal{O}(1),
\end{align}
where we adopt a notation consistent with Sec.~\ref{sec:analytic}.

Taking the relevant derivatives of the expressions for $\log\left(\frac{-\bar\Box}{\mu^2}\right)R^{(1)}$ and $\log\left(\frac{-\bar\Box}{\mu^2}\right)R^{\alpha~(1)}_{~\beta}$ derived in Appendix~\ref{app:weak_field} and substituting in \eqref{eq:Tmunu_weak} yields the components of the RSET.
In the near-surface limit, the weak field approximation gives the following predictions for the leading order divergences. On the interior,
\begin{align}
\label{eq:Tab_weak_int}
    &\langle \hat T^{t}_{~t}\rangle^-_\textrm{ren}=\frac{-6\bar R^{\tau~(1)}_{~\tau}+\bar R^{(1)}}{2880\pi^2\varepsilon^3}-\left(\xi -\tfrac{1}{6}\right)^2\frac{ \bar R^{(1)}}{8 \pi^2  \varepsilon^3}+\mathcal{O}(\varepsilon^{-2}),\nonumber\\
    &\langle \hat T^{\theta}_{~\theta}\rangle^-_\textrm{ren}=\frac{-6 \bar R^{\theta~(1)} _{~\theta}+  \bar R^{(1)}}{2880 \pi ^2 \varepsilon^3}-\left(\xi -\tfrac{1}{6}\right)^2\frac{ \bar  R^{(1)}}{8\pi^2  \varepsilon ^3}+\mathcal{O}(\varepsilon^{-2}),\nonumber\\
    &\langle \hat T^{r}_{~r}\rangle^-_\textrm{ren}=\frac{-6\bar R^{r~(1)}_{~r}+\bar R^{(1)}}{2880 \pi ^2\, r_0 \,\varepsilon ^2}-\left(\xi -\tfrac{1}{6}\right)^2\frac{ \bar  R^{(1)}}{8\pi^2\, r_0\,  \varepsilon^2}+\mathcal{O}(\varepsilon^{-1}),
\end{align}
and on the exterior
\begin{align}
\label{eq:Tab_weak_ext}
    &\langle \hat T^{t}_{~t}\rangle^+_\textrm{ren}=\frac{-6 \bar R^{\tau~(1)}_{~\tau}+\bar R^{(1)}}{2880\pi^2\varepsilon^3}-\left(\xi -\tfrac{1}{6}\right)^2\frac{ \bar  R^{(1)}}{8\pi^2  \varepsilon ^3}+\mathcal{O}(\varepsilon^{-2}),\nonumber\\
    &\langle \hat T^{\theta}_{~\theta}\rangle^+_\textrm{ren}=\frac{-6 \bar R^{\theta~(1)} _{~\theta}+\bar R^{(1)}}{2880 \pi ^2 \varepsilon ^3}-\left(\xi -\tfrac{1}{6}\right)^2\frac{ \bar  R^{(1)}}{8\pi^2  \varepsilon ^3}+\mathcal{O}(\varepsilon^{-2}),\nonumber\\
    &\langle \hat T^{r}_{~r}\rangle^+_\textrm{ren}=\frac{6 \bar R^{r~(1)}_{~r}-\bar R^{(1)}}{2880 \pi ^2 \,r_0 \,\varepsilon ^2}+\left(\xi -\tfrac{1}{6}\right)^2\frac{ \bar  R^{(1)}}{8\pi^2 \,r_0\,  \varepsilon ^2}+\mathcal{O}(\varepsilon^{-1}).
\end{align}
where the quantities $\bar R^{(1)}$, $\bar R^{\tau~(1)}_{~\tau}$, $\bar R^{r~(1)}_{~r}$ and $\bar R^{\theta~(1)}_{~\theta}$ are defined in~\eqref{eq:linear_bar_R}.
Both the interior and exterior components satisfy, at leading order in $\varepsilon$, the conservation equation with respect to flat spacetime.

To compare these weak-field approximations to the WKB approximations of the previous section, we must expand the WKB expressions in \eqref{eq:vpolshell}-\eqref{eq:ext} in the small compactness parameter $M/r_0$, assuming this to be small. This ensures we are evaluating the WKB approximation in the same domain of validity as the weak-field approximation. When we do this, we find agreement between the two approximations for both the RVP and the RSET components.

\section{Results}\label{sec:results}

We have been able to calculate the exact values of $\langle\hat{\phi}^{2}\rangle_\textrm{ren}$ and $\langle\hat{T}^{\alpha}_{~\beta}\rangle_\textrm{ren}$ at $r=0$ and their leading order divergences as the surface of the shell is approached. In this Section we apply the extended-coordinate prescription to numerically compute the RVP and the RSET in the bulk of the spacetime. We will first validate numerically the near-shell expansions obtained in Section~\ref{sec:analytic}. After this, we will obtain, for a minimally coupled field, the RVP and RSET generated by the thin shell as its surface approaches its own Schwarzschild radius. 
Finally, we consider a shell very close to the black hole limit and analyze how $\langle\hat{\phi}^{2}\rangle_\textrm{ren}$ and $\langle\hat{T}^{\alpha}_{~\beta}\rangle_\textrm{ren}$ compare to the values of the Boulware vacuum of a Schwarzschild black hole for minimally and non-minimally coupled fields as we move away from the shell. 

In all the numerical results shown in what follows, we have obtained the Euclidean modes guaranteeing a constant Wronskian up to 25 decimal places. This accuracy is enough to obtain reliable numerical results. We considered the wide range of $\omega$ values $\omega M \in \left[10^{-10},20\right]$, and summed up to $l = 30$ unless specified otherwise.
The accuracy of the obtained RSETs was verified through conservation, which we checked to be approximately $3-6$ orders of magnitude smaller than the individual RSET components up to close to the thin shell, where our numerical results poorly estimate the divergence --- as expected --- since it depends exclusively on the high-$\omega$ and high-$l$ behavior that is being truncated away.

\subsection{Numerical verification of surface divergences}\label{Subsec:SurfDivs}
\begin{figure*}
        \centering
        \includegraphics[width=\linewidth]{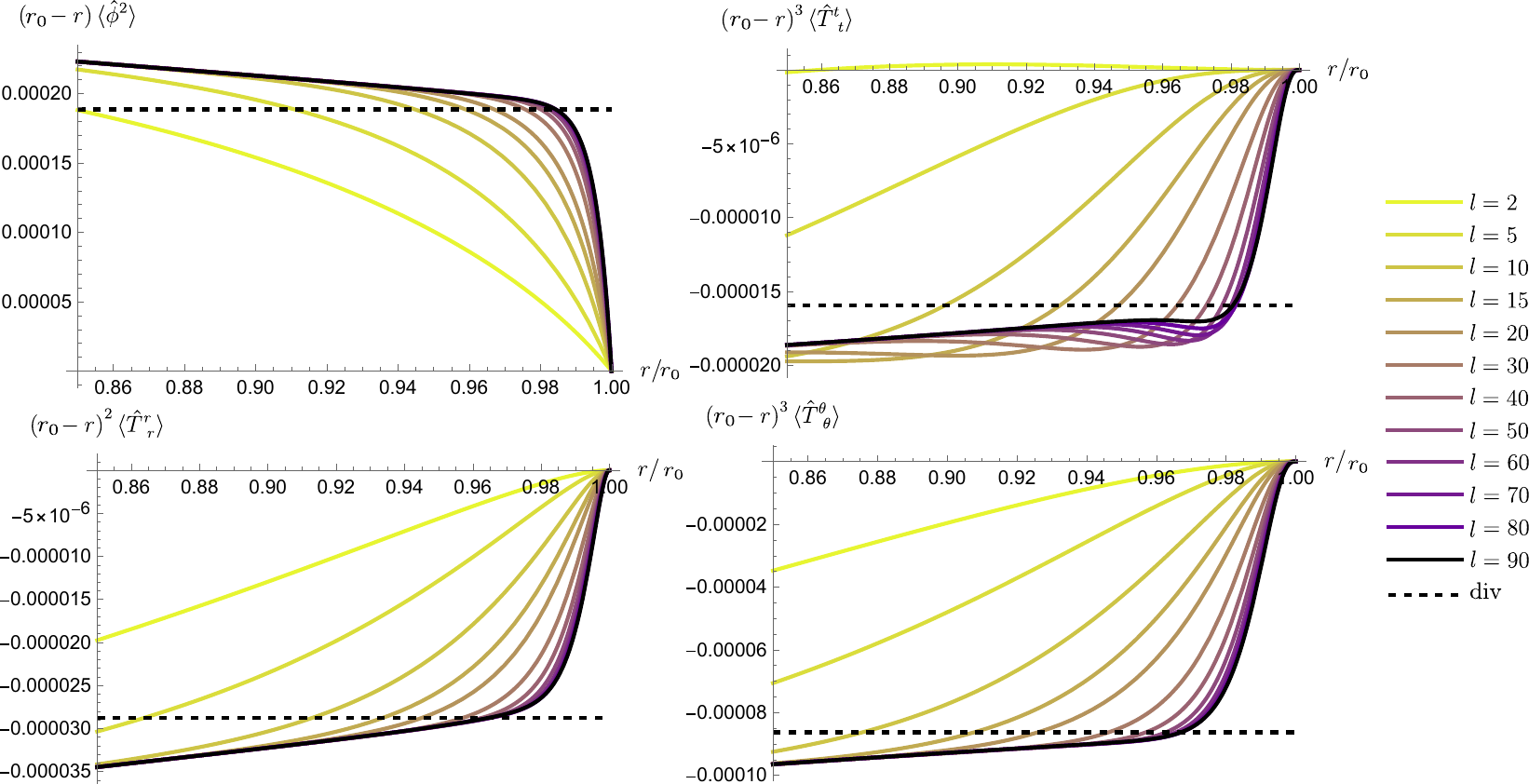}
        \caption{Renormalized vacuum polarization and renormalized stress-energy tensor of a massless, minimally coupled scalar field inside a spherical thin shell of radius $r_{0}=3M$. The colored curves are purely numerical results obtained by summing up to increasing $\ell$ values and integrating up to $\omega M=20$. The black dashed lines correspond to the coefficient of the leading-order analytic divergence estimated through the WKB approximation. A tendency to approximate these values is visible for the vacuum polarization and all RSET components.}
        \label{Fig:IntDiv}
\end{figure*}

\begin{figure*}
        \centering
        \includegraphics[width=\linewidth]{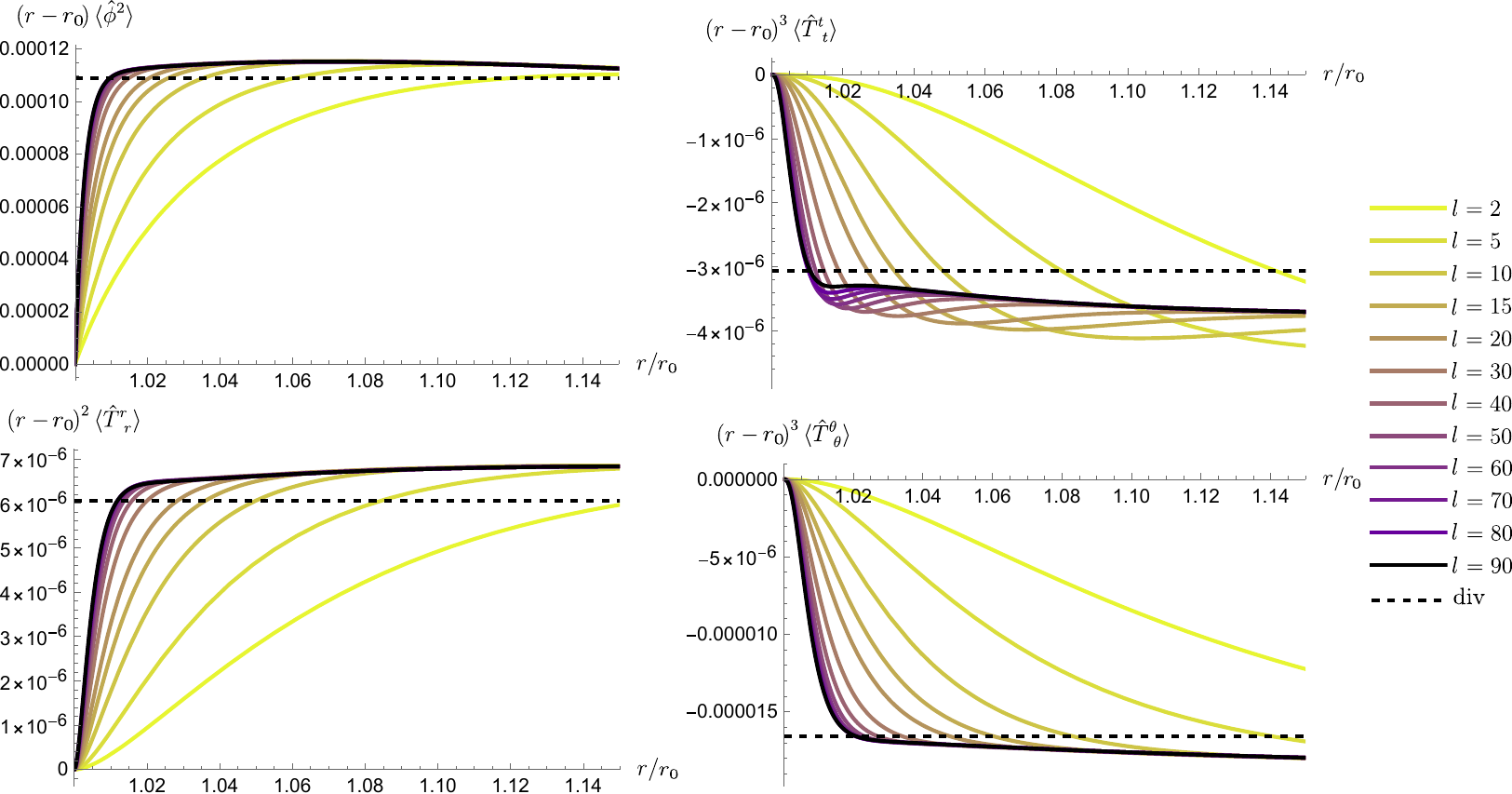}
        \caption{Renormalized vacuum polarization and renormalized stress-energy tensor of a massless, minimally coupled scalar field outside a spherical thin shell of radius $r_{0}=3M$. The colored curves are purely numerical results obtained by summing up to increasing $l$ values and integrating up to $\omega M=20$. The black dashed lines correspond to the coefficient of the leading-order analytic divergence estimated through the WKB approximation. A tendency to approximate these values is visible for the vacuum polarization and all RSET components.}
        \label{Fig:ExtDiv}
    \end{figure*}

Fig.~\ref{Fig:IntDiv} and~\ref{Fig:ExtDiv} show the RVP and RSET of a massless and minimally coupled scalar field on both sides of a spherical thin shell of radius $r_{0}=3M$. We have rescaled these quantities to subtract their respective divergent behaviors, and compared them with the analytic coefficients~\eqref{eq:int} and~\eqref{eq:ext} obtained in Subsec.~\ref{Subsec:WKB}. As we incorporate more $l$ modes to our $l$-sums, we observe that the contribution from higher multipoles becomes increasingly negligible far from the surface. This makes it difficult to verify that numerical results reproduce the correct divergent behavior, as this would require computing tremendous amounts of modes. Nonetheless, we take the results in Figures~\ref{Fig:IntDiv} and~\ref{Fig:ExtDiv} as strong numerical indications in favor of the divergent behavior identified in Eqs.~\eqref{eq:int} and~\eqref{eq:ext}. We chose to show results for a shell of $R=3M$ because bigger shells require less $l$ modes to observe the surface divergences approximated clearly. We have checked that more compact shells show a similar agreement between numerical results and analytic expressions.

\subsection{Varying the compactness}\label{Subsec:Comp}
\begin{figure*}
        \centering
        \includegraphics[width=\linewidth]{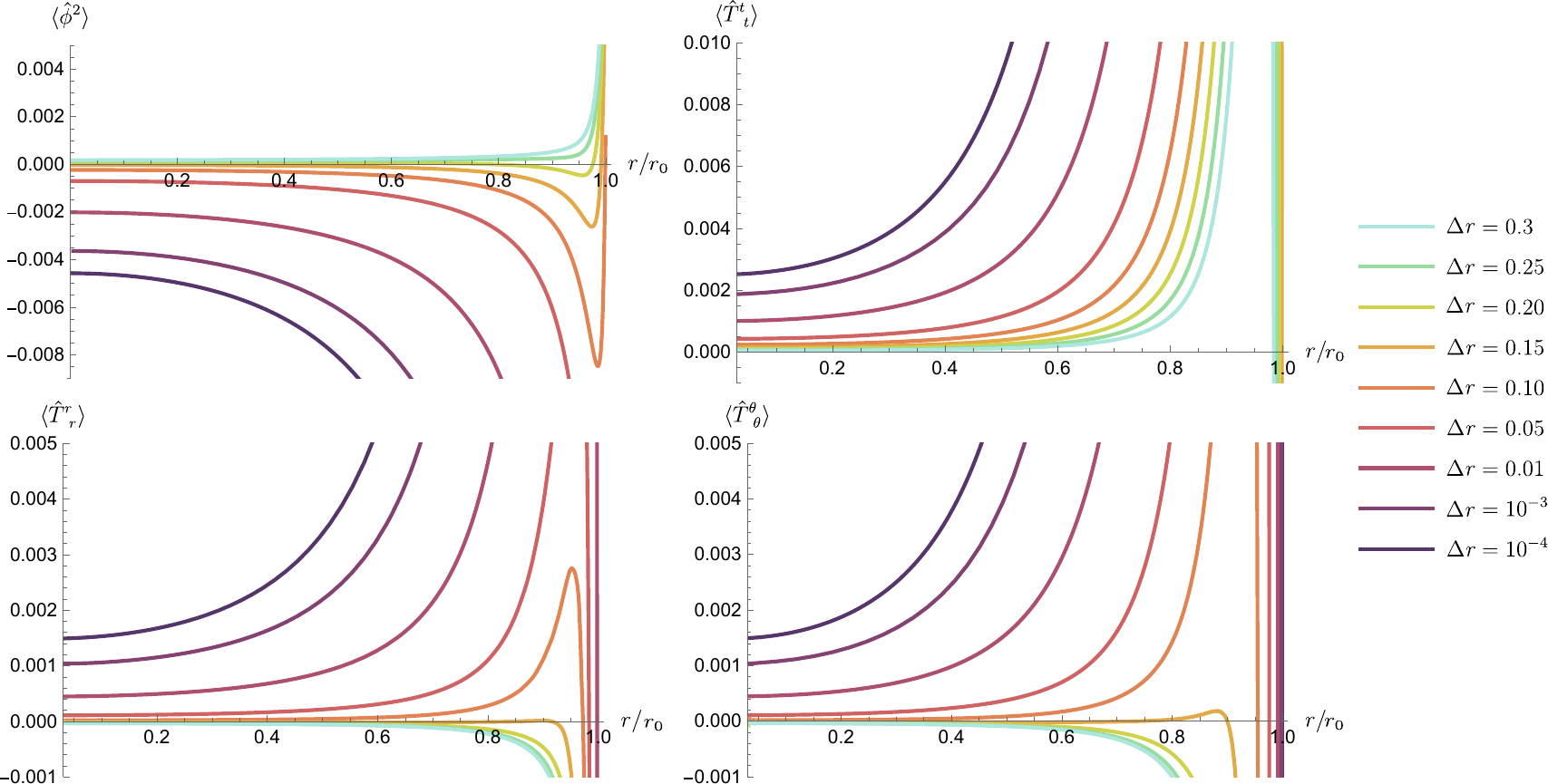}
        \caption{Renormalized vacuum polarization and renormalized stress-energy tensor of a massless, minimally coupled scalar field inside  spherical thin shells of different radius $\Delta r =r_{0}-2M$. These are purely numerical results obtained by summing up to $l=30$ and integrating up to $\omega M=20$. Results near the surface of the shell are missing divergent high-frequency contributions that would make them agree with our analytic results. }
        \label{Fig:CompInt}
    \end{figure*}

\begin{figure*}
        \centering
        \includegraphics[width=\linewidth]{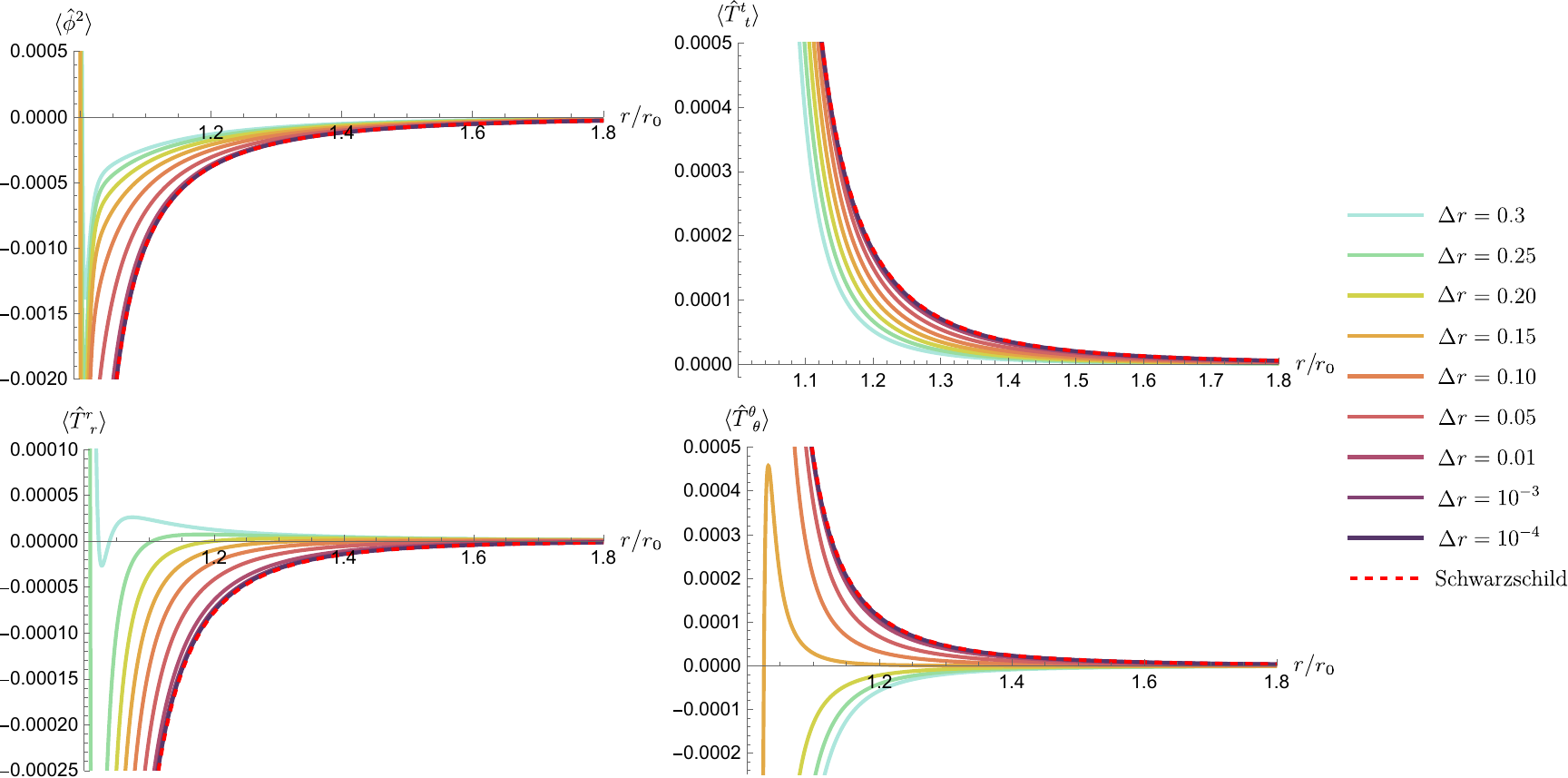}
        \caption{Renormalized vacuum polarization and renormalized stress-energy tensor of a massless, minimally coupled scalar field outside  spherical thin shells of different radius $\Delta r =r_{0}-2M$. These are purely numerical results obtained by summing up to $l=30$ and integrating up to $\omega M=20$. Results near the surface of the shell are missing divergent high-frequency contributions that would make them agree with our analytic results. }
        \label{Fig:CompExt}
    \end{figure*}

Figures~\ref{Fig:CompInt} and~\ref{Fig:CompExt} show the RVP and RSET of a massless and minimally coupled scalar field as the shell is made increasingly more compact. We have fixed $M = 1$ and considered shells of radii \mbox{$R=\left\{2.30,2.25,2.20,2.15,2.10,2.05,2.01,2.001,2.0001\right\}$}. Our results are not to be trusted close to the surface (essentially, where the curves display near-surface turning points), since the numerics are missing relevant high $l$ and $\omega$ contributions. On the other hand, our results are completely reliable in the bulk region of the spacetime. 
    
From the interior (Fig.~\ref{Fig:CompInt}), we find that the central magnitude of the RVP and RSET increases with the compactness of the shell, reproducing exactly the $r=0$ values calculated from Eqs.~\eqref{eq:phisq} and~\eqref{eq:RSET} in the black hole limit. For a minimally coupled field, the RVP changes sign from positive to negative as the shell is made more compact. The opposite happens for the radial and angular pressures. The interior energy density is negative always. An observer sitting at the center of the shell will thus perceive some stress-energy contributions coming exclusively from quantum vacuum polarization effects. These are exclusively non-local, and arise due to the fact that the spacetime is not globally Minkowski. Any contribution that depends on radial derivatives of the metric will vanish inside the shell, but contributions to the RVP and RSET that are generated due to the presence of an external Schwarzschild region give a non-zero contribution that cannot be renormalized away. Naturally, the presence of these contributions will curve the (classically) Minkowskian background and generate subsequent, curvature-dependent vacuum polarization effects, although these would contribute only as an $\mathcal{O}\left(\hbar^2\right)$ effect. We will look more closely at this point in Subsection~\ref{subsec:backreaction}.

In the exterior region, represented in Fig.~\ref{Fig:CompExt}, we observe that the vacuum polarization and the energy density are always negative, whereas the pressures have sign changes depending on the compactness. Recall that, near the shell, our numerical results are to be discarded in favor of the analytic approximations. Together with the results for the thin shell, we have plotted the resulting values for the RVP and RSET of a Schwarzschild black hole in the Boulware state (for $\xi=0$). We observe that these values are very close to the vacuum polarization generated by a thin shell with $r_{0}=2.001M$. Both curves are nearly undistinguishable for the range of the $r$ coordinate plotted in Fig.~\ref{Fig:CompExt}. We will extensively discuss this aspect in Subsec.~\ref{Subsec:Boul}.

\subsection{Backreaction effects inside the shell}
\label{subsec:backreaction}
In this Subsection, we will compute the backreaction effects produced by the RSET in the region interior to the shell. The analysis can be extended to the exterior in a straightforward way, but the simplicity of the interior region already allows us to calculate the backreacted metric analytically near the surface and at the center of the shell. The backreacted metric is computed via the 
semiclassical Einstein equations~\eqref{eq:semieqs}, which we recall here (in a slightly modified notation) for convenience
\begin{equation}
   \label{eq:semiclassicaleqs} G^{-1}\mathsf{G}_{\alpha\beta}+\Lambda\,\mathsf{g}_{\alpha\beta}+\alpha_{1}\,\mathsf{H}^{(1)}_{\alpha\beta}+\alpha_{2}\,\mathsf{H}^{(2)}_{\alpha\beta}=8\pi\left(\mathsf{T}_{\alpha\beta}^{\textrm{(cl)}}+\epsilon\,\langle \hat{\mathsf{T}}_{\alpha\beta}\rangle\right)
\end{equation}
where $\epsilon$ on the right-hand side is to keep track of the quantum corrections so $\mathcal{O}(\epsilon)\sim\mathcal{O}(\hbar)$. Different renormalizations of the quantum stress-energy tensor are degenerate with different renormalizations of the constants $\alpha_{1}$, $\alpha_{2}$, $\Lambda$ and  $G^{-1}$~\cite{Sanders:2020osl}. The terms $\mathsf{H}^{(1)}_{\alpha\beta}$ and $\mathsf{H}^{(2)}_{\alpha\beta}$ are the geometrical tensors defined by
\begin{align}
    \label{eq:Htensors}
    \mathsf{H}^{(1)}_{\alpha\beta}&=-2\Box \mathsf{R}_{\alpha\beta}+\tfrac{2}{3}\nabla_{\alpha}\nabla_{\beta}\mathsf{R}+\tfrac{1}{3}\mathsf{g}_{\alpha\beta}\Box \mathsf{R}-\tfrac{1}{3}\mathsf{g}_{\alpha\beta}\mathsf{R}^{2}\nonumber\\
	&+\tfrac{4}{3}\mathsf{R}\,\mathsf{R}_{\alpha\beta}+(\mathsf{R}_{\lambda\rho}\mathsf{R}^{\lambda\rho})\mathsf{g}_{\alpha\beta}-4 \mathsf{R}_{\alpha\lambda\beta\rho}\mathsf{R}^{\lambda\rho},\nonumber\\
    \mathsf{H}_{\alpha\beta}^{(2)}
	&=2\nabla_{\alpha}\nabla_{\beta}\mathsf{R}-2\mathsf{g}_{\alpha\beta}\Box \mathsf{R}+\tfrac{1}{2}\mathsf{R}^{2}\mathsf{g}_{\alpha\beta}-2 \mathsf{R}\,\mathsf{R}_{\alpha\beta}.
\end{align}
It is clear that $\mathsf{H}^{(1)}_{\alpha\beta}=\mathsf{H}^{(2)}_{\alpha\beta}=0$ in Ricci-flat spacetimes ($\mathsf{R}_{\alpha\beta}=\mathsf{R}=0$). Hence, they do not contribute to the equations at linear order in $\epsilon$ both inside and outside the shell, notably simplifying the analysis.  

We now expand the metric as $\mathsf{g}_{\alpha\beta}=g_{\alpha\beta}+\epsilon h_{\alpha\beta}$. At first order in $\epsilon$, we have
\begin{align}
   \label{eq:perturbationeqn}
   \delta G_{\alpha\beta}+\delta G^{-1}\,G_{\alpha\beta}+\delta\Lambda\,g_{\alpha\beta}=8\pi\,\langle \hat{T}_{\alpha\beta}\rangle_{\textrm{ren}}, 
\end{align}
where we have written the renormalized constants in the form $\Lambda+\epsilon \delta \Lambda$  (with the background value $\Lambda=0$)
and $G+\epsilon \delta G$. There is an additional ambiguity in the form of a renormalization lengthscale entering the RSET~\cite{AHS1995}, but it can also be checked to vanish in the case under consideration. The perturbation of the Einstein tensor is given explicitly by
\begin{align}
  \label{eq:perturbationdef}
  -2\,\delta G_{\alpha\beta}=\Box \bar{h}_{\alpha\beta}+g_{\alpha\beta}\nabla^{\lambda}\nabla^{\rho}\bar{h}_{\lambda\rho}-2\nabla^{\lambda}\nabla_{(\alpha}\bar{h}_{\beta)\lambda}\nonumber\\
	-g_{\alpha\beta}R^{\lambda\rho}\bar{h}_{\lambda\rho}+R\,\bar{h}_{\alpha\beta}
\end{align}
and where $\bar{h}_{\alpha\beta}=h_{\alpha\beta}-\tfrac{1}{2}g_{\alpha\beta}g^{\lambda\rho}h_{\lambda\rho}$ is the trace-reversed metric perturbation. The derivatives in (\ref{eq:perturbationdef}) are with respect to the background metric $g_{\alpha\beta}$. Inside the shell, where the metric $g_{\alpha\beta}$ is Minkowski, we have $G_{\alpha\beta}=0$, hence the second term in~\eqref{eq:perturbationeqn} is absent. We further assume $\delta \Lambda=0$ for simplicity. The resulting $\mathcal{O}(\epsilon)$ equations inside the shell are
\begin{equation}
    \delta G_{\alpha\beta}=8\pi\langle \hat{T}_{\alpha\beta}\rangle_{\textrm{ren}}.
\end{equation}

Following~\cite{York1984} and~\cite{arrecheaetal2025}, we write the full metric coefficients as 
\begin{align}
     \mathsf{g}_{tt}=-e^{2\psi(r)}\left(1-2m(r)/r\right),\quad \mathsf{g}_{rr}=\left(1-2m(r)/r\right)^{-1},
\end{align}
where $m(r)$ denotes the Misner-Sharp mass, and then we expand these metric functions as
\begin{equation}
    \psi(r)=\frac{1}{2}\log{\left(1-\frac{2M}{r_{0}}\right)}\left(1+\epsilon \mathcal{R}(r)\right),\quad m(r) = \epsilon \mathcal{M}(r).
\end{equation}
We obtain the following $tt$ and $rr$ components of the linearized semiclassical Einstein equations
\begin{align}\label{eq:perteinstein}
-\frac{2\epsilon\mathcal{M}'}{r^2}=
&
8\pi \epsilon\langle \hat{T}^{t}_{~t}\rangle+\mathcal{O}\left(\epsilon^2\right),\nonumber\\
\frac{\epsilon\left(-2\mathcal{M}'+r \mathcal{R}'\right)}{r^2}=
&
8\pi \epsilon \langle \hat{T}^{r}_{~r}\rangle+\mathcal{O}\left(\epsilon^2\right).
\end{align}
These can be combined to produce a differential equation for $\mathcal{R}$ 
\begin{equation}
    \frac{\epsilon \mathcal{R}'}{r}=8\pi \epsilon \left(\langle \hat{T}^{r}_{~r}\rangle-\langle \hat{T}^{t}_{~t}\rangle\right)+\mathcal{O}\left(\epsilon^2\right)
\end{equation}
We are going to solve these equations in various limits: near the surface of the shell $r\to r_{0}$, for shells of any compactness, and at the center of the shell $r=0$ when its surface approaches the black hole limit $r_{0}\to 2M$.

The $tt$ component of the RSET obeys the near-shell expansion
\begin{equation}\label{eq:ttnearshell}
    \langle \hat{T}^{t}_{~t}\rangle=\frac{\langle \hat{T}^{t}_{~t}\rangle^{-}_{3}}{\left(r_{0}-r\right)^3}+\mathcal{O}\left(r_{0}-r\right)^{-2},
\end{equation}
where $\langle \hat{T}^{t}_{~t}\rangle^{-}_{3}$ is the coefficient of the leading-order divergence in~\eqref{eq:int}. Recall the divergence in the $\langle \hat{T}^{r}_{~r}\rangle^{-}_{3}$ component is subleading. Replacing~\eqref{eq:ttnearshell} in~\eqref{eq:perteinstein}, we find
\begin{align}\label{eq:nearsurf}
    \mathcal{R}(r)
    &
    =-\frac{4\pi r_{0}}{(r_{0}-r)^2}\langle \hat{T}^{t}_{~t}\rangle^{-}_{3}+\mathcal{O}\left(r_{0}-r\right)^{-1},\nonumber\\
    \mathcal{M}(r)
    &
    =-\frac{2\pi r_{0}^2}{(r_{0}-r)^{2}}\langle \hat{T}^{t}_{~t}\rangle^{-}_{3}+\mathcal{O}\left(r_{0}-r\right)^{-1}.
\end{align}
When inserted in the Ricci scalar, these functions yield a behavior $R\propto \left(r_{0}-r\right)^{-4}$, indicating the presence of a curvature singularity at $r=r_{0}$ generated by the backreaction of the RSET. Of course, since the RSET is infinite, the validity of truncating the semiclassical equations to linear order in $\hbar$ is put into question. Hence, the metric functions~\eqref{eq:nearsurf} should be interpreted as the proof that static thin shells break down the perturbative character of the semiclassical expansion.

Next, we turn our attention to $r=0$, where backreaction effects are perturbative and under control. To obtain simple, analytic results, and to maximize semiclassical effects, we further assume the shell is approaching the black hole limit $r_{0}\to 2M$ so that we can neglect the mode integrals in the $r=0$ RSET expressions~\eqref{eq:RSET}.
Inserting these expressions in~\eqref{eq:perteinstein}, we find the following leading-order solutions near $r=0$,
\begin{align}
    \mathcal{R}(r)
    &
    \simeq\frac{\pi r^2}{45r_{0}^4}\left[\frac{\pi^2+2880\mathscr{I}}{24}-2\left(\pi^2+180\mathscr{I}\right)\left(\xi-\frac{1}{6}\right)\right],\nonumber\\
    \mathcal{M}(r)
    &
    \simeq
    -\frac{\pi r^3}{90r_{0}^4}\left[\frac{\pi^2-1440\mathscr{I}}{24}+\left(\pi^2+720\mathscr{I}\right)\left(\xi-\frac{1}{6}\right)\right],
\end{align}
where we have neglected higher order terms in $r$. The Ricci scalar of this metric is
\begin{equation}
    R=-\frac{\pi^3\epsilon }{60 r_{0}^4}-\frac{144\pi \mathscr{I}}{r_{0}^4}\left(\xi-\frac{1}{6}\right)\epsilon+\mathcal{O}\left(r\right)^2,
\end{equation}
which is finite at $r=0$, and whose sign depends on the field coupling $\xi$. Notice that the resulting metric has $\mathsf{g}_{tt} \mathsf{g}_{rr}\neq \text{const.}$ near $r=0$, which means that the metric cannot be described locally by a static slicing of de Sitter or Anti de Sitter spacetimes. These results indicate that the vacuum polarization generated inside the shell behaves like an anisotropic and inhomogeneous cloud whose stress-energy contribution diverges as the surface is approached. 

\subsection{Exterior of the shell and universality of the Boulware vacuum}\label{Subsec:Boul}
At the classical level the exterior to any spherical mass is described by the Schwarzschild metric~\cite{Jebsen2005}. This is no longer true in semiclassical gravity where, even though the exterior to a Schwarzschild black hole and a spherical thin shell are identical at the classical level, the Boulware vacuum gets polarized differently by both backgrounds. This is a consequence of the non-local character of vacuum polarization, which is sensitive to global aspects of the solution such as the presence of a horizon or a thin shell. The non-uniqueness of the exterior region has been denoted as quantum hair~\cite{Calmet:2021stu,Perrucci:2024qrr}. Measuring how the vacuum polarization exterior to a compact object deviates from the one exterior to a black hole allows us to test directly whether the exterior to spherical black hole mimickers has universal characteristics in the semiclassical theory, or depends on the characteristics of matter instead. With the tools developed in this work, we can test this hypothesis directly. A thorough exploration would extend beyond the scope of this paper, thus we will present two example cases for different field couplings and discuss their features.

\begin{figure*}
    \centering
        \includegraphics[width=\linewidth]{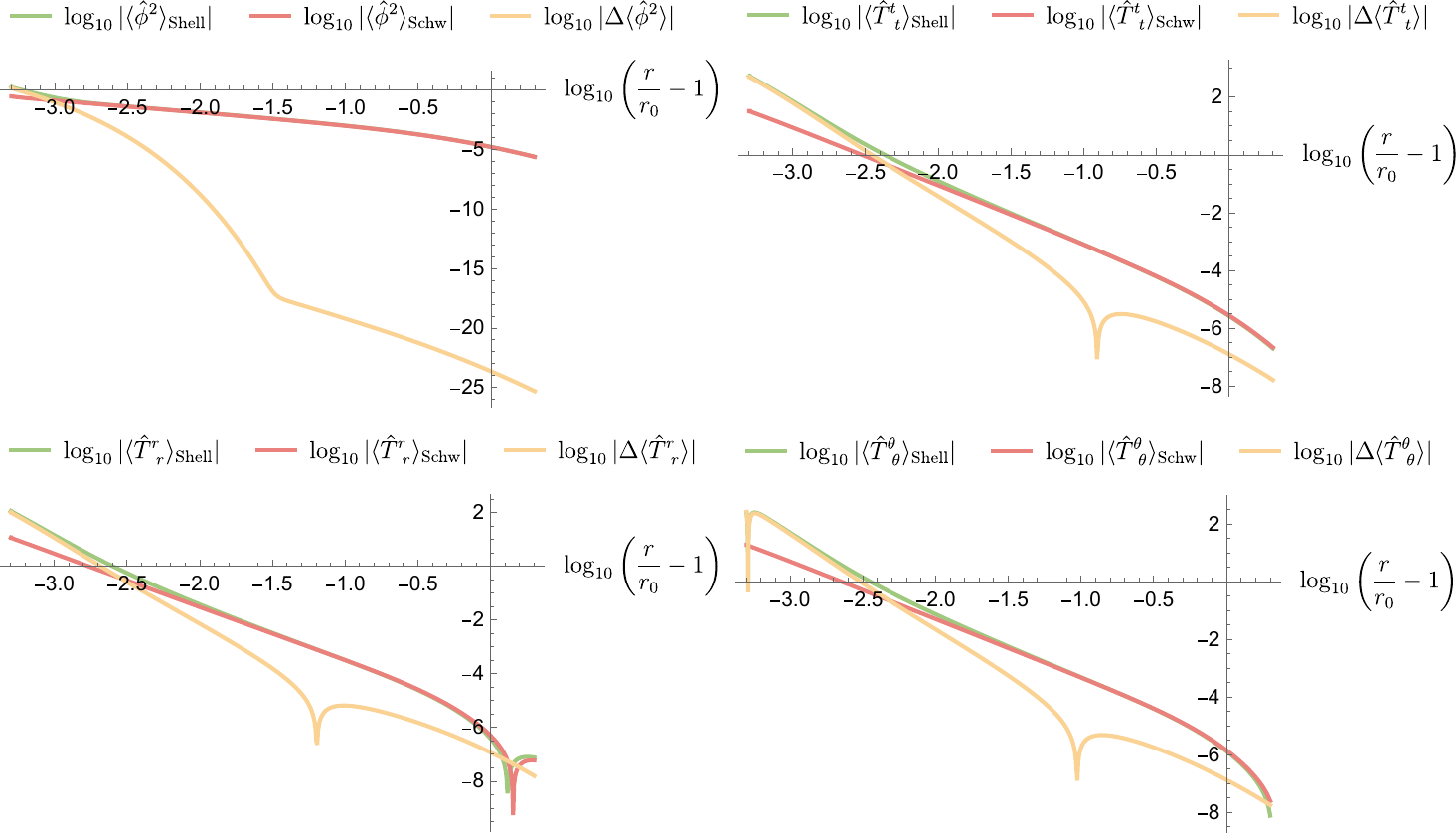}
        \caption{Renormalized vacuum polarization and renormalized stress-tensor of a massless, minimally coupled scalar field outside a shell of $r_{0}=2.001M$ (in green) and outside the event horizon of the Schwarzschild black hole (in red). Their difference decays quickly (notice the logarithmic scale of the horizontal axis) as one moves away from the shell, consistent with the far-field universality of the Boulware vacuum for minimally coupled fields found in~\cite{Anderson:2006gu}. These results have been produced integrating up to $\omega M=20$, summing up to $l=60$, and subtracting third-order regularization parameters.}
        \label{Fig:Difs}
\end{figure*}

\begin{figure*}
    \centering
        \includegraphics[width=\linewidth]{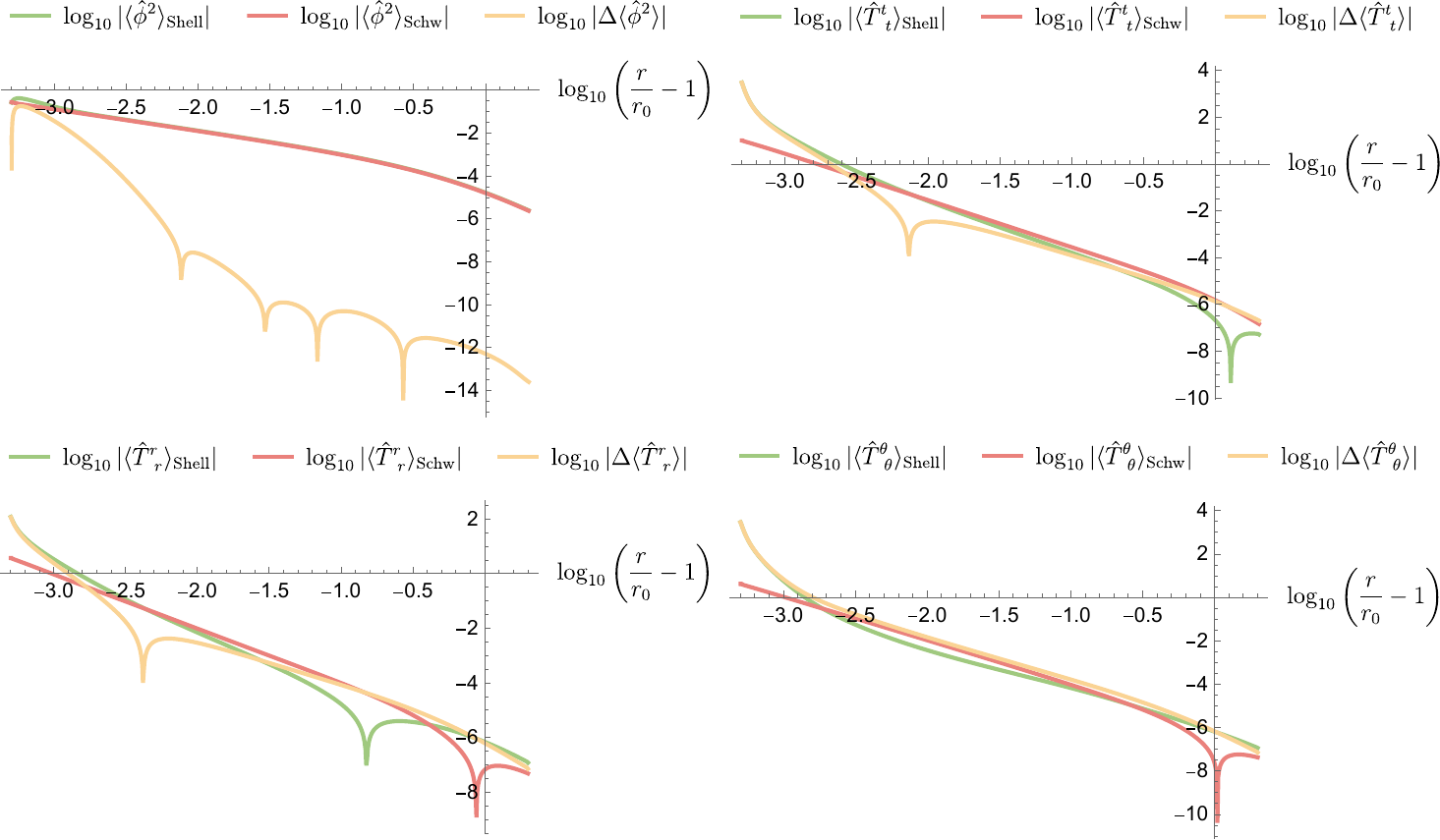}
        \caption{Renormalized vacuum polarization and renormalized stress-tensor of a massless scalar field with $\xi=1/8$ outside a shell of $r_{0}=2.001M$ (in green) and outside the event horizon of the Schwarzschild black hole (in red). Contrary to the $\xi=0$ case, the RSET differences of non-minimally coupled fields do not decay rapidly, showing discrepancies comparable, in magnitude, to the RSET components themselves at intermediate radii. The difference in the vacuum polarization is negligible instead. These results have been produced integrating up to $\omega M=20$, summing up to $l=60$, and subtracting third-order regularization parameters.}
        \label{Fig:Difsxi1o8}
\end{figure*}

Figure~\ref{Fig:Difs} compares the individual components (in logarithmic scale) of the RVP and RSET of a massless, minimally coupled field of a  shell of radius $r_{0}=2.001M$ and those of a Schwarzschild black hole. We can clearly observe that both quantities approach each other quickly as one moves away from the surface of the shell. For the RVP, both curves are visibly indistinguishable already at $\log{\left(r/r_{0}-1\right)}\approx-6$, whereas for the RSET this happens at around $\log{\left(r/r_{0}-1\right)}\approx-3$. The presence of the thin shell with a Minkowski interior thus leaves a subtle trace on the vacuum polarization at intermediate distances from the shell when $\xi=0$. As the shell is approached, however, its characteristic divergence dominates the behavior of quantum observables, which deviate from their values in Schwarzschild. These results can be compared with the case of a field with $\xi=1/8$, shown in Figure~\ref{Fig:Difsxi1o8}. In that case, the values of the RVP also coincide far from the shell, but the individual components of the RSET decay differently, showing a tendency for the shell RSET to decay slower with $r$. 

The far-field limit of the Boulware vacuum outside spherical masses has been previously analyzed in~\cite{percival2022bosonic,Satz2005vacuum,Anderson:2006gu,Carlson:2010yw}, indicating that quantum hair vanishes asymptotically for minimally coupled fields, but stays for non-minimally coupled fields, something consistent with our numerical results.

We have explored a moderate range of the  space of parameters of thin shell configurations. In the last Section of this paper, we suggest directions that might be worth exploring in the future.

\section{Discussion and conclusions}
\label{sec:conclusions}

In this article, we presented a detailed analysis of the features of the renormalized vacuum polarization (RVP) and renormalized stress-energy tensor (RSET) for a quantum scalar field in the Boulware vacuum, propagating in a spacetime with a static, spherical thin shell matching a Minkowski interior to a Schwarzschild exterior. 
We treated the Boulware state as a zero temperature state, defined on the Wick-rotated Riemannian manifold. Such treatment, which we detailed at the beginning of this paper, is convenient because the Euclidean Green's function possesses a simpler singularity structure, being singular only in the coincident limit, and thereby simplifying the development of a renormalization procedure for computing the RVP and the RSET.  This prescription, which was generalized to Boulware states in Ref.~\cite{arrecheaetal2025}, enables us to express the geometrical short-distance divergence encoded in the Hadamard parametrix as a mode-sum in the same form as the Euclidean Green's function, thus allowing a mode-by-mode renormalization that can be efficiently implemented in situations involving self-gravitating matter fluids.

Having outlined the renormalization approach, we have then presented Israel's thin shell formalism and applied it to a spherical thin shell joining Minkowski and Schwarzschild spacetimes. Then, we showed how the Euclidean modes, necessary for the computation of the Euclidean Green's function of the shell,  are most efficiently obtained as a linear combination of the Minkowski and Schwarzschild modes, with the coefficients computed by matching conditions at the shell. This formulation also explicitly isolates the boundary terms, which encode the presence of the shell and hence possible resulting divergences, in the Green's function. The methodology we have introduced in this work can be generalized, in a straightforward way, to other static material configurations with Schwarzschild exteriors.

To estimate how renormalized expectation values diverge at the shell, we adopted a WKB approximation for the Euclidean modes. We found the power and sign of these divergences to be in agreement with flat spacetime calculations~\cite{PhysRevD.20.3063}.
As an independent consistency check, we also implemented a weak-field approximation~\cite{Satz2005vacuum} valid in the regime of small shell compactness, $M/r_0\ll1$. 
The obtained results were in agreement with the WKB ones at leading order in $M/r_0$.

We also analyzed the RVP and RSET at the center of the shell. Through a radial point-splitting of the Euclidean Green's function, 
we derived semi-analytical expressions for these quantities at $r=0$, which only require the numerical computation of the $l=0,1$ shell modes at the surface $p_{\omega l}^{\textrm{shell}}(r_0)$ and $q_{\omega l}^{\textrm{shell}}(r_0)$. Moreover, these representations become fully analytical in the black hole limit, and describe bounded semiclassical effects. 

Having determined the analytical properties of the Boulware vacuum at the boundaries of the spacetime, we then carried out an extensive numerical analysis, evaluating the RVP and RSET throughout the spacetime over a wide range of parameters. We first provided a numerical verification of the divergent behavior obtained with the analytic methods, evaluated the RVP and RSET for thin shells of different compactness, and showed how the RSET components decay differently at large distances depending on the Ricci coupling of the field. We also performed backreaction calculations near $r=r_{0}$ and $r=0$ to estimate the quantum corrections to the background metric.

The above analyses not only set the grounds for future explorations of the properties of Boulware states, but also hint at relevant properties of the quantum vacuum in situations of extreme compactness. 

Several clear and valuable extensions can be considered at this point. In order to better disentangle the black hole limit from the thin shell limit, it would be beneficial to consider shells with finite thickness. Unfortunately, deriving smooth thick shell models having the right distributional stress-energy tensor in the thin shell limit is a sophisticated problem, but some directions are being explored by the authors at this moment. What could be gained from such models is a clearer examination of the universality of the Boulware state near the black hole limit. 

Of course, given the generality of our treatment, a straightforward extension would be to consider different shell interiors and evaluate their respective black hole limits. Spherical shells enclosing a static de Sitter  (gravastars~\cite{Mazur:2001fv}) or Anti de Sitter (AdS-stars~\cite{Arrechea:2025vxp}) interior have been proposed as black hole mimickers supported by vacuum polarization effects. If the effective stress-energy tensors of these models (or some modified version of them that replaces the thin surface layer by a smooth region~\cite{Cattoen:2005he}) could be obtained, even if just qualitatively, from the Boulware RSET of a scalar field, that alone would reinforce such proposals.

The extension of these analysis to charged static shells would enable one to analyze vacuum polarization effects in situations near extremality. If the charge-to-mass ratio of the shell is such that its surface can approach the gravitational radius of the extremal Reissner-Nordstr\"om black hole, a presumably regular black hole limit could be attained. The regularity of this limit stems from the numerical evidence of the regularity of the Boulware state in extremal Reissner-Nordstr\"om black holes~\cite{AndersonEBH,arrecheaetal2025}. 

The simplicity of the interior geometry of the shell causes every curvature-dependent vacuum polarization effect to vanish altogether. However, this situation is not generic, which leads us to the last interesting scenario that we would like to mention: compact stars in equilibrium. In general relativity, compact stars composed of standard matter satisfy upper compactness limits, all of them bounded by the Buchdahl limit~\cite{PhysRev.116.1027}. These limits are reached when the pressures of the fluid diverge at their center. We expect that as this limit is approached there will be strong vacuum polarization effects that will react back on the background~\cite{Arrecheaetal2023,Reyes2023,Matsuo:2026isn}, either taming this growth of the pressure (relaxing the compactness bound), or stimulating it (making the bound more stringent). This direction is currently being explored by the authors and we expect to report on it soon.

\section{Acknowledgements}
L.P. acknowledges funding from Taighde Éireann - Research Ireland under Grant number GOIPG/2024/4003.

The authors want to thank Elizabeth Winstanley and Sam Dolan for illuminating discussions on early stages of this investigation.

\begin{widetext}
\section*{Appendix}
\appendix
\section{Hadamard coefficients}
\label{app:HCoeff}

Below we list the coefficients $\mathcal{D}_{ab}^{(\mathrm{r})}$, $\mathcal{D}_{ab}^{(\mathrm{p})}$, $\mathcal{T}_{ab}^{(\mathrm{l})}$, $\mathcal{T}_{ab}^{(\mathrm{r})}$ and $\mathcal{T}_{ab}^{(\mathrm{p})}$  to 2nd order. In the supplemental material, we provide a \textit{Mathematica} Notebook containing these expressions.
\begin{align*}
 	&\mathcal{D}_{00}^{(\mathrm{r})}=2,\nonumber\\
 	&\mathcal{D}_{10}^{(\mathrm{r})}=\frac{r^2 h f'^2+r f \left(r f' h'+2 h \left(f'-r f''\right)\right)+f^2 \left(-2 r h'+4 h^2-4 h\right)}{24 r^2 f h^2},\nonumber\\
 	&\mathcal{D}_{11}^{(\mathrm{r})}=\frac{r^2 f'^2-4 r f f'-4 f^2 (h-1)}{24 r^2 h},\nonumber\\
 	&\mathcal{D}_{20}^{(\mathrm{r})}=\frac{1}{11520 r^4 f^2 h^4}\Bigg[47 r^4 h^2 f'^4-8 r^3 f h f'^2 \left(h \left(14 r f''+5 f'\right)-7 r f' h'\right)\nonumber\\
    &\qquad-4 r f^3 \left(57 r^2 f' h'^2-6 r h \left(9 r f'' h'+f' \left(4 r h''-13 h'\right)\right)+20 h^3 \left(r f''-3 f'\right)+h^2 \left(f' \left(72-10 r h'\right)+8 r \left(3 r f^{(3)}-4 f''\right)\right)\right)\nonumber\\
    &\qquad+r^2 f^2 \left(57 r^2 f'^2 h'^2-12 r h f' \left(9 r f'' h'+f' \left(2 r h''+3 h'\right)\right)+h^2 \left(36 r^2 f''^2-4 f'^2+8 r f' \left(6 r f^{(3)}+11 f''\right)\right)\right)\nonumber\\
    &\qquad+4 f^4 \left(57 r^2 h'^2+h^2 \left(76-20 r h'\right)+8 r h \left(10 h'-3 r h''\right)+4 h^4-80 h^3\right)\Bigg],\nonumber\\
 	&\mathcal{D}_{21}^{(\mathrm{r})}=\frac{1}{5760 r^4 f h^3}\Bigg[9 r^4 h f'^4+r^3 f f'^2 \left(11 r f' h'-2 h \left(11 r f''+3 f'\right)\right)+2 r^2 f^2 f' \left(-33 r f' h'+20 h^2 f'+h \left(44 r f''-68 f'\right)\right)\nonumber\\
    &\qquad+4 r f^3 \left(33 r f' h'+h \left(f' \left(82-5 r h'\right)-22 r f''\right)+10 h^2 \left(r f''-7 f'\right)\right)-8 f^4 \left(h \left(26-5 r h'\right)+11 r h'+14 h^3-40 h^2\right)\Bigg],\nonumber\\
 	&\mathcal{D}_{22}^{(\mathrm{r})}=\frac{\left(-r^2 f'^2+4 r f f'+4 f^2 (h-1)\right)^2}{1152 r^4 h^2},\nonumber\\
    &\mathcal{D}_{11}^{(\mathrm{p})}=\frac{f h'-h f'}{12 r f h^2},\nonumber\\
    &\mathcal{D}_{21}^{(\mathrm{p})}=\frac{1}{2880 r^3 f^2 h^4}\Bigg[5 r^2 h^2 f'^3+2 r f h f' \left(7 r f' h'+11 h \left(f'-r f''\right)\right)-2 f^3 \left(10 h^2 h'+57 r h'^2+2 h \left(7 h'-12 r h''\right)\right)\nonumber\\
    &\qquad+f^2\left(57 r^2 f' h'^2+20 h^3 f'-4 h^2 \left(5 f'-6 r^2 f^{(3)}\right)-2 r h \left(27 r f'' h'-4 f' \left(5 h'-3 r h''\right)\right)\right.\Bigg],\nonumber\\
 	&\mathcal{D}_{22}^{(\mathrm{p})}=\frac{1}{2880 r^4 f^2 h^4}\Bigg[-9 r^2 h^2 f'^2+4 r f h \left(-7 r f' h'-5 h^2 f'+h \left(3 r f''+2 f'\right)\right)\nonumber\\
    &\qquad+f^2 \left(57 r^2 h'^2+4 h^2 \left(5 r h'-3\right)-8 r h \left(3 r h''+h'\right)+12 h^4\right)\Bigg],\nonumber\\
     &\mathcal{T}_{00}^{(\mathrm{l})}=\frac{1}{24 r^2 f^2 h^2}\Bigg[(6 \xi -1) r^2 h f'^2+(6 \xi -1) r f \left(r f' h'-2 h \left(r f''+2 f'\right)\right)\nonumber\\
    &\qquad+4 f^2 \left((6 \xi -1) r h'+h^2 \left(6 \xi +3 \mu ^2 r^2-1\right)+(1-6 \xi ) h\right)\Bigg],\nonumber\\
\end{align*}

\begin{align*}
 	&\mathcal{T}_{10}^{(\mathrm{l})}=\frac{1}{17280 r^4 f^4 h^5}\Bigg[2 \Bigg(4 \left(135 r^4 \mu ^4+60 r^2 (9 \xi -1) \mu ^2+540 \xi ^2-120 \xi +2\right) h^5-80 \left(3 r^2 (9 \xi -1) \mu ^2+54 \xi ^2-21 \xi +2\right) h^4\nonumber\\
    &\qquad\qquad+4 \left(540 \xi ^2-300 \xi +5 r \left(9 r^2 (12 \xi -1) \mu ^2+216 \xi ^2-60 \xi +4\right) h'+38\right) h^3\nonumber\\
    &\qquad\qquad-2 r \left(\left(2160 \xi ^2-960 \xi +94\right) h'+9 r \left(10 (6 \xi -1) h''+r (20 \xi -3) h^{(3)}\right)\right) h^2\nonumber\\
    &\qquad\qquad+9 r^2 h' \left(40 \left(6 \xi ^2+5 \xi -1\right) h'+13 r (20 \xi -3) h''\right) h+126 r^3 (3-20 \xi ) h'^3\Bigg) f^4\nonumber\\
    &\qquad-2 r \Bigg(20 \left(\left(9 r^2 (12 \xi -1) \mu ^2+216 \xi ^2-60 \xi +4\right) f'+3 r \left(r^2 (18 \xi -1) \mu ^2+6 \xi  (6 \xi -1)\right) f''\right) h^4\nonumber\\
    &\qquad\qquad-2 \Big(f' \left(2160 \xi ^2-960 \xi +15 r \left(r^2 (18 \xi -1) \mu ^2+6 \xi  (6 \xi -1)\right) h'+94\right)\nonumber\\
    &\qquad\qquad+9 r \left(2 \left(60 \xi ^2+10 \xi -3\right) f''+r \left((60 \xi -11) f^{(3)}+2 r (5 \xi -1) f^{(4)}\right)\right)\Big) h^3\nonumber\\
    &\qquad\qquad+r \Big(r \left(72 r (5 \xi -1) f'' h''+h' \left(\left(2160 \xi ^2+1680 \xi -373\right) f''+108 r (5 \xi -1) f^{(3)}\right)\right)\nonumber\\
    &\qquad\qquad+9 f' \left(8 \left(75 \xi ^2-5 \xi -1\right) h'+r \left((100 \xi -17) h''+2 r (5 \xi -1) h^{(3)}\right)\right)\Big) h^2\nonumber\\
    &\qquad\qquad+r^2 h' \left(171 r (1-5 \xi ) h' f''+f' \left(117 r (1-5 \xi ) h''-2 \left(540 \xi ^2+915 \xi -172\right) h'\right)\right) h+126 r^3 (5 \xi -1) f' h'^3\Bigg) f^3\nonumber\\
    &\qquad+r^2 h \Bigg(60 \left(r^2 (18 \xi -1) \mu ^2+6 \xi  (6 \xi -1)\right) f'^2 h^3+5 r^2 \left(54 \xi ^2-222 \xi +43\right) f'^2 h'^2\nonumber\\
    &\qquad\qquad+2 \left(72 \left(15 \xi ^2-15 \xi +2\right) f'^2+r \left(\left(2160 \xi ^2-2640 \xi +401\right) f''+90 r (1-5 \xi ) f^{(3)}\right) f'+10 r^2 \left(54 \xi ^2-42 \xi +7\right) f''^2\right) h^2\nonumber\\
    &\qquad\qquad+2 r f' \left(9 f' \left(4 (40 \xi -7) h'+5 r (5 \xi -1) h''\right)-5 r \left(108 \xi ^2-219 \xi +41\right) h' f''\right) h\Bigg) f^2\nonumber\\
    &\qquad-2 r^3 h^2 f'^2 \left(r \left(-270 \xi ^2+570 \xi -107\right) f' h'+2 h \left(\left(540 \xi ^2-705 \xi +107\right) f'+r \left(270 \xi ^2-570 \xi +107\right) f''\right)\right) f\nonumber\\
    &\qquad+r^4 \left(270 \xi ^2-930 \xi +179\right) h^3 f'^4\Bigg],\nonumber\\
 	&\mathcal{T}_{11}^{(\mathrm{l})}=\frac{1}{17280 r^4 f^3 h^5}\Bigg[10 (33 \xi -16) r^4 h^3 f'^4-2 r^3 f h^2 f'^2 \left(2 (44-75 \xi ) r f' h'+h \left(4 (75 \xi -44) r f''+21 (30 \xi -7) f'\right)\right)\nonumber\\
    &\qquad-r^2 f^2 h \Bigg(2 (89-165 \xi ) r^2 f'^2 h'^2+60 \mu ^2 r^2 h^3 f'^2+r h f' \left((420 \xi -307) r f'' h'+f' \left(9 (20 \xi -9) r h''+4 (330 \xi -79) h'\right)\right)\nonumber\\
    &\qquad\qquad+h^2 \left(8 (15 \xi +8) r^2 f''^2+(38-480 \xi ) f'^2-6 r f' \left(3 (20 \xi -9) r f^{(3)}+(300 \xi -83) f''\right)\right)\Bigg)\nonumber\\
    &\qquad+r f^3 \Bigg(-126 r^3 f' h'^3+r^2 h h' \left(171 r f'' h'+f' \left(117 r h''+2 (330 \xi -7) h'\right)\right)+40 h^4 \left(3 \mu ^2 r^3 f''+f' \left(30 \xi +3 \mu ^2 r^2-5\right)\right)\nonumber\\
    &\qquad\qquad-2 r h^2 \left(r \left(36 r f'' h''+h' \left(54 r f^{(3)}+(211-600 \xi ) f''\right)\right)-9 f' \left((80 \xi -8) h'+r \left((1-20 \xi ) h''-r h^{(3)}\right)\right)\right)\nonumber\\
    &\qquad\qquad+4 h^3 \left(f' \left(-15 \mu ^2 r^3 h'+60 \xi -4\right)+9 r \left((6-40 \xi ) f''+r \left(r f^{(4)}+(7-20 \xi ) f^{(3)}\right)\right)\right)\Bigg)\nonumber\\
    &\qquad+2 f^4 \Bigg(126 r^3 h'^3+4 h^3 \left(15 r h' \left(6 \xi +\mu ^2 r^2-1\right)+420 \xi -76\right)+r^2 h h' \left((199-1320 \xi ) h'-117 r h''\right)\nonumber\\
    &\qquad\qquad+6 r h^2 \left((4-60 \xi ) h'+3 r \left(r h^{(3)}+(40 \xi -6) h''\right)\right)+8 h^5 \left(30 \xi +15 \mu ^2 r^2-2\right)-40 h^4 \left(48 \xi +3 \mu ^2 r^2-8\right)\Bigg)\Bigg],
    \end{align*}
    
    \begin{align*}
    &\mathcal{T}_{10}^{(\mathrm{r})}=\frac{1}{1152 r^4 f^2 h^3}\left(-r^2 f'^2+4 r f f'+4 f^2 (h-1)\right) \Bigg((6 \xi -1) r^2 h f'^2+(6 \xi -1) r f \left(r f' h'-2 h \left(r f''+2 f'\right)\right)\nonumber\\
    &\qquad\qquad+4 f^2 \left((6 \xi -1) r h'+h^2 \left(6 \xi +3 \mu ^2 r^2-1\right)+(1-6 \xi ) h\right)\Bigg),\nonumber\\
    &\mathcal{T}_{10}^{(\mathrm{p})}=\frac{1}{288 r^4 f^2 h^3}\Bigg[(h-1) \Big((6 \xi -1) r^2 h f'^2+(6 \xi -1) r f \left(r f' h'-2 h \left(r f''+2 f'\right)\right)\nonumber\\
    &\qquad\qquad+4 f^2 \left((6 \xi -1) r h'+h^2 \left(6 \xi +3 \mu ^2 r^2-1\right)+(1-6 \xi ) h\right)\Big)\Bigg],\nonumber\\
 	&\mathcal{T}_{11}^{(\mathrm{p})}=-\frac{1}{288 r^4 f^2 h^3}\Bigg[\left(r f'+2 f (h-1)\right) \Big((6 \xi -1) r^2 h f'^2+(6 \xi -1) r f \left(r f' h'-2 h \left(r f''+2 f'\right)\right)\nonumber\\
    &\qquad\qquad+4 f^2 \left((6 \xi -1) r h'+h^2 \left(6 \xi +3 \mu ^2 r^2-1\right)+(1-6 \xi ) h\right)\Big)\Bigg].\nonumber\\
 \end{align*}
\section{Regularization parameters}
\label{app:reg_par}
In this Appendix, we present the expressions for the regularization parameters $\Psi_{\omega l}(a,b|r)$ and $\chi_{\omega l}(a,b|r)$ appearing in the mode-sum ansatzes for the direct and tail parts of the Hadamard parametrix, respectively, and obtained by inverting Eqs.~(\ref{eq:directmodesum}) and (\ref{eq:logterms}). These regularization parameters are distributions and not ordinary integrable functions. Their explicit representation factors into terms that involve delta distributions and terms that involve hypergeometric functions. In particular, we obtain 
\begin{align}
   & \Psi_{\omega l}(a,b|r)=\mathcal{I}_{\omega l}(a,b|r)
    -\frac{(2r^{2})^{a}}{f^{a+b+1}}\frac{(-1)^{a+l}}{b!}\sum_{p=1}^{a}\frac{2^{a-p}(a-p)!(a-p+b)!}{(a-p+l+1)!(a-p-l)!}\left(\frac{f}{2r^{2}}\right)^{p}\delta^{(2p-2)}(\omega),\\
& \chi_{\omega l}(a,b\,|r)=\mathcal{J}_{\omega l}(a,b\,|r)+\sqrt{\pi}(2r^{2}\rho^{2})^{a-b}(a-b)!\,2^{2a}(-1)^{l}\nonumber\\
   & \qquad\qquad\times\Bigg\{\sum_{k=0}^{a}\frac{k!(a-k)!\delta^{(2a-2k)}(\omega)\left[\psi(a-b+1) -\psi(a-b-k+1)+\psi(a-k+1)+\psi(k-a+\tfrac{1}{2})+2\log(\tfrac{2\rho}{\lambda})\right]}{(a-k-b)!(2a-2k)!\Gamma(k-a+\tfrac{1}{2})(k+l+1)!(k-l)!(2\rho^{2})^{k}}\nonumber\\
   &\qquad\qquad-2\sum_{k=0}^{a}\frac{(a-k)!k!}{(a-k-b)!\Gamma(k-a+\tfrac{1}{2})(k+l+1)!(k-l)!(2\rho^{2})^{k}} \sum_{\substack{p=0\\p\ne a-k}}^{\infty}\frac{\delta^{(2p)}(\omega)\lambda^{2p-2a+2k}}{(2p)!(2p-2a+2k)}
    \Bigg\},
\end{align}
where $\mathcal{I}_{\omega l}(a,b|r)$ and $\mathcal{J}_{\omega l}(a,b|r)$ are given by
\begin{align}
\label{eq:directregparam}
    &\mathcal{I}_{\omega l}(a,b|r)=\frac{(2r^{2})^{a-1/2}}{f^{a+b+1/2}}\frac{2^{a-2}\sqrt{2 \pi}(-1)^{a+l}}{\Gamma(b+1)}\Bigg\{\Gamma(a+\tfrac{1}{2})\Gamma(a+b+\tfrac{1}{2}){}_{2}\hat{F}_{3}\left(\begin{array}{c|}\{a+\tfrac{1}{2},a+b+\tfrac{1}{2}\}\\\{a+l+\tfrac{3}{2}, a-l+\tfrac{1}{2}, \tfrac{1}{2}\}\end{array}\,\,\frac{\omega^{2}r^{2}}{f}\right)\nonumber\\
   &\qquad\qquad -\Gamma(a+1)\Gamma(a+b+1)\frac{|\omega|r}{\sqrt{f}}\,{}_{2}\hat{F}_{3}\left(\begin{array}{c|}\{a+1,a+b+1\}\\\{\tfrac{3}{2},a+l+2,a-l+1\}\end{array}\,\,\frac{\omega^{2}r^{2}}{f}\right)\Bigg\},\\
    &\mathcal{J}_{\omega l}(a,b\,|r)=\sqrt{\pi}(2r^{2}\rho^{2})^{a-b}(a-b)!2^{2a}(-1)^{l}\Bigg\{-\sum_{k=0}^{a-b}\frac{\theta\left(\omega^{2}-\lambda^{2}\right)(a-k)!k!|\omega|^{2k-2a-1}}{(a-k-b)!\Gamma(k-a+\tfrac{1}{2})(k+l+1)!(k-l)!(2\rho^{2})^{k}}\nonumber\\
    &\qquad\qquad-\frac{(-1)^{b}}{(2\rho^{2})^{a+1}}|\omega|(a+1)!b!\,{}_{2}\hat{F}_{3}\left(\begin{array}{c|}\{a+2,b+1\}\\\{\tfrac{3}{2},a-l+2,a+l+3\}\end{array}\,\,\frac{\omega^{2}}{2\rho^{2}}\right)\nonumber\\
    &\qquad\qquad+\frac{(-1)^{b}}{2^{a+1/2}\rho^{2a+1}}\Gamma(a+\tfrac{3}{2})\Gamma(b+\tfrac{1}{2}){}_{2}\hat{F}_{3}\left(\begin{array}{c|}\{a+\tfrac{3}{2},b+\tfrac{1}{2}\}\\\{\tfrac{1}{2},a-l+\tfrac{3}{2},a+l+\tfrac{5}{2}\}\end{array}\,\,\frac{\omega^{2}}{2\rho^{2}}\right)\Bigg\},
\end{align}
and ${}_{2}\hat{F}_{3}$ is the regularized general hypergeometric function \cite{NIST:DLMF}. The functions $\mathcal{I}_{\omega l}(a,b|r)$ and $\mathcal{J}_{\omega l}(a,b|r)$ can be equivalently expressed in terms of the generalized incomplete gamma function $\Gamma(c,z_0,z_1)=\int_{z_0}^{z_1}t^{c-1}e^{-t}dt$ as
\begin{align}
    &\mathcal{I}_{\omega l}(a,b\,|r)=\lim_{x\to-1}\frac{2^l (-1)^{a+b} \left(2 r^2\right)^{a-\frac{1}{2}}}{2\, b! f^{a+b+\frac{1}{2}}} \Bigg\{\sum _{k=1}^b \sum _{j=0}^l \frac{(-1)^{j+1} j! \binom{l}{j} \binom{\frac{1}{2} (l+j-1)}{l}}{2(-k+j+1)!} \frac{\partial ^{b-k}}{\partial x^{b-k}}\left((1-x)^{a+b-\frac{1}{2}} e^{-|\omega|  \sqrt{\frac{2 (1-x) r^2}{f}}}\right)\nonumber\\
    &\quad+\sum _{k=b}^l \sum _{j=0}^{k-b} \frac{k! \binom{l}{k} \binom{\frac{1}{2} (k+l-1)}{l} (-1)^{-2 b+k-j}}{(k-b)!} \binom{k-b}{j} \left(|\omega|  \sqrt{\frac{2 r^2}{f}}\right)^{-2 a-2 k+2 j-1} \Gamma \left(2 a+2 k-2 j+1,0,\frac{2 r |\omega|}{\sqrt{f}} \right)\Bigg\},\\
     &\mathcal{J}_{\omega l}(a,b\,|r)=-\lim_{|\omega'|\to|\omega|}\frac{f^{a-b}}{4 \pi } \sum _{k=0}^{a-b} 2^{l+2} (-1)^{a-k} \binom{a-b}{k} \frac{\partial ^{2 a-2 k}}{\partial |\omega'|^{2 a-2 k}}\sum _{n=0}^l \sum _{j=0}^n \frac{\pi}{|\omega'|}  (-1)^{n-j} \binom{l}{n} \binom{\frac{1}{2} (l+n-1)}{l} \binom{n}{j}\nonumber\\
    &\qquad\qquad\qquad\times\left(\frac{2 r^2}{f}\right)^k \left(\frac{\Gamma \left(2 (k-j+n+1),0, \frac{2 |\omega'| r}{\sqrt{f}}\right)}{\left(\frac{\sqrt{2} |\omega'| r}{\sqrt{f}}\right)^{2 (k-j+n+1)}}-\frac{2^{k-j+n} \theta (\lambda^2-\omega^2 )}{k-j+n+1}\right).
\end{align}
In the supplemental material, we include a \textit{Mathematica} Notebook containing these expressions.
Using these expressions, a mode-sum representation of the vacuum polarization is given by Eq.~(\ref{eq:vacpolren}) with
\begin{align}
\label{eq:k_oml_m}
    k_{\omega l}^{(m)}(r)=\sum_{a=0}^{m}\sum_{b=0}^{a}\mathcal{D}_{ab}^{(\textrm{r})}(r)\mathcal{I}_{\omega l}(a,b|r)+\sum_{a=1}^{m-1}\sum_{b=0}^{a-1}\mathcal{T}_{ab}^{(\textrm{r})}(r)\mathcal{I}_{\omega l}(a+1,b|r)+\sum_{a=0}^{m-1}\sum_{b=0}^{a}\mathcal{T}_{ab}^{(\textrm{l})}(r)\mathcal{J}_{\omega l}(a,b\,|r).
\end{align}

\section{Weak field approximation}
\label{app:weak_field}

In this Appendix, following the procedure outlined in Ref.~\cite{boasso2025vacuum}, we derive the terms $\log\left(\frac{-\bar\Box}{\nu^2}\right)R^{(1)}$ and $\log\left(\frac{-\bar\Box}{\nu^2}\right)R^{\alpha~(1)}_{~\beta}(x)$, necessary to compute the VP and RSET in the context of the weak field approximation. Hereafter, we drop the $^{(1)}$ superscript for notational simplicity; all quantities are understood to be linear in $M/r_0$.

The terms involving the Ricci tensor are most easily obtained by considering the Cartesian components of $R_{\alpha\beta}$, which are derived through the tensor transformation law
\begin{equation}
	R'^{\alpha}_{~\beta} = \frac{\partial x'^{\alpha}}{\partial x^{\rho}}\frac{\partial x^{\lambda}}{\partial x'^{\beta}}R^{\rho}_{~\lambda},
\end{equation}
where the Cartesian coordinates are $x'{}^{\alpha}=\{\tau,x,y,z\}$. We then express the Cartesian components in terms of spherical coordinates $x{}^{\alpha}=\{\tau,r,\theta,\phi\}$
\begin{align}
    R'^{x}_{~x}&=\sin^2\theta\cos^2\phi \,R^r_{~r}+\cos^2\theta\cos^2\phi\, R^\theta_{~\theta}+\sin^2\phi\, R^\phi_{~\phi},\nonumber\\
    R'^{y}_{~y}&=\sin^2\theta\sin^2\phi \,R^r_{~r}+\cos^2\theta\sin^2\phi\, R^\theta_{~\theta}+\cos^2\phi\, R^\phi_{~\phi},\nonumber\\
    R'^z_{~z}&=R^\theta_{~\theta}+\cos^2\theta(R^r_{~r}-R^\theta_{~\theta}),\nonumber\\
    R'^x_{~y}&=\sin^2\theta\sin\phi\cos\phi\, R^r_{~r}+\cos^2\theta\sin\phi\cos\phi \,R^\theta_{~\theta}\nonumber\\
    &\quad-\sin\phi\cos\phi R^\phi_{~\phi},\nonumber\\
    R'^x_{~z}&=\sin\theta\cos\theta\cos\phi(R^r_{~r}-R^\theta_{~\theta}),\nonumber\\
    R'^y_{~z}&=\sin\theta\cos\theta\sin\phi(R^r_{~r}-R^\theta_{~\theta}).
\end{align}
Introducing the Ricci scalar and tensor into Eq.~\eqref{eq:log_op} and expressing the integrals in spherical coordinates, yields representations of the nonlocal operators for which the integrations can be carried out straightforwardly
  \begin{align}
    \log\left(\frac{-\bar\Box}{\mu^2}\right)R &= -\frac{1}{2\pi}\int_{0}^{\infty}\mathrm{d}r'\,r'{}^{2}\int_{0}^{\pi}\mathrm{d}\theta'\,\sin\theta'\int_{0}^{2\pi}\mathrm{d}\phi'\,\frac{R(r',\theta',\phi')}{(r^{2}+r'{}^{2}-2rr'\cos\theta')^{3/2}}\label{eq:log_op_R}\\
    \log\left(\frac{-\bar\Box}{\mu^2}\right)R'^{\alpha}_{~\beta} &= -\frac{1}{2\pi}\int_{0}^{\infty}\mathrm{d}r'\,r'{}^{2}\int_{0}^{\pi}\mathrm{d}\theta'\,\sin\theta'\int_{0}^{2\pi}\mathrm{d}\phi'\,\frac{R'^{\alpha}_{~\beta}(r',\theta',\phi')}{(r^{2}+r'{}^{2}-2rr'\cos\theta')^{3/2}}.\label{eq:log_op_Rab} 
\end{align}  
Computing the $\phi'$ integral yields for the Ricci scalar
\begin{align}
    \frac{1}{\pi}\int_{0}^{2\pi}\mathrm{d}\phi'\,R(r',\theta',\phi')  =2 R,
\end{align}
and for the diagonal components of the Ricci tensor
\begin{align}
	&\frac{1}{\pi}\int_{0}^{2\pi}\mathrm{d}\phi'\,R'^\tau_{~\tau}(r',\theta',\phi')  =2 R^\tau_{~\tau},\nonumber\\
    &\frac{1}{\pi}\int_{0}^{2\pi}\mathrm{d}\phi'\,R'^x_{~x}(r',\theta',\phi')  =\frac{1}{\pi}\int_{0}^{2\pi}\mathrm{d}\phi'\,R'^y_{~y}(r',\theta',\phi')=R^r_{~r}+R^\phi_{~\phi}+(R^\theta_{~\theta}-R^r_{~r})\cos^{2}\theta'  ,\nonumber\\
    &\frac{1}{\pi}\int_{0}^{2\pi}\mathrm{d}\phi'\,R'^z_{~z}(r',\theta',\phi')  =2 R^\theta_{~\theta}-2(R^\theta_{~\theta}-R^r_{~r})\cos^{2}\theta'.
\end{align}
All the off-diagonal components of the Ricci tensor vanish upon integration in $\phi'$, because of spherical symmetry.
To compute the $\theta'$ integral we change the integration variable to $u=\cos\theta'$ and obtain
\begin{align}
	g_{1}(r´) &:=\int_{-1}^{1}\mathrm{d}u\,(r^{2}+r'{}^{2}-2rr'\,u)^{-3/2} =\left\{
	\begin{aligned}
		&\frac{2}{r(r-r')(r+r')} \quad&\text{ if }r'<r, \\
		&\frac{2}{r'(r'-r)(r+r')} \quad&\text{ if }r'> r,
	\end{aligned}\right.\label{eq:theta_int1}\\
	g_{2}(r´) &:=\int_{-1}^{1}u^2\mathrm{d}u\,(r^{2}+r'{}^{2}-2rr'\,u)^{-3/2} =\left\{
	\begin{aligned}
		&\frac{2(r^{2}+2r'{}^{2})}{3r^3(r-r')(r+r')} \quad&\text{ if }r'<r, \\
		&\frac{2(2r^{2}+r'{}^{2})}{3r'{}^{3}(r'-r)(r+r')} \quad&\text{ if }r'> r.
	\end{aligned}\right.
    \label{eq:theta_int2}
\end{align}
Putting together Eqs.~\eqref{eq:log_op_R}--\eqref{eq:theta_int2}, we obtain
\begin{align}
&\log\left(\frac{-\bar\Box}{\mu^2}\right) R =-\int_{0}^{\infty}\mathrm{d}r'{r'}^{2}R\,g_1,\nonumber\\
	&\log\left(\frac{-\bar\Box}{\mu^2}\right) R'^{\tau}_{~\tau} =-\int_{0}^{\infty}\mathrm{d}r'{r'}^{2}R^\tau_{~\tau}\,g_1,\nonumber\\
    &\log\left(\frac{-\bar\Box}{\mu^2}\right) R'^{x}_{~x} =-\frac12\int_{0}^{\infty}\mathrm{d}r'{r'}^{2}\left[(R^r_{~r}+R^\phi_{~\phi})\,g_1+(R^\theta_{~\theta}-R^r_{~r})\,g_2\right],\nonumber\\
    &\log\left(\frac{-\bar\Box}{\mu^2}\right) R'^{z}_{~z} =-\int_{0}^{\infty}\mathrm{d}r'{r'}^{2}\left[R^\theta_{~\theta}\,g_1-(R^\theta_{~\theta}-R^r_{~r})\,g_2\right].
\end{align}
Introducing Eq.s~\eqref{eq:ricci_tens_distr_linear} and~\eqref{eq:ricci_scal_distr_linear} into the previous expressions, the radial integrals are easily computed using the properties of the delta distribution. For instance, for the Ricci scalar, using Eq.~\eqref{eq:theta_int1}, we can write
\begin{align}
 \log\left(\frac{-\bar\Box}{\mu^2}\right)R =-\bar R \int_{0}^{r}\mathrm{d}r'\,r'{}^{2}\,\delta(r'-r_0)\frac{2}{r(r-r')(r+r')}-\bar R \int_{r}^{\infty}\mathrm{d}r'\,r'{}^{2}\,\delta(r'-r_0)\frac{2}{r'(r'-r)(r+r')}.
\end{align}

In the interior of the shell, where $r<r_0$, the first integral vanishes. Conversely, in the exterior, where $r>r_0$, the second integral vanishes. Hence, we find
\begin{align}
 \log\left(\frac{-\bar\Box}{\mu^2}\right)R =\begin{cases} -\frac{2 \bar  R \,r_0}{(r_0-r)(r+r_0)}, \qquad r<r_0,\\
-\frac{2 \bar R\,r_0^2}{r(r-r_0)(r+r_0)}, \qquad r>r_0. \end{cases}
\label{eq:logR_results}
\end{align}
A similar approach can be used to perform the radial integrals of the Ricci tensor components. The resulting tensor is spherically symmetric and diagonal in Cartesian coordinates, $\log\left(\frac{-\bar\Box}{\mu^2}\right) R'^{\alpha}_{~\beta}=\textrm{diag}(a,c,c,b)$. In spherical coordinates, it simply transforms to $\log\left(\frac{-\bar\Box}{\mu^2}\right) R^{\alpha}_{~\beta}=\textrm{diag}(a,b,c,c)$. The components in spherical coordinates are then
\begin{align}
    \log\left(\frac{-\bar\Box}{\mu^2}\right) R^{\tau}_{~\tau}=\left\{
    \begin{aligned}
        &-r_0\,\bar{R}^\tau_{~\tau}\frac{2}{(r_0-r)(r+r_0)},\quad\;\;\; r<r_0,\\
        &-r_0^2\,\bar{R}^\tau_{~\tau}\frac{2}{r(r-r_0)(r+r_0)},\quad r>r_0,
    \end{aligned}
    \right.
\end{align}
\begin{align}
    \log\left(\frac{-\bar\Box}{\mu^2}\right) R^{r}_{~r}=\left\{
    \begin{aligned}
        &-\frac43 \bar R^\theta_{~\theta}\frac{1}{r_0}-\frac23\bar{R}^r_{~r}\frac{2 r^2+r_0^2}{r_0\,(r_0-r)(r+r_0)},\quad\;\;\; r<r_0,\\
         &-\frac43 \bar R^\theta_{~\theta}\frac{r_0^2}{r^3}-\frac23\bar{R}^r_{~r}\frac{r_0^2}{r^3}\frac{2r_0^2+r^2}{(r-r_0)(r+r_0)},\quad r>r_0,
    \end{aligned}
    \right.
\end{align}
\begin{align}
    \log\left(\frac{-\bar\Box}{\mu^2}\right) R^{\theta}_{~\theta}=\left\{
    \begin{aligned}
        &-\frac23\bar{R}^\theta_{~\theta}\frac{2 r_0^2+r^2}{r_0\,(r_0-r)(r+r_0)}-\frac23 \bar R^r_{~r}\frac{1}{r_0},\quad\;\;\; r<r_0,\\
        &-\frac23\bar{R}^\theta_{~\theta}\frac{r_0^2}{r^3}\frac{r_0^2+2r^2}{(r-r_0)(r+r_0)}-\frac23 \bar R^r_{~r}\frac{r_0^2}{r^3},\quad r>r_0.
    \end{aligned}
    \right.
\end{align}
\end{widetext}

\bibliographystyle{apsrev4-1}
	\bibliography{bib}
\end{document}